\documentclass[reprint, amsmath, amssymb, aps, prl, floatfix, superscriptaddress
]{revtex4-2}
\pdfoutput=1

\usepackage{soul}

\usepackage[T1]{fontenc} 
\usepackage{lmodern}
\usepackage{amsmath}
\usepackage{array}
\usepackage{dsfont}
\usepackage{mathtools}
\usepackage{graphicx}
\usepackage{dcolumn}
\usepackage{bm}
\usepackage{hyperref}
\usepackage{physics}
\usepackage{empheq}
\usepackage{color}
\usepackage{multirow}
\usepackage{orcidlink}
\usepackage{afterpage}
\usepackage{mathrsfs}
\usepackage{tikz}
\usetikzlibrary{shapes}
\usetikzlibrary{tikzmark,calc}

\allowdisplaybreaks

\begin{document}

\title{Elastic Response and Instabilities of Anomalous Hall Crystals}

\author{F\'elix Desrochers \orcidlink{0000-0003-1211-901X}}
\email{felix.desrochers@mail.utoronto.ca}
\affiliation{%
Department of Physics, University of Toronto, Toronto, Ontario M5S 1A7, Canada
}%
\author{Mark R. Hirsbrunner \orcidlink{0000-0001-8115-6098}}
\affiliation{%
Department of Physics, University of Toronto, Toronto, Ontario M5S 1A7, Canada
}%
\author{Joe Huxford \orcidlink{0000-0002-4857-0091}}
\affiliation{%
Department of Physics, University of Toronto, Toronto, Ontario M5S 1A7, Canada
}%
\author{Adarsh S. Patri \orcidlink{0000-0002-7845-7823}}
\affiliation{
Department of Physics and Astronomy \& Stewart Blusson Quantum Matter Institute, University of British Columbia, Vancouver BC, Canada, V6T 1Z4
}
\author{T. Senthil}
\affiliation{
Department of Physics, Massachusetts Institute of Technology, Cambridge, Massachusetts 02139, USA
}
\author{Yong Baek Kim}%
\email{ybkim@physics.utoronto.ca}
\affiliation{%
 Department of Physics, University of Toronto, Toronto, Ontario M5S 1A7, Canada
}%

\date{\today}

\begin{abstract}
Anomalous Hall crystals (AHCs) are exotic phases of matter that simultaneously break continuous translation symmetry and exhibit the quantum anomalous Hall effect. AHCs have recently been proposed to explain the observation of an integer quantum anomalous Hall phase in a multilayer graphene system. Despite intense theoretical and experimental interest, little is known about the mechanical properties of AHCs. We study the elastic properties of AHCs first by using a continuum model with quadratic dispersion and uniform Berry curvature. We find using time-dependent Hartree-Fock that the stiffness of the AHC is an order of magnitude smaller than that of the WC, which we attribute to the finite Chern number of the AHC preventing exponential localization of the charge density. By modifying the dispersion relation to include a local minimum modeled after that of rhombohedral pentalayer graphene (R5G), we find that deformations away from the triangular lattice minimize the AHC's kinetic energy, which overwhelms the small stiffness and triggers a mechanical instability. Using a microscopic model of R5G, we observe a similar mechanical instability over an experimentally relevant parameter regime. We conclude that the topologically limited stiffness of AHCs makes them susceptible to mechanical instabilities, an important consideration when interpreting experiments in terms of AHCs.
\end{abstract}

\maketitle

\textit{Introduction}.---
It has long been known that strong interactions in electronic systems can spontaneously break continuous translation symmetry, leading to the formation of Wigner crystals~ \cite{wigner1934interaction, pines1952collective, bohm1953collective, tsui2024direct}. In the presence of external magnetic fields, such systems can also exhibit the quantum Hall effect, forming what is called a Hall crystal~\cite{halperin1986compatibility, kivelson1986cooperative, kivelson1987cooperative, tesanovic1989hall, tsui2024direct}. Comparatively little is understood about related phases that exhibit the quantum Hall effect with no external field, spontaneously breaking both translation and time-reversal symmetry. These systems, dubbed anomalous Hall crystals (AHCs), have become a topic of intense theoretical study~\cite{dong2024theory, zhou2024fractional, dong2024anomalous, soejima2024anomalous, dong2024stability, kwan2023moire, yu2024moire, tan2024parent, tan2024wavefunction, patri2024extended, zheng2024sublattice, zhou2024new, zeng2024berry} in the wake of recent experimental results on moir\'e platforms~\cite{lu2024fractional, lu2025extended, xie2024even, waters2024interplay, aronson2024displacement, choi2024electric}. In particular, the excitement follows from reports of the integer and fractional quantum anomalous Hall (IQAH/FQAH) effects in rhombohedral pentalayer graphene (R5G) slightly misaligned with a hexagonal boron nitride substrate (i.e., a R5G/hBN moir\'e heterostructure)~\cite{lu2024fractional}.

The IQAH is seen in these experiments when the first conduction band is filled ($\nu=1$ with respect to the moir\'e unit cell). The origin of this IQAH state is quite unconventional, as numerical studies show that the non-interacting band structure is metallic for experimentally relevant parameters. The isolated $|\mathcal{C}|=1$ Chern band only appears with the inclusion of the Coulomb interaction~\cite{arbeitman2024Moire, arbeitman2024Moire, kwan2023moire, huang2024self, Huang2025, guo2024theory, dong2024theory, dong2024anomalous, dong2024stability, soejima2024anomalous, zhou2024fractional}. Experimentally, the IQAH and FQAH phases are observed when the system is subjected to a strong displacement field that polarizes the conduction electrons away from the moir\'e potential induced by the hBN substrate. The spatial separation between the hBN and the conduction electrons calls into question the role of the moir\'e potential in stabilizing the IQAH effect. Indeed, Hartree-Fock (HF) calculations support the presence of an AHC in the absence of a moir\'e potential, wherein strong interactions break translation symmetry to induce the formation of a Chern band~\cite{dong2024anomalous, zhou2024fractional, dong2024theory, dong2024stability, arbeitman2024Moire, kwan2023moire}.

Despite the large body of recent work on AHCs, little is yet known about their mechanical properties. This is somewhat surprising, as it is clear even from earlier considerations that the elastic response of AHCs likely differs dramatically from that of conventional WCs~\cite{bonsall1977some, cote1991collective, chitra2001pinned, fogler2000dynamical, cote2008dynamical}. A conventional WC is a triangular lattice of exponentially localized charges whose localization increases with the interaction strength. In contrast, the finite Chern number of AHCs presents a topological obstruction to forming exponentially localized orbitals~\cite{marzari2012maximally, brouder2007exponential, li2024constraints}, suggesting that real-space density modulations, and thus the mechanical stiffness, may be weaker in AHCs than WCs. A further consequence of this obstruction is that the semi-classical arguments for the stability of the triangular lattice in WCs cannot be applied to AHCs~\cite{bonsall1977some}. To the contrary, recent theoretical works on R5G hinted that the triangular lattice AHC phase may be unstable to unit cell doubling deformations, both via study of the collective modes obtained through time-dependent Hartree-Fock~\cite{kwan2023moire} and by direct comparison with calculations on enlarged unit cells~\cite{zhou2024new}. However, a more comprehensive perspective beyond these specific instabilities is pressingly needed. 

\begin{figure}   
    \includegraphics[width=1.00\linewidth]{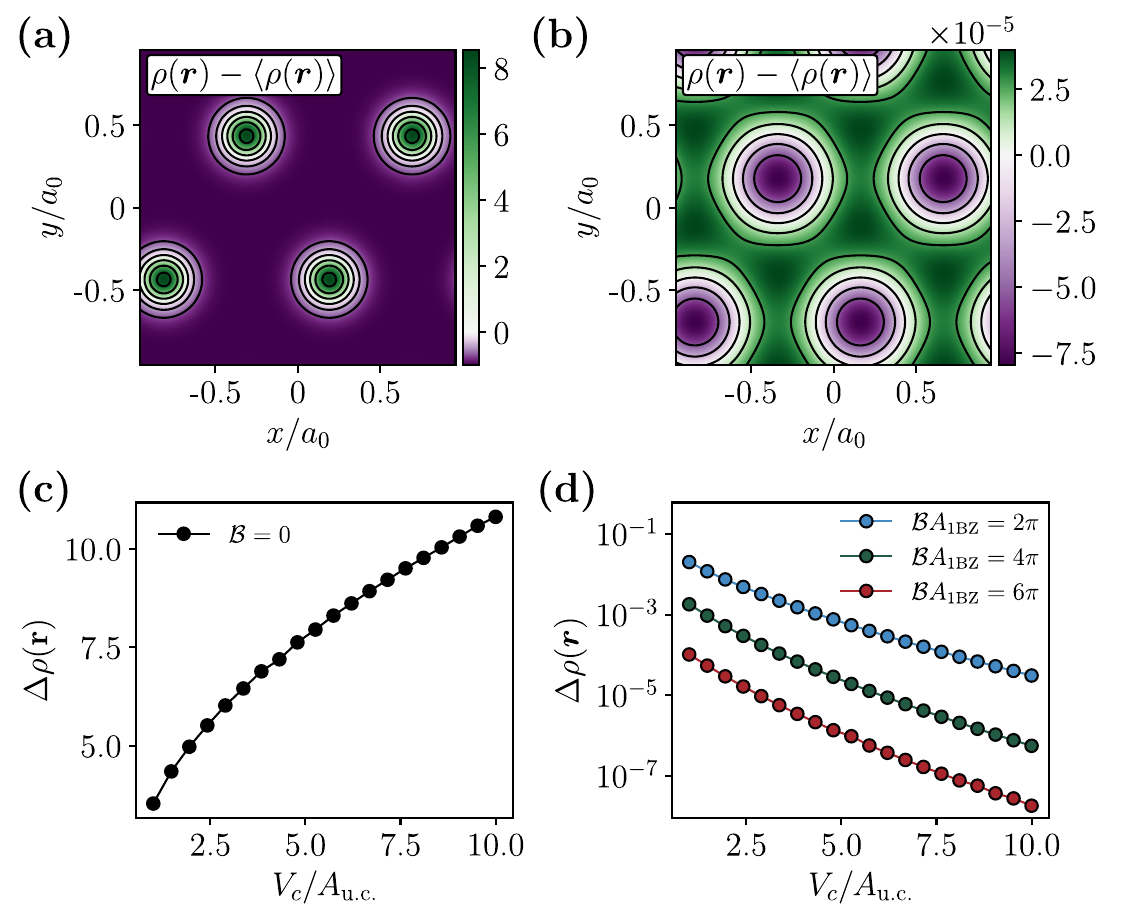}
    \caption{
    Representative real space charge density modulations for (a) the WC and (b) the $\mathcal{C}=1$ AHC, arising in the ideal parent band with $V_{c}/A_{\text{u.c.}}=7.63$. Maximum charge density variation $\Delta\rho(\boldsymbol{r})\equiv\max[\rho(\boldsymbol{r})] - \min[\rho(\boldsymbol{r})]$ in the HF ground state obtained by keeping the 97 closest reciprocal lattice points and with $n_1=23$ for (c) a WC $(\mathcal{B}=0)$ and (d) AHCs with $\mathcal{B}A_{1\text{BZ}}=2\pi$, $4\pi$, and $6\pi$.} 
    \label{fig:figure_1}
\end{figure}

In this letter, we study the elastic response of AHCs to lattice deformations. We first study AHCs in a simple ideal parent band continuum model of interacting electrons with a quadratic dispersion and constant Berry curvature~\cite{tan2024parent}. Using time-dependent Hartree-Fock (tdHF) numerics and analytical calculations based on a variational AHC ansatz~\cite{tan2024parent}, we conclude that the mechanical stiffness of the AHC is an order of magnitude weaker than the WC. This result indicates that small perturbations may drive the stiffness to a negative value, meaning that the AHC is susceptible to mechanical instabilities. To explore this possibility, we compute the stiffness of the parent band model with the quadratic dispersion replaced by one that emulates the low-energy dispersion of R5G. We show that the presence of a minimum in the dispersion near the Brillouin zone (BZ) boundary allows lattice deformations to reduce the kinetic energy. This change slightly reduces the stiffness of the WC, but makes the AHC mechanically unstable over a wide range of interaction strengths. We also apply a similar analysis to a realistic continuum model of R5G, finding again that the triangular lattice AHC is mechanically unstable over an experimentally relevant parameter regime. We conclude with a discussion of the implications of these results and important topics for future research.

\textit{Model}.---We first consider a minimal Hamiltonian that describes spin- and valley-polarized electrons projected into a single continuum parent band, $\mathcal{H}=\mathcal{H}_0 + \mathcal{H}_{\mathrm{int}}$, where the kinetic term $\mathcal{H}_0=\sum_{\boldsymbol{k}} c_{\boldsymbol{k}}^{\dagger} \mathcal{E}(\boldsymbol{k}) c_{\boldsymbol{k}}$ has a quadratic dispersion $\mathcal{E}(\boldsymbol{k})=|\boldsymbol{k}|^2/2 m$. The $c_{\boldsymbol{k}}^{\dagger}$ operator creates an electron with unbounded momentum $\boldsymbol{k}$ in the parent band (i.e., $c_{\boldsymbol{k}}^{\dagger}\ket{0} = |\boldsymbol{k}\rangle=e^{i \boldsymbol{k} \cdot \boldsymbol{r}}\left|s_{\boldsymbol{k}}\right\rangle$, with $\ket{s_{\boldsymbol{k}}}$ describing internal degrees of freedom).
The electrons interact through a band-projected density-density term of the form $\mathcal{H}_{\mathrm{int}}=\frac{1}{2 A} \sum_{\boldsymbol{k}_1 \boldsymbol{k}_2 \boldsymbol{k}_3 \boldsymbol{k}_4} \tilde{V}_{\boldsymbol{k}_1 \boldsymbol{k}_2 \boldsymbol{k}_3 \boldsymbol{k}_4} c_{\boldsymbol{k}_1}^{\dagger} c_{\boldsymbol{k}_2}^{\dagger} c_{\boldsymbol{k}_3} c_{\boldsymbol{k}_4}$,
where $A$ is the area of the system and $\tilde{V}_{\boldsymbol{k}_1 \boldsymbol{k}_2 \boldsymbol{k}_3 \boldsymbol{k}_4}= V\left(\boldsymbol{k}_1-\boldsymbol{k}_4\right) \mathcal{F}\left(\boldsymbol{k}_1, \boldsymbol{k}_4\right) \mathcal{F}\left(\boldsymbol{k}_2, \boldsymbol{k}_3\right) \delta_{\boldsymbol{k}_1+\boldsymbol{k}_2-\boldsymbol{k}_3-\boldsymbol{k}_4}$. We consider the unscreened Coulomb potential $V(\boldsymbol{q})=V_c/|\boldsymbol{q}|$, and the form factors $\mathcal{F}(\boldsymbol{k},\boldsymbol{q})$ entering the projected Coulomb interaction are formally given by $\mathcal{F}\left(\boldsymbol{k}, \boldsymbol{q}\right) = \braket{s_{\boldsymbol{k}}}{s_{\boldsymbol{q}}}$. They encode the quantum geometry of the band and are taken to be
\begin{align}
    \mathcal{F}\left(\boldsymbol{k}, \boldsymbol{q}\right)=\exp[-\frac{\mathcal{B}}{4}\left(\left|\boldsymbol{k} - \boldsymbol{q}\right|^2+2 i \boldsymbol{k}\times \boldsymbol{q}\right)], \label{eq:form_factor_parent_band}
\end{align}
where $\boldsymbol{k} \times \boldsymbol{q} \equiv k_x q_y-k_y q_x$~\cite{tan2024parent}. This choice of form factor corresponds to a band with uniform Berry curvature $\mathcal{B}(\boldsymbol{k}) = \mathcal{B}$ and a Fubini-Study metric $g_{\mu \nu}^{\mathrm{FS}}(\boldsymbol{k}) = \frac{1}{2} \mathcal{B} \delta_{\mu \nu}$ that saturates the trace $\operatorname{Tr}\left[g_{\mu \nu}^{\mathrm{FS}}(\boldsymbol{k})\right] \geq |\mathcal{B}(\boldsymbol{k})|$ condition bound~\cite{parameswaran2013fractional, yu2024quantum}. We note that the parent band form factor~\eqref{eq:form_factor_parent_band} is the same as for the lowest Landau level (LLL) with magnetic length $\ell_{B}^2=\mathcal{B}$, making the parent band model a dispersive analog of the LLL with unrestricted momentum~\cite{tan2024parent}.

\begin{figure*}[t]
    \centering
    \includegraphics[width=1.0\textwidth]{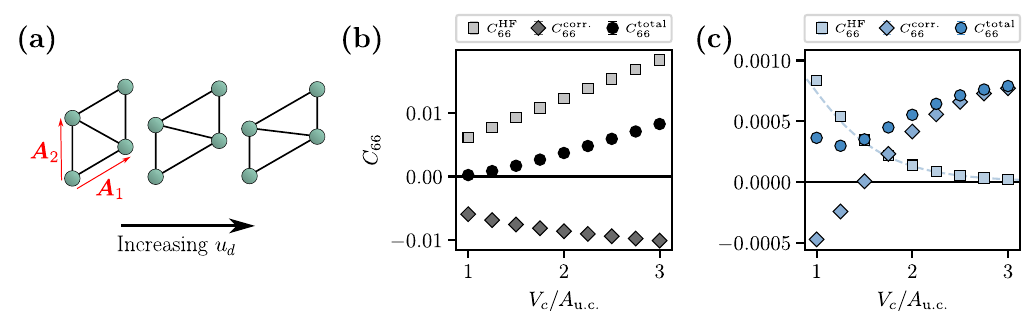}\vspace{-3mm}
    \caption{(a) A depiction of the dilation deformation of the triangular lattice employed to compute the stiffness. The stiffness of (b) the WC with $\mathcal{B}=0$ and (c) the AHC with $\mathcal{B}=2\pi$, both as a function of the interaction strength, with the HF and correlation energy contributions depicted individually, in addition to the total stiffness. The dashed line corresponds to a perturbative expression for the HF ground state energy exact in the large-$V_c$ limit.}
\label{fig:fig_2}
\end{figure*}

When $\mathcal{B}=0$, the parent band model describes the usual two-dimensional electron gas and exhibits a transition from a Fermi liquid to a WC for strong interactions. In contrast, if a sufficiently large Berry flux threads the first BZ formed by the resulting crystal, the Fermi liquid instead transitions to an AHC with Chern number given by the integer nearest to $\mathcal{B} A_{\text{1BZ}}/2\pi$, where $A_{\text{1BZ}}$ is the area of the first BZ. This nearest integer rounding of the Berry curvature can be understood in terms of a Berry-flux quantization condition~\cite{dong2024stability} (see supplemental material~\footnote{See Supplemental Material [INSERT URL] for details on the elastic modulus tensor, Hartree-Fock and time-dependent Hartree-Fock implementations, analytical AHC energy derivations in the strong interaction regime, and additional numerical results for rhombohedral pentalayer graphene, which includes Refs.~\cite{kresse1996efficient, pulay1980convergence, pulay1982improved, rohwedder2011analysis, khalafSoftModesMagic2020, kwanExcitonBandTopology2021, kwanMoireFractionalChern2023, roweEquationsofMotionMethodExtended1968, jain2025elementary, wolfQuasibosonApproximationYields2024, Bateman1953, Park2023topological, zhang2010band, jung2014abinitio, jung2015origin}.}). The unit cell and first BZ area of the AHC are determined by the electronic density, such that there is one electron per unit cell (i.e., filling unity $\nu=1$).

This idealized model is a useful approximation for spin- and valley-polarized systems with a low electronic density, in which the atomic BZ is irrelevant, and the Berry curvature perceived by electrons near the band edge is approximately constant. Although highly simplified, it offers an analytically tractable model that can be compared with numerical calculations employing more realistic models. In what follows, we set $m=1/2$ and the length of the reciprocal lattice vectors to unity (i.e., $|\boldsymbol{G}_{1,\triangle}|=1$) such that energy is measured in units of $|\boldsymbol{G}_{1,\triangle}|^2/2 m$.

We plot in Fig.~\ref{fig:figure_1} (a, b) the ground state charge density of the WC $(\mathcal{B}A_{\text{1BZ}}=0)$ and AHC $(\mathcal{B}A_{\text{1BZ}}=2\pi)$ phases of the parent band model obtained via self-consistent Hartree-Fock (see supplemental material~\cite{Note1}). The charge density variation of the WC is orders of magnitude larger than that of the AHCs, confirming that the topological obstruction to constructing maximally localized Wannier functions drastically reduces real-space density modulations of the AHCs. Furthermore, the overall spatial patterns obtained are dissimilar: the WC forms a triangular network of localized charges (Fig.~\ref{fig:figure_1}(a)), whereas the AHCs form a honeycomb structure (Fig.~\ref{fig:figure_1}(b))~\cite{zheng2024sublattice, dong2024anomalous, zhou2024fractional}. In Fig.~\ref{fig:figure_1}(c, d), we plot the charge density variation as a function of $V_c$ for the WC $(\mathcal{B}A_{\text{1BZ}}=0)$ and a range of AHC phases $(\mathcal{B}A_{\text{1BZ}}=2\pi,\,4\pi,\,6\pi)$, showing that the charge becomes more localized in the WC as the interaction strength increases, but the opposite occurs in the AHCs. The low-charge-density modulation is further exacerbated for larger Chern numbers. These findings indicate that WCs and AHCs may have disparate mechanical properties. In the next section, we specifically study how the mechanical stiffness of each phase differs.

\textit{Parent band stiffness}.---To compute the mechanical stiffness of WCs and AHCs in the parent band model, we assume the system crystallizes in a triangular lattice with basis vectors $\boldsymbol{A}_1 = 2\pi(1, 1/\sqrt{3})$ and $\boldsymbol{A}_2 = 2\pi(0,2/\sqrt{3})$, such that the lattice site positions are $\boldsymbol{R}= m \boldsymbol{A}_1 + n \boldsymbol{A}_2$ ($m,n\in\mathbb{Z}$). We apply deformations to the lattice of the form $\boldsymbol{R}'=\boldsymbol{R} + \boldsymbol{u}(\boldsymbol{r}) = m \boldsymbol{A}_1' + n\boldsymbol{A}_2'$, where $\boldsymbol{A}_1'$ and $\boldsymbol{A}_2'$ are the basis vectors of the deformed lattice, and study how the ground state energy per electron varies as a function of the deformation. The second-order derivatives of the ground state energy per electron with respect to deformations can be directly related to the elastic coefficients that appear in the usual long-wavelength description of deformable media~\cite{cote2008dynamical, landau2012theory}. The $C_6$ point group symmetry of the triangular lattice restricts the elastic modulus matrix to have only two independent elements, $C_{12}$ and $C_{66}$. We do not consider the area non-preserving deformations required to access $C_{12}$, and instead study the response to area-preserving dilations of the form $\boldsymbol{A}_1'=(1 + u_d) \boldsymbol{A}_1$ and $\boldsymbol{A}_2'=(1+u_d)^{-1} \boldsymbol{A}_2$ (see Fig.~\ref{fig:fig_2}(a)). From these deformations we obtain the $C_{66}$ elastic coefficient as $C_{66} = 3n_0/16 \left(\partial^2 f/\partial u_d^2\right)$, where $n_0=N/A$ is the electron density and $f=E_0/N$ is the ground state energy per electron (see supplemental material~\cite{Note1}). Since we consider only a single elastic coefficient, we refer to $C_{66}$ as the stiffness.

Figs.~\ref{fig:fig_2} (b) and (c) present the stiffness of topologically trivial (WC) and non-trivial (AHC) crystals arising in the parent band model at $\mathcal{B}A_{\text{1BZ}}=0$ and $\mathcal{B}A_{\text{1BZ}}=2\pi$, respectively. We compute the ground state energy via HF, incorporating the Madelung energy to alleviate finite-size effects arising from the long-range Coulomb interaction~\cite{Madelung}. The momentum cutoff required for the energy to converge in this approach grows rapidly as $V_c$ is increased, so we supplement this with a perturbative approach (dashed lines in Fig.~\ref{fig:fig_2} (c, e)) that is valid at large interaction strengths (see supplemental material~\cite{Note1}). We also obtain the correlation energy by further calculating the spectrum of excitations above the ground state via time-dependent Hartree-Fock (tdHF) (see supplemental material~\cite{Note1}). To clarify the role that the correlation energy plays, we separately compute the contributions to the stiffness coming from the HF and correlation energies, $C_{66}^{\text{HF}}$ and $C_{66}^{\text{corr}}$, respectively, in addition to the total stiffness $C_{66}$.

First, analyzing the stiffness at the HF level, we see the stiffness of the WC, shown in Fig.~\ref{fig:fig_2}(b), increases with $V_c$, as is classically expected~\cite{bonsall1977some, cote2008dynamical}. In contrast, the stiffness of the AHC with $\mathcal{B}A_{\text{1BZ}} = 2\pi$ is orders of magnitude weaker than for the WC, as demonstrated in panel (c) of Fig.~\ref{fig:fig_2}. More strikingly, the stiffness of the AHCs unexpectedly decreases asymptotically to zero as the interaction strength increases. 

This implies that the energy difference between different lattices also approaches zero, which we confirm by computing the HF ground state energy difference between the triangle and square lattices for AHCs in the End Matter.  This is in stark contrast to the same energy comparison for the Wigner crystal, where the energy difference increases with the interaction strength (see End Matter). From an elastic point of view, the AHC thus becomes more ``fluid-like'' as interactions increase. We emphasize that even in this limit where all lattice shapes are degenerate, the system remains an electronic crystal with a finite charge gap, distinct from a Fermi liquid. We also point out that the stiffness is greater for larger Chern number AHCs because the band-projected interaction $\tilde{V}_{\boldsymbol{k}_1\boldsymbol{k}_2\boldsymbol{k}_3\boldsymbol{k}_4}$ is more strongly suppressed at larger Berry curvature as a result of the Gaussian prefactor in the form factors~\eqref{eq:form_factor_parent_band}.

This decaying stiffness can be qualitatively understood by noting that the trace condition violation, bandwidth, and Berry curvature variation all decrease with stronger interactions in the AHC~\cite{tan2024parent}. Indeed, the energetics of the parent band model is dominated by the Fock term, which is minimized when the trace condition violation of the filled HF band is reduced~\cite{tan2024parent, abouelkomsanquantum2023}. Therefore, strong interactions drive the system to the \emph{ideal flatband} limit~\cite{claassen2015position, wang2021exact, ledwidth2023vortexability, estienne2023ideal}. This, combined with the vanishing Berry curvature fluctuations, indicates that the emergent HF ground state closely resembles a filled Landau level, i.e., a quantum Hall fluid \cite{parameswaran2013fractional, Qi2011, Parameswaran2012, Roy2014}. However, going beyond mean-field, stronger interactions will couple the HF ground state to other states more strongly, and the true ground state will no longer be described by a filled Landau level Slater determinant.

Moving on to consider the contribution of the correlation energy to the stiffness, $C_{66}^{\text{corr.}},$ we again see the WC and AHC behave drastically differently. For the WC, plotted in Fig.~\ref{fig:fig_2} (b), $C_{66}^{\text{corr.}}$ is negative and decreases with increasing interaction strength. In contrast, $C_{66}^{\text{corr.}}$ for the AHC changes sign from negative to positive and appears to level off as $V_c$ increases, shown in Fig.~\ref{fig:fig_2} (c). The reduction of the stiffness for the WC is expected, as HF famously underestimates the interaction strength at which the two-dimensional electron gas crystallizes, and the ground state at these intermediate interaction strengths should actually be a liquid~\cite{PhysRevB.39.5005, rapisarda1996diffusion}. However, it appears that HF instead underestimates the stiffness of the AHC, which is stabilized by correlations to a finite value at large $V_c$.

Although the AHC does not flow to a state with vanishing stiffness at large interaction strengths when accounting for correlations, the total stiffness of the AHC is still an order of magnitude smaller than that of the WC. As such, we can conclude that the simple intuition provided in the introduction holds when considering quantum fluctuations: the topological obstruction limits the stiffness of AHCs. This limited stiffness may mean that small perturbations away from the ideal parent band model can drive a mechanical instability. We next study the stiffness of AHCs in the context of rhombohedral pentalayer graphene~\cite{dong2024anomalous, soejima2024anomalous, dong2024theory, dong2024stability, zhou2024fractional, kwan2023moire}.

\begin{figure}
    \centering
    \includegraphics[width=0.95\linewidth]{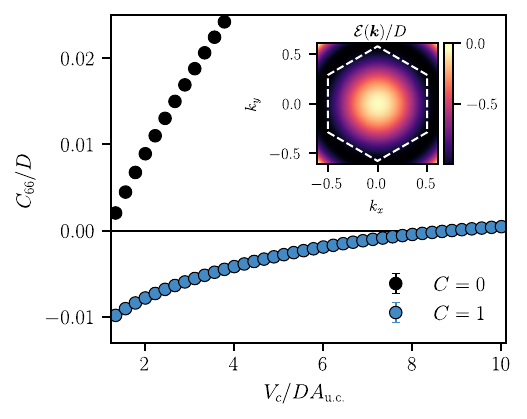}
    \caption{The stiffness of the WC with $\mathcal{B}A_{\text{1BZ}}=0$ (black) and the AHC with $\mathcal{B}A_{\text{1BZ}}=2\pi$ (blue) arising in the parent band model with a Mexican hat dispersion as a function of interaction strength. The minimum of the dispersion is located at $|\boldsymbol{k}|=0.6$ (i.e., $k_0=0.6$), and both $V_c$ and $C_{66}$ are scaled by $D$. The inset shows the dispersion as a function of momentum, with the 1BZ depicted by the dashed lines.}
\label{fig:figure_3}
\end{figure}

\textit{Kinetic driven instability}.---One of the ways that R5G differs from the ideal parent band model is in its dispersion, which possesses a distinct local minimum along a ring located at some finite momentum, as shown in Fig.~\ref{fig:figure_3} and the End Matter. We can study the effect of dispersion on stiffness in isolation by modifying the dispersion of the parent band while keeping the form factors unchanged. To do so, we replace the quadratic dispersion of the parent band model with $\mathcal{E}(\boldsymbol{k})=D\left(\left(|\boldsymbol{k}|/k_0 \right)^4 - \left(|\boldsymbol{k}|/k_0\right)^2\right)$ which possesses a local minimum of depth $D$ at $|\boldsymbol{k}|=k_0$. The band minimum in R5G occurs just outside the 1BZ for displacement fields that realize the IQAH, so we set $k_0=0.6|\boldsymbol{G}|$. We show in the supplemental material that the kinetic energy can be reduced by deforming the lattice to access a larger portion of the dispersion minimum, which may drive an instability. Indeed, we see that the AHC is unstable over a large parameter range while the WC remains stable, as shown by the HF stiffness plotted in Fig.~\ref{fig:figure_3}. This result indicates that minimizing the kinetic energy overwhelms the small stiffness of the AHC except for large values of $V_c/D$, where the potential energy dominates the physics. We note that whenever the mechanical stiffness at the HF level is negative, the tdHF spectrum will be complex. Therefore, we cannot compute the stiffness arising from the correlation energy for this system. To confirm the pertinence of the above analysis, we present in the End Matter a numerical calculation of the AHC stiffness based on a microscopic model of R5G in an experimentally relevant parameter regime. Our analysis reveals a similar instability of the triangular lattice driven by kinetic energy.

\textit{Discussion}.---We showed that AHCs in the ideal parent band model have a much weaker mechanical stiffness than conventional WCs, confirming the intuition that the topological obstruction of the Chern number limits the stiffness. The small stiffness indicates that mechanical instabilities of triangular lattice AHCs may arise beyond the ideal limit. Indeed, we confirm the presence of such an instability in a toy model with a Mexican hat dispersion and in a realistic microscopic model of rhombohedral multilayer graphene with a strong displacement field. Despite the specificity of the models we studied, our results are quite general, relying only on the finite Chern number of the AHC. The weak mechanical stiffness of AHCs also indirectly implies a low speed of sound and an overall low-energy phonon spectrum. These low-energy collective modes may have a sizeable entropic contribution at finite temperatures that could be important for understanding the thermal crossover (or transition) from the IQAH to the FQAH in R5G/hBN~\cite{lu2025extended, xie2024integer, huang2024impurity, dassarma2024thermal, shavit2024Entropy, wei2025edge, patri2024extended}.

Several other important questions remain to be addressed in future studies. The recent observation of the IQAH over an extended range of filling and displacement fields in R5G/hBN~\cite{lu2025extended} further motivates the study of density-varying deformations beyond the area-preserving transformations we focused on. Studying the response of AHCs to such deformations, either in the ideal limit or with more realistic models, would clarify the competition between the elastic and commensuration energies in the presence of a periodic potential, which is crucial for interpreting the experiment~\cite{patri2024extended}. We also note that it may be possible to tune the stability of the putative AHC in R5G by varying the displacement field, given its kinetic origin and the displacement-field dependence of R5G's dispersion. Furthermore, investigating such distortions should help evaluate the possibility of stabilizing fractional anomalous Hall crystals, as recently proposed in the parent-band model~\cite{tan2024wavefunction}.

\begin{acknowledgments}
We thank Trithep Devakul, Tixuan Tan, Junkai Dong, Ashvin Viswanath, Daniel Parker, Tomohiro Soejima, Yves H. Kwan, and B. Andrei Bernevig for insightful discussions. We acknowledge support from the Natural Sciences and Engineering Research Council of Canada (NSERC) and the Center of Quantum Materials at the University of Toronto. Computations were performed on the Cedar and Fir clusters hosted by the Digital Research Alliance of Canada. F.D. is further supported by the Vanier Canada Graduate Scholarship (CGV-186886). A.S.P. is supported by NSERC, CIFAR, and the Gordon and Betty Moore Foundation’s EPiQS Initiative through Grant No. GBMF11071 at the University of British Columbia.
\end{acknowledgments}

%

\clearpage
\onecolumngrid
\section*{End Matter}
\twocolumngrid

\begin{figure}[b]
    \centering
    \includegraphics[width=1.0\linewidth]{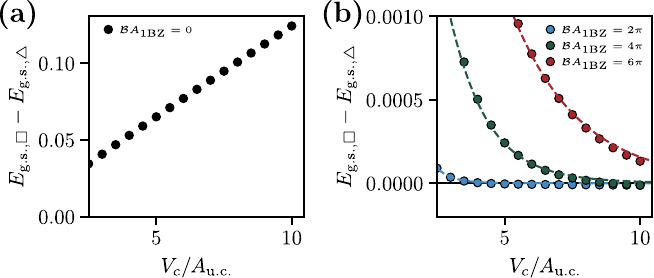}\vspace{-3mm}
    \caption{The difference between the HF ground state energy of the (a) WC with $\mathcal{B}=0$ and (b) AHCs with $\mathcal{B}=2\pi,$ $4\pi,$ and $6\pi$ on the square and triangular lattices, as a function of interaction strength. The dashed lines in (b) utilize the same perturbative expansion as in Fig.~\ref{fig:fig_2}(c).}
\label{fig:fig_energy_diff_lattice}
\end{figure}

\textit{Energy difference between lattices}.---We have mentioned in the main text that the energy difference between AHCs with different lattices decreases with interaction strength for the ideal parent band. To show this convincingly, we present in Fig.~\ref{fig:fig_energy_diff_lattice}(b) the HF ground state energy difference between the triangle and square lattices for AHCs with $\mathcal{B}A_{\text{1BZ}} = 2\pi$, $4\pi$, and $6\pi$. We see that the energy difference decreases monotonically with interactions for $\mathcal{C}=1$, $2$, and $3$. We note that the square lattice even appears lower in energy than the triangular lattice (Fig.~\ref{fig:fig_2}(e)) at large interaction strengths, but this is an artifact arising from the finite momentum cutoff. Indeed, the perturbative calculation predicts that the triangular lattice always remains stable and lower in energy than the square lattice for large $V_c$. Computing the stiffness of the square lattice also shows that it is unstable to shear deformations towards the triangular lattice, for both the WC and AHC. This behavior of the energy difference for AHC is in stark contrast with the same comparison for the Wigner crystal (Fig.~\ref{fig:fig_energy_diff_lattice}(a)). In this case, the energy difference grows with interaction strength.

\begin{figure}
\includegraphics[width=1.00\linewidth]{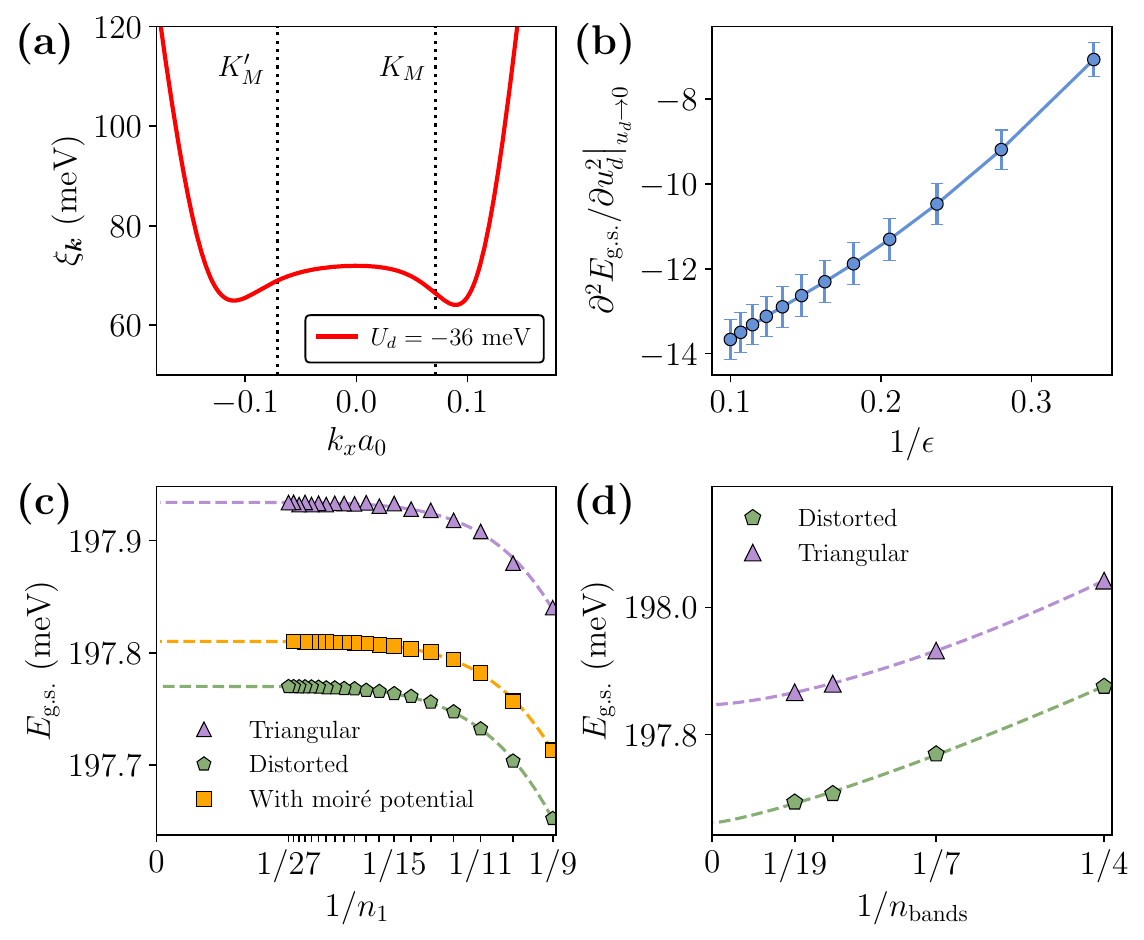}
\caption{(a) Continuum R5G dispersion in a strong displacement field corresponding to an interlayer potential difference of $U_d=-36$~meV. The vertical dashed lines indicate the location of the $K$ point of the 1BZ that spontaneously emerges when the triangular lattice crystal forms. (b) Dilation stiffness of the R5G AHC as a function of the inverse dielectric constant calculated with $n_1=23$, $U_d=-36$~meV, and $n_{\text{bands}}=7$. Error bars are from uncertainties in evaluating the second-order derivative. (c) Convergence with respect to system size $n_1 \cross n_1$ (for $n_{\text{bands}}=7$) of the ground state energy per conduction electron on undistorted ($u_d=0$) and distorted ($u_d=-0.15$) triangular lattices for $U_d=-36$~meV and $\epsilon=8.07$. The same convergence is also shown in the presence of a moiré potential induced by a hexagonal boron nitride substrate on the bottom layer with twist angle $\theta=0.77^\circ$. (d) Convergence of the same quantity as (c) but with respect to the number of conduction bands $n_{\text{bands}}$ (for $n_1=23$).
}
\label{fig:figure_4}
\end{figure}

\textit{Rhombohedral pentalayer graphene}.---To further study possible instabilities of AHCs, we compute the HF stiffness of the AHC arising in R5G using an explicit microscopic model (see supplemental material~\cite{Note1} for details). We model the displacement field as a layer potential $U_d$ and use a dual-gated screened interaction
$V_c^{\mathrm{sc}}(\boldsymbol{q})= e^2 \tanh \left(|\boldsymbol{q}| d_s\right)/(2 \epsilon_0 \epsilon |\boldsymbol{q}|)$, where $\epsilon$ is the dielectric constant and $d_s$ the distance separating the metallic gates. We focus on the experimentally relevant parameter regime, i.e., a strong displacement field $U_d=-36$~meV, electronic density consistent with a filled moiré conduction band ($\nu=1$), and a twist angle of $\theta=0.77^\circ$. Our HF calculations assume spin-valley polarization, retain only the lowest $n_{\text{bands}}$ conduction bands, and use the triangular lattice with an orientation that minimizes the ground state energy as the starting point. This orientation is found to respect the $C_3$ symmetry of the underlying microscopic model (see supplemental material~\cite{Note1} for details).

Fig.~\ref{fig:figure_4}(b) shows the evolution of $C_{66}^{\text{HF}}$ with the interaction strength (controlled by the inverse dielectric constant $1/\epsilon$). The stiffness is negative for the entire range of $\epsilon$ considered, signaling the mechanical instability of the triangular lattice. To verify that this instability is not due to finite-size effects, we compare the ground state energy of the triangular lattice ($u_d=0$) and a $C_3$-symmetry breaking dilated triangular lattice ($u_d=-0.15$) for a range of system sizes (Fig.~\ref{fig:figure_4}(c)) and number of conduction bands (Fig.~\ref{fig:figure_4}(d)). The ground state energy of the distorted lattice is always smaller, even when extrapolated to the limit of infinite system size or number of bands. This confirms the mechanical instability of the previously assumed triangular lattice AHC in R5G for an experimentally relevant parameter regime within the HF approximation. In the supplemental material, we consider how the kinetic, Hartree, and Fock energies change as functions of $u_d$, and further show that lowering the displacement field such that the local minimum of the dispersion disappears stabilizes the triangular lattice AHC. We note that even when including a moiré potential induced by a hexagonal boron nitride substrate on the bottom layer (with a twist angle of $\theta = 0.77^\circ$) and assuming a weakly pinned triangular AHC aligned with the moiré lattice, the deformed lattice configuration remains energetically favorable (Fig.~\ref{fig:figure_4}(c)). Taken together, these results confirm that the instability is driven by the kinetic energy, as predicted by our analysis of the modified parent band model.

\end{document}


\title{Supplemental Material for \linebreak
``Elastic Response and Instabilities of Anomalous Hall Crystals''}

\author{F\'elix Desrochers \orcidlink{0000-0003-1211-901X}}
\email{felix.desrochers@mail.utoronto.ca}
\affiliation{%
Department of Physics, University of Toronto, Toronto, Ontario M5S 1A7, Canada
}%
\author{Mark R. Hirsbrunner \orcidlink{0000-0001-8115-6098}}
\affiliation{%
Department of Physics, University of Toronto, Toronto, Ontario M5S 1A7, Canada
}%
\author{Joe Huxford \orcidlink{0000-0002-4857-0091}}
\affiliation{%
Department of Physics, University of Toronto, Toronto, Ontario M5S 1A7, Canada
}%
\author{Adarsh S. Patri \orcidlink{0000-0002-7845-7823}}
\affiliation{
Department of Physics and Astronomy \& Stewart Blusson Quantum Matter Institute, University of British Columbia, Vancouver BC, Canada, V6T 1Z4
}
\author{T. Senthil}
\affiliation{
Department of Physics, Massachusetts Institute of Technology, Cambridge, Massachusetts 02139, USA
}
\author{Yong Baek Kim}%
\email{ybkim@physics.utoronto.ca}
\affiliation{%
 Department of Physics, University of Toronto, Toronto, Ontario M5S 1A7, Canada
}%

\date{\today}

\maketitle

\tableofcontents

\setcounter{secnumdepth}{3}
\setcounter{equation}{0}
\renewcommand{\theequation}{S\arabic{equation}}
\setcounter{table}{0}
\renewcommand{\thetable}{S\arabic{table}}
\setcounter{figure}{0}
\renewcommand{\thefigure}{S\arabic{figure}}

\section{Mechanical response} \label{si_sec:Mechanical_response_schfa}

In this section, we comment on how the shear and dilation stiffnesses defined in the main text are related to the usual elastic coefficients that appear in the long wavelength description of a deformable medium. The deformation energy in the continuum limit can be written as 
\begin{align}
        \Delta E = \frac{1}{2} C_{abcd} \varepsilon_{ab} \varepsilon_{cd},
\end{align}
where the symmetric strain tensor is defined as
\begin{align}
    \varepsilon_{ab}(\boldsymbol{r}) &= \frac{1}{2}\left( \pdv{u_a(\boldsymbol{r})}{r_b} + \pdv{u_{b}(\boldsymbol{r})}{r_{a}} \right)
\end{align}
with the displacement vector $\boldsymbol{u}(\boldsymbol{r})$ and $a, b\in\{x,y\}$. The elastic modulus tensor (or stiffness tensor) $C_{abcd}$ must satisfy the generic symmetry constraints $C_{abcd}=C_{bacd}=C_{abdc}=C_{cdab}$, such that there are only six independent components ($C_{xxxx},C_{yyyy},C_{xyxy},C_{xxyy},C_{xxxy},C_{yyxy}$) in two-dimensions. Using Voigt notation, the deformation energy can then be written concisely as~\cite{landau2012theory} 
\begin{align}
    \Delta E &= \frac{1}{2} \int d^2 r \mqty(\varepsilon_{xx}, \varepsilon_{yy}, 2\varepsilon_{xy}) \mqty(C_{11} & C_{12} & C_{16} \\ C_{12} & C_{22} & C_{26} \\ C_{16} & C_{26} & C_{66}) \mqty(\varepsilon_{xx} \\ \varepsilon_{yy} \\ 2\varepsilon_{xy}),
\end{align}
where $C_{11}=C_{xxxx}$, $C_{22}=C_{yyyy}$, $C_{12}=C_{xxyy}$, $C_{16}=C_{xxxy}$, $C_{26}=C_{yyxy}$, and $C_{66}=C_{xyxy}$. The $C_6$ point group symmetry imposes that~\cite{landau2012theory, cote2008dynamical}
\begin{subequations}    
\begin{align}
    C_{26} &= C_{16} = 0 \\
    C_{11} &= C_{22} = 2C_{66} + C_{12}.
\end{align}
\end{subequations}
 The elastic energy can then be written using only two stiffness coefficients
\begin{align}
    \Delta E &= \frac{1}{2} \int d^2 r \mqty(\varepsilon_{xx}, \varepsilon_{yy}, 2 \varepsilon_{xy}) \mqty(2C_{66} + C_{12} & C_{12} & 0\\ C_{12} & 2C_{66} + C_{12} & 0 \\ 0 & 0 & C_{66}) \mqty(\varepsilon_{xx} \\ \varepsilon_{yy} \\ 2 \varepsilon_{xy}) \label{eq:deformation_energy_triangle}.
\end{align}
We also note that the lattice structure is stable if the elastic modulus matrix is positive definite. That is, the triangular lattice is stable if
\begin{subequations}
    \begin{align}
        C_{66} > 0
    \end{align}
and 
    \begin{align}
            C_{12} + C_{66} &>0.
    \end{align}
\end{subequations}

The coefficients can be extracted by computing the ground state energy as a function of the deformation strength for specific distortions. Below, we derive the explicit relation between the elastic coefficients and the curvature of the deformation energy for shear and area-preserving dilations. 

We first discuss how to parameterize lattice deformations. The basis vectors for a generic two-dimensional lattice can be written as 
\begin{subequations}
    \begin{align}
        \boldsymbol{A}_1 &= a_0 \eta \mqty(\sin(\varphi), \cos(\varphi)) \\
        \boldsymbol{A}_2 &= a_0 \mqty(0,1).
    \end{align}
\end{subequations}
For the triangular lattice, we have $\eta=1$ and $\varphi=\pi/3$. The associated basis vectors of the reciprocal lattice are
\begin{subequations}
    \begin{align}
        \boldsymbol{G}_1 &= \frac{2\pi}{a_0\eta} \left( \csc(\varphi),0 \right) \\
        \boldsymbol{G}_2 &=  \frac{2\pi}{a_0} \mqty(-\cot(\varphi), 1).
    \end{align}
\end{subequations}
Suppose the initial lattice sites $\boldsymbol{R}=m \boldsymbol{A}_1 + n \boldsymbol{A}_2$ ($m,n\in\mathbb{Z}$) are displaced by $\boldsymbol{u}(\boldsymbol{r})$. The new sites of the deformed lattice are $\boldsymbol{R}'=m\boldsymbol{A}_1 + n \boldsymbol{A}_2 + \boldsymbol{u}(\boldsymbol{r})$, which can also be expressed as $\boldsymbol{R}'=m\boldsymbol{A}_1' + n \boldsymbol{A}_2'$, where we have introduced the basis vectors for the deformed lattice
\begin{subequations}
    \begin{align}
        \boldsymbol{A}_1' &= a_0' \eta' \mqty(\sin(\varphi'), \cos(\varphi')) \\
        \boldsymbol{A}_2' &= a_0' \mqty(0,1).
    \end{align}
\end{subequations}
The corresponding reciprocal lattice vectors of the deformed lattice are 
\begin{subequations}
    \begin{align}
        \boldsymbol{G}_1' &= \frac{2\pi}{a_0'\eta'} \left( \csc(\varphi'),0 \right) \\
        \boldsymbol{G}_2' &=  \frac{2\pi}{a_0'} \mqty(-\cot(\varphi'), 1). 
    \end{align}
\end{subequations}
A deformation can thus be parameterized by the evolution of $a_0'$, $\eta'$, and $\varphi'$ as a function of the deformation strength $u_0$. For instance, a shear deformation of the form $u_x(\boldsymbol{r})=0$ and $u_{y}(\boldsymbol{r})=u_s x$ (for a lattice site $\boldsymbol{R}$, $x$ is defined as $\hat{x}\cdot\mathbf{R}$) leads to 
\begin{subequations}
\begin{align}
    a_0^{\prime}&=a_0 \\
    \eta^{\prime}&=\eta \sqrt{1 + 2 u_s \sin (\varphi) \cos (\varphi) + u_s^2 \sin ^2(\varphi)} \\
    \sin \left(\varphi^{\prime}\right)&=\frac{\sin (\varphi)}{\sqrt{1 + 2 u_s \sin (\varphi) \cos (\varphi) + u_s^2 \sin ^2(\varphi)}}.
\end{align}
\end{subequations}
Using the above parameterization, the symmetric strain tensor components are $\varepsilon_{xx}=\varepsilon_{yy} =0$ and $\varepsilon_{xy}=u_s/2$. Making this replacement in Eq.~\eqref{eq:deformation_energy_triangle}, the deformation energy for a shear deformation is $\Delta E_{\text{shear}} = A u_s^2 C_{66}/2$. Defining the deformation energy per electron as $f = \Delta E/N$, we then see that the shear stiffness defined in the main text is related to $C_{66}$ by
\begin{align}
    \left. \pdv[2]{f_{\text{shear}}}{u_s} \right|_{u_s\to 0} = n_0^{-1} C_{66},
\end{align}
where the electronic density is $n_0 = N/A$.

In addition to shear deformations, we also study area-preserving dilations of the form
\begin{equation}
    \begin{aligned}
        \boldsymbol{A}_1' &= (1+u_d) \boldsymbol{A}_1 \\
        \boldsymbol{A}_2' &= (1+u_d)^{-1} \boldsymbol{A}_2,
    \end{aligned}
\end{equation}
that can be parameterized by 
\begin{subequations}
\begin{align}
    a_0^{\prime} &= a_0/(1+u_d) \\
    \eta^{\prime} &= (1+u_d)^2\eta  \\
    \sin \left(\varphi^{\prime}\right) &= \sin (\varphi).
\end{align}
\end{subequations}
The displacement vector then takes the form
\begin{align}
    u_x(\boldsymbol{r}) = u_d x, \quad u_{y}(\boldsymbol{r}) = \frac{u_d (2 + u_d)\cot{\varphi}}{1 + u_d}  x - \frac{u_d}{1+u_d} y,
\end{align}
such that the symmetric strain tensor components are
\begin{subequations}
    \begin{align}
        \varepsilon_{xx} &= u_d \\
        \varepsilon_{yy} &= -\frac{u_d}{1+u_d} \\
        \varepsilon_{xy} &= \frac{u_d (2 + u_d)}{2(1 + u_d)} \cot(\varphi).
    \end{align}
\end{subequations}
The corresponding deformation energy is 
\begin{align}
    \Delta E_{\text{dilation}} &= A \frac{u_d^2}{2 (1+u_d)^2} \left( (C_{12} + C_{66})u_d^2 + C_{66}(2 + u_d)^2 \csc^2(\varphi)\right),
\end{align}
which yields
\begin{align}
    \left. \pdv[2]{f_{\text{dilation}}}{u_d} \right|_{u_d\to 0} = 4 n_0^{-1} \csc^2(\varphi) C_{66} = \frac{16}{3 n_0} C_{66}.
\end{align}
Consequently, the curvature of the deformation energy for area-preserving dilations is also determined by $C_{66}$ for the triangular lattice.

\section{Hartree-Fock calculations of the parent band model} \label{si_sec:Hartree_fock}

\subsection{Hartree-Fock decoupling} \label{si_subsec:Hartree_fock_decoupling}

The Hartree-Fock approximation is a variational approach over the space of Slater determinant states. It amounts to a mean-field treatment of the quartic interaction term that leads to the Hartree and Fock terms
\begin{subequations}\label{eq:hf_term_parent_band}
    \begin{align}
        \mathcal{H}_{\mathrm{H}} &= \frac{1}{A} \sum_{\substack{\boldsymbol{k}_1 \boldsymbol{k}_2 \\\boldsymbol{g}_1 \boldsymbol{g}_2 \boldsymbol{g}_3 \boldsymbol{g}_4}} V\left(\boldsymbol{g}_1-\boldsymbol{g}_4 \right)\mathcal{F}\left(\boldsymbol{k}_1+\boldsymbol{g}_1, \boldsymbol{k}_1+\boldsymbol{g}_4\right) \mathcal{F}\left(\boldsymbol{k}_2+\boldsymbol{g}_2, \boldsymbol{k}_2+\boldsymbol{g}_3\right) \nonumber \\
        &\hspace{3cm} \times \delta(\boldsymbol{g}_1 + \boldsymbol{g}_2 - \boldsymbol{g}_3 -\boldsymbol{g}_4) \mathcal{P}_{\boldsymbol{g}_1 \boldsymbol{g}_4}(\boldsymbol{k}_1) c_{\boldsymbol{k}_2 \boldsymbol{g}_2}^{\dagger} c_{\boldsymbol{k}_2 \boldsymbol{g}_3}  \\ 
        \mathcal{H}_{\mathrm{F}} &= -\frac{1}{A} \sum_{\substack{\boldsymbol{k}_1 \boldsymbol{k}_2 \\\boldsymbol{g}_1 \boldsymbol{g}_2 \boldsymbol{g}_3 \boldsymbol{g}_4}} V\left(\boldsymbol{k}_1+\boldsymbol{g}_1-\boldsymbol{k}_2-\boldsymbol{g}_4 \right) \mathcal{F}\left(\boldsymbol{k}_1+\boldsymbol{g}_1, \boldsymbol{k}_2+\boldsymbol{g}_4\right) \mathcal{F}\left(\boldsymbol{k}_2+\boldsymbol{g}_2, \boldsymbol{k}_1+\boldsymbol{g}_3\right) \nonumber \\
        &\hspace{3cm} \times \delta(\boldsymbol{g}_1+\boldsymbol{g}_2-\boldsymbol{g}_3 -\boldsymbol{g}_4) 
        \mathcal{P}_{\boldsymbol{g}_1 \boldsymbol{g}_3}(\boldsymbol{k}_1) c_{\boldsymbol{k}_2 \boldsymbol{g}_2}^{\dagger} c_{\boldsymbol{k}_2 \boldsymbol{g}_4},
    \end{align}
\end{subequations}
where the density matrix
\begin{align}
    \mathcal{P}_{\boldsymbol{g}_1\boldsymbol{g}_2}\left(\boldsymbol{k}\right)= \left\langle c_{\boldsymbol{k} \boldsymbol{g}_{1}}^{\dagger} c_{\boldsymbol{k} \boldsymbol{g}_{2}}\right\rangle \label{eq:density_matrix_scc}
\end{align}
is in a one-to-one correspondence with Slater determinant states.

Following the approach used in Refs. \cite{tan2024parent, tan2024wavefunction}, we remove the long-ranged part of the Coulomb interaction by excluding $V(\boldsymbol{q} = 0)$ from the momentum sum (and do the same for the sum over $\boldsymbol{g}$ in the Hartree term). The $\boldsymbol{q} = 0$ gives a contribution $V(0)(N^2 -N)/2A$, which is irrelevant in our study since we always keep the electronic density constant. We follow this prescription throughout our analysis using Hartree-Fock and the variational ansatz. 

\subsection{Details about the Hartree-Fock numerics} \label{si_subsec:Hartree_fock_numerics}

\begin{figure}
 \includegraphics[width=0.95\linewidth]{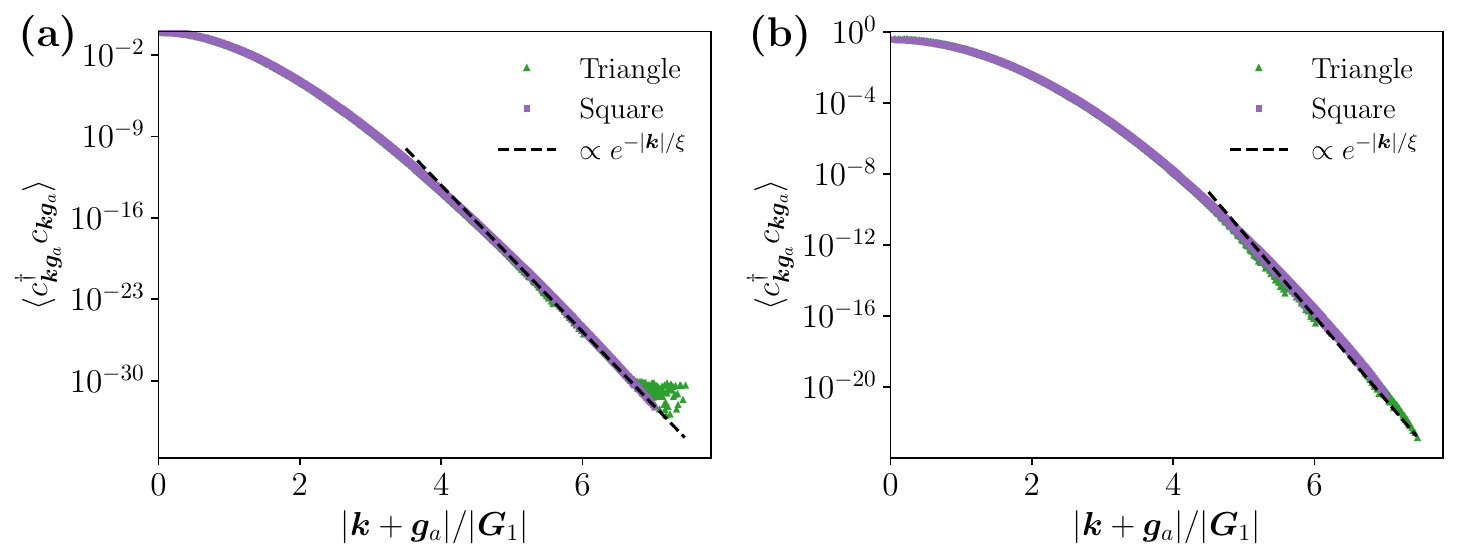}
\caption{Momentum space occupation for the square and triangular lattice with (a) $V_{c}/A_{\text{u.c.}} = 2$ and (b) $V_{c}/A_{\text{u.c.}} = 8$. Simulations are done with $n_1=n_2=15$ by keeping 125 and 129 reciprocal lattice points for the triangular and square lattice, respectively. The occupation is $\mathcal{O}(1)$ at the first Brillouin zone center and decays exponentially for large momentum. \label{si_fig:density_matrix_exp_decay}}
\end{figure}

To find the optimal density matrix, one has to solve for $\mathcal{P}_{\boldsymbol{g}_1\boldsymbol{g}_2}\left(\boldsymbol{k}\right)$ self-consistently. In our case, we numerically solve the self-consistency equation~\eqref{eq:density_matrix_scc}. To do so, the first Brillouin zone is discretized as
\begin{align}
    \boldsymbol{k}=\frac{i}{n_1} \boldsymbol{G}_1+\frac{j}{n_2} \boldsymbol{G}_2 
\end{align}
where $i\in\{0,1,2,\ldots n_1-1\}$, and $j\in\{0,1,2,\ldots n_2-1\}$. The kinetic, Hartree, and Fock terms are constructed by including all reciprocal lattice points $\boldsymbol{g}=a\boldsymbol{G}_1 + b \boldsymbol{G}_2$ ($a,b\in \mathbb{Z}$) within a cutoff $|\boldsymbol{g}| \le \Lambda |\boldsymbol{G}_1|$. Our simulations include the $n_{\Gamma}=97$ closest reciprocal lattice points ($\Lambda\approx 5$). With these values, we find good convergence of the self-consistency conditions and ground state energy (see Sec.~\ref{si_subsec:Hartree_fock_finite_size_scaling}). As illustrated in Fig.~\ref{si_fig:density_matrix_exp_decay}(a), the density matrix occupation for the furthermost reciprocal lattice points with such cutoffs in the crystalline phases is usually less than $10^{-30}$ for $V_c/A_{\text{u.c.}}=2$. It decays more slowly when interactions increase (Fig.~\ref{si_fig:density_matrix_exp_decay}(b)).

\begin{figure}
\includegraphics[width=0.95\linewidth]{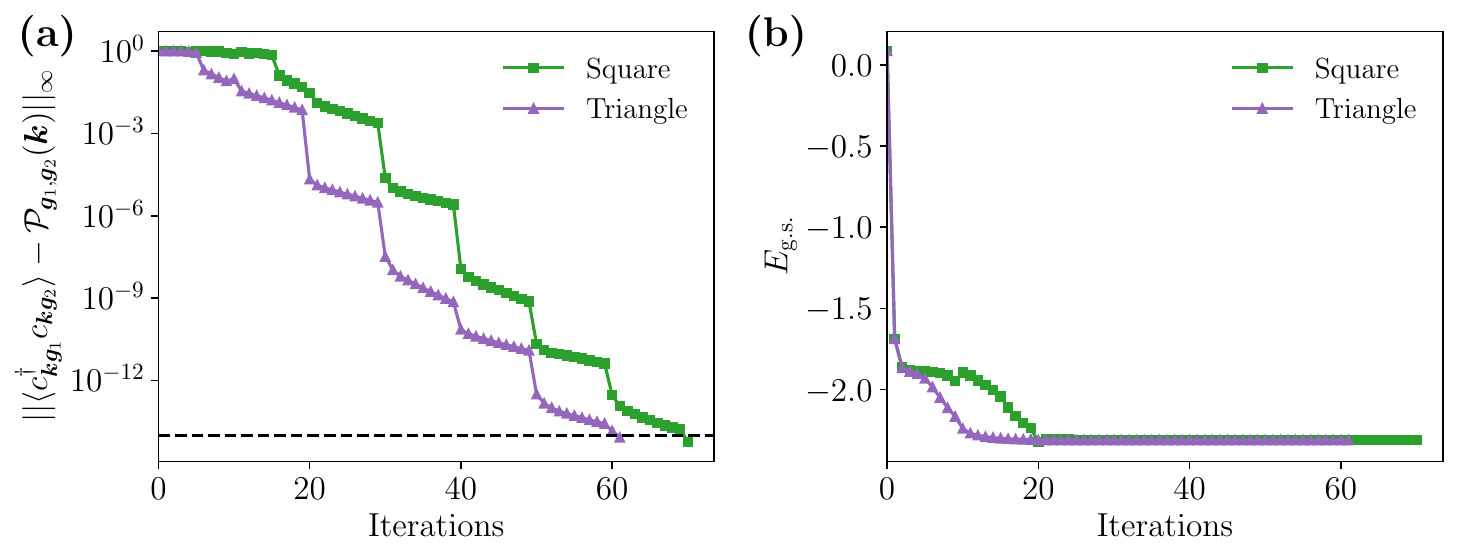}
\caption{Typical evolution of (a) the residual norm and (b) ground state energy per particle (excluding the Madelung energy) when solving the self-consistency equations using periodic Pulay mixing for the square and triangle lattices. The results are in the AHC phase with $\mathcal{B}=2\pi/A_{\text{1BZ}}$, $V_{c}/A_{\text{u.c.}} = 2$ and $n_1=n_2=21$. \label{si_fig:convergence_scc}}
\end{figure}

To solve the self-consistency conditions, we randomly initialize a density matrix $\mathcal{P}^{(0)}_{\boldsymbol{g}_1\boldsymbol{g}_2}\left(\boldsymbol{k}\right)$ and update it using periodic Pulay mixing, a method also known as periodic direct inversion of the iterative subspace (DIIS) \cite{pulay1980convergence, pulay1982improved, kresse1996efficient,rohwedder2011analysis}. At every iteration, the residual $\rho^{(n)}_{\boldsymbol{g}_1\boldsymbol{g}_2}(\boldsymbol{k})$ is evaluated
\begin{align}
    \rho^{(n)}_{\boldsymbol{g}_1\boldsymbol{g}_2}(\boldsymbol{k}) &= \left\langle c_{\boldsymbol{k} \boldsymbol{g}_{1}}^{\dagger} c_{\boldsymbol{k} \boldsymbol{g}_{2}}\right\rangle^{(n)} - \mathcal{P}^{(n)}_{\boldsymbol{g}_1\boldsymbol{g}_2}(\boldsymbol{k}),
\end{align}
where $\left\langle A \right\rangle^{(n)}$ denotes an average computed from the ground state of the Hartree-Fock Hamiltonian~\eqref{eq:hf_term_parent_band} with density matrix $\mathcal{P}^{(n)}_{\boldsymbol{g}_1\boldsymbol{g}_2}(\boldsymbol{k})$. 
The density matrix used for the next iteration is then computed using simple mixing
\begin{align}
    \mathcal{P}^{(n+1)}_{\boldsymbol{g}_1\boldsymbol{g}_2}(\boldsymbol{k}) &= \mathcal{P}^{(n)}_{\boldsymbol{g}_1\boldsymbol{g}_2}(\boldsymbol{k}) + \alpha_{\text{mixing}} \rho^{(n)}_{\boldsymbol{g}_1\boldsymbol{g}_2}(\boldsymbol{k}),
\end{align}
where $\alpha_{\text{mixing}} \in(0,1]$. However, after every $k_{\text{diis}}$ steps, the new density matrix is instead evaluated using DIIS. It is given by a linear combination of the $n_{\text{diis}}$ previous steps 
\begin{align}
    \mathcal{P}^{(n+1)}_{\boldsymbol{g}_1\boldsymbol{g}_2}(\boldsymbol{k}) = c_n \mathcal{P}^{(n)}_{\boldsymbol{g}_1\boldsymbol{g}_2}(\boldsymbol{k}) + c_{n-1} \mathcal{P}^{(n-1)}_{\boldsymbol{g}_1\boldsymbol{g}_2}(\boldsymbol{k}) + \ldots + c_{n-n_{\text{diis}}} \mathcal{P}^{(n-n_{\text{diis}})}_{\boldsymbol{g}_1\boldsymbol{g}_2}(\boldsymbol{k})
\end{align}
that minimizes the Euclidian norm of $\sum_{i=0}^{n_{\text{diis}}} c_{n-i} \rho^{(n-i)}_{\boldsymbol{g}_1\boldsymbol{g}_2}(\boldsymbol{k})$ subject to the normalization constraint $\sum_{i=0}^{n_{\text{diis}}}c_{n-i}=1$. This is achieved by solving the linear system of equations
\begin{align}
\mqty(
    B_{n,n} & B_{n,n-1} &  \ldots & B_{n, n-n_{\text{diis}}} & -1 \\
    B_{n-1,n} & B_{n-1,n-1}  & \ldots & B_{n-1, n-n_{\text{diis}}} & -1 \\
    \vdots & \vdots & \ddots & \vdots & \vdots \\
    B_{n-n_{\text{diis}},n} & B_{n-n_{\text{diis}},n-1}  & \ldots & B_{n-n_{\text{diis}},n-n_{\text{diis}}} & -1 \\
    1 & 1 & 1 & \ldots &  0
)
\mqty(
    c_{n} \\ c_{n-1} \\c_{n-2} \\\vdots \\c_{n-n_{\text{diis}}} \\\lambda
) =
\mqty(
    0 \\ 0 \\ 0 \\ \vdots \\ 0 \\ 1
),
\end{align}
where 
\begin{align}
    B_{i,j}= \sum_{\boldsymbol{k} \boldsymbol{g}_1 \boldsymbol{g}_2} (\rho^{(i)}_{\boldsymbol{g}_1\boldsymbol{g}_2}(\boldsymbol{k}))^* \rho^{(j)}_{\boldsymbol{g}_2\boldsymbol{g}_1}(\boldsymbol{k}).
\end{align}
The iteration is stopped when the infinity (or maximum) norm of the residual array is smaller than a threshold $\lambda_{\text{thresh}}$
\begin{align}
    \norm{\rho^{(n)}_{\boldsymbol{g}_1\boldsymbol{g}_2}(\boldsymbol{k})}_{\infty} = \max(|\rho^{(n)}_{\boldsymbol{g}_1\boldsymbol{g}_2}(\boldsymbol{k})|) \le \lambda_{\text{thresh}}.
\end{align}
In this work, we use $\alpha_{\text{mixing}} =0.9$, $k_{\text{diis}}=10$, $n_{\text{diis}}=5$ and a threshold of $\lambda_{\text{thresh}}=10^{-14}$. A typical evolution of the residual norm and ground state energy when solving the self-consistency conditions is shown in Fig.~\ref{si_fig:convergence_scc}.

\subsection{Convergence} \label{si_subsec:Hartree_fock_finite_size_scaling}

\begin{figure}
\includegraphics[width=0.99\linewidth]{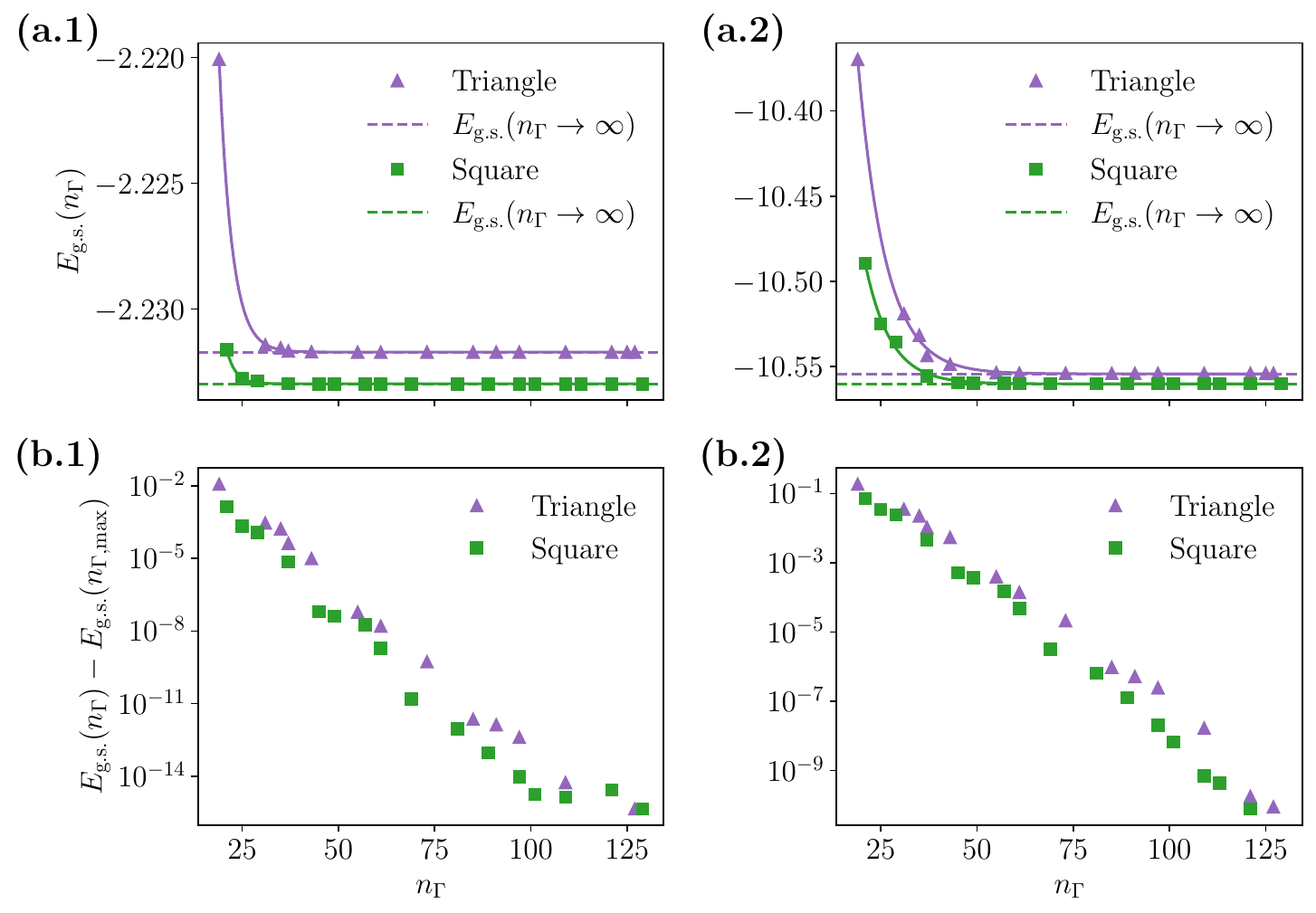}
\caption{Scaling of the ground state energy per particle as a function of the number of reciprocal lattice points included $n_\Gamma$ in the anomalous Hall crystal phase ($\mathcal{B}=2\pi/A_{1\text{BZ}}$) for the triangular and square lattices on a (a) linear and (b) logarithmic scale for (1) $V_{c}/A_{\text{u.c.}}=2$ and (2) $V_{c}/A_{\text{u.c.}}=8$. Simulations are done with $n_1=n_2=15$. The square lattice only appears more stable than the triangle because of the finite $n_1$ and because we did not include the Madelung energy. \label{si_fig:convergence_gs_energy_cutoff}}
\end{figure}

Our numerical approach is limited in accuracy by the finite momentum cutoff $\Lambda$, or equivalently, the finite number of reciprocal lattice points $n_\Gamma$ included. Fig.~\ref{si_fig:convergence_gs_energy_cutoff} shows the evolution of the ground state energy per particle as $n_\Gamma$ increased for the triangular and square lattice in the AHC phase with $\mathcal{B}A_{\text{1BZ}}=2\pi$. We see that the ground state energy converges quickly with $n_\Gamma$. More precisely, as is clear from the panels (b.1)-(b.2) that are displayed on a logarithmic scale, the ground state energy per particle decays exponentially with $n_\Gamma$ as 
\begin{align}
    E_{\text{g.s.}}(n_{\Gamma}) &= E_{\text{g.s.}}(n_{\Gamma}\to\infty) - C e^{-D n_{\Gamma}}. \label{eq:finite_size_scaling_cutoff}
\end{align}

From the figure, it can be observed that one needs to include more reciprocal lattice points to achieve a similar convergence of the ground state energy at larger interaction strengths. This is simply because the density matrix decays more slowly in momentum space for larger $V_c$ (see Fig.~\ref{si_fig:density_matrix_exp_decay}). For $n_{\Gamma}=97$ (which is the number of reciprocal lattice points included for both the triangle and square lattice in the main text), the energy difference with the infinite cutoff extrapolated energy (i.e., $E_{\text{g.s.}}(n_{\Gamma}\to\infty)$) is on the order of $10^{-14}$ for $V_c/A_{\text{u.c.}}=2$ and $10^{-9}$ for $V_c/A_{\text{u.c.}}=8$. Those energy differences are significantly smaller than those involved in the comparison between the triangular and square lattice ground state energies presented in the main text. The cutoffs used should thus be sufficiently large so as not to affect the reliability of our conclusions.

\begin{figure}
\includegraphics[width=0.4\linewidth]{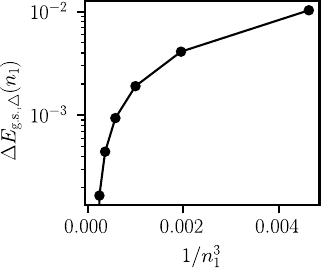}
\caption{The change in the HF ground state energy as the system size increases, scaling as $1/n_1^3$. The black line is a guide to the eye, and $\Delta E_{\mathrm{g.s.}}(n_1) = E_{\mathrm{g.s.}}(n_1 + 1) - E_{\mathrm{g.s.}}(n_1)$ \label{si_fig:convergence_gs_energy_n1}}
\end{figure}

Next, calculations of the ground state energy also suffer from finite-size effects arising from the finite linear system size $n_1$. This typically leads to the ground state energy converging slowly as $1/n_1$, due to the long-range Coulomb interaction. To alleviate these effects, we incorporate the Madelung energy~\cite{Madelung}. Incorporating the Madelung contribution yields excellent convergence even for small system size, as seen in Fig.~\ref{si_fig:convergence_gs_energy_n1}.

\subsection{Hartree-Fock Phase diagram} \label{si_subsec:Hartree_fock_phase_diagram}

\begin{figure*}
    \centering
    \includegraphics[width=1.0\textwidth]{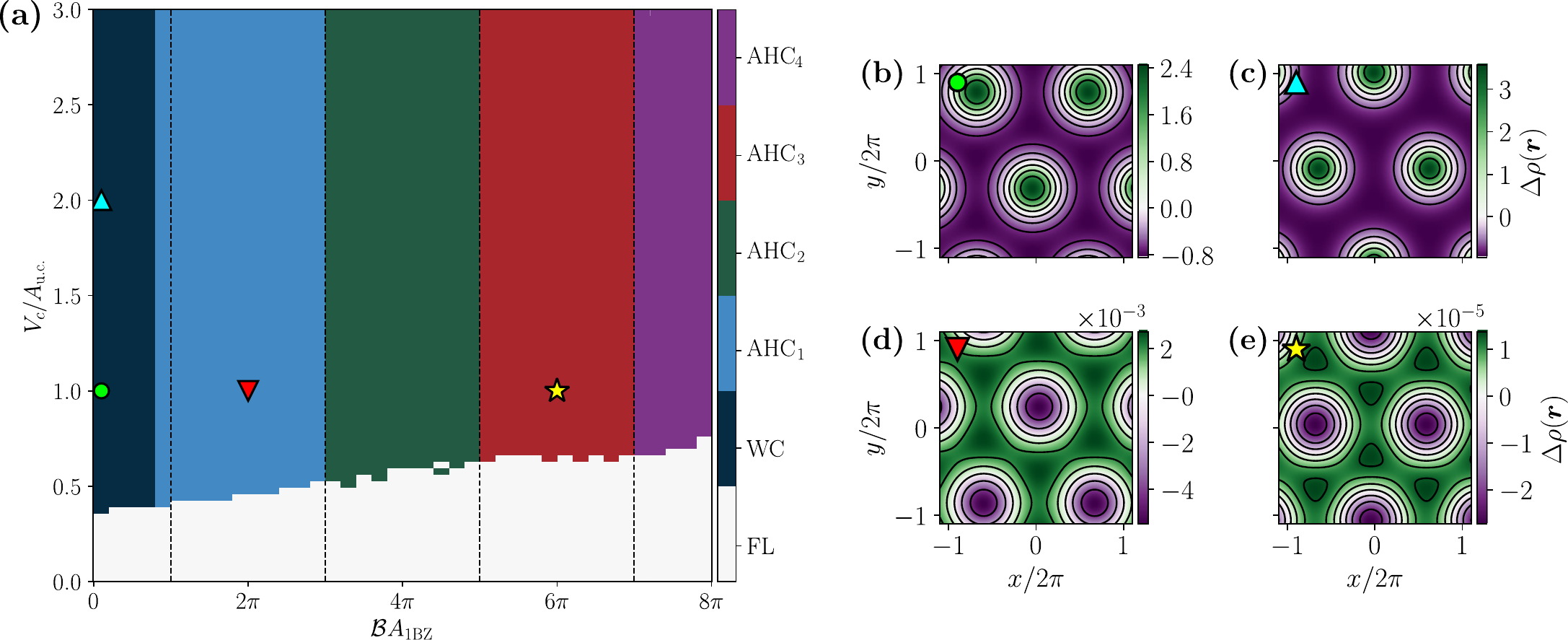}
    \caption{(a) Hartree-Fock phase diagram of the parent band model obtained by keeping $n_{\Gamma}=61$ reciprocal lattice points and a linear system size of $n_1=21$. Triangular lattice AHC with Chern number $\mathcal{C}$ are denoted by AHC$_{\mathcal{C}}$. Representative real space charge density variation for the (b)-(c) WC, (d) AHC$_{1}$, and (e) AHC$_{3}$.} 
\label{si_fig:phase_diagram}
\end{figure*}

As supplemental results, we present in Fig.~\ref{si_fig:phase_diagram} a large HF phase diagram obtained by keeping the $n_{\Gamma}=61$ closest reciprocal lattice points and a finite system size of $21\times 21$. This phase diagram shows the transition from the Fermi liquid (FL) to the WC/AHC as the interaction is increased. It also shows the transition from the WC to the AHC with $\mathcal{C}=1$ and between AHC with different Chern numbers. Those transitions happen when the closest integer to $\mathcal{B}A_{\text{1BZ}}/(2\pi)$ changes. 
This `rounding' of the Berry curvature to the nearest integer was previously addressed in Ref.~\cite{dong2024stability}, where the Fock energy term is recast into a momentum space analog of a narrow superconducting ring in a background magnetic field, with the crystal order parameter and Berry curvature of the parent band taking the role of the superconducting order parameter and magnetic field, respectively. The subsequent `rounding' of the Berry curvature is understood as the momentum-space analog of the flux-quantization condition.

\section{Time-dependent Hartree-Fock} \label{si_sec:tdHF}
Time-dependent Hartree-Fock (tdHF) is a method for studying neutral particle-hole excitations above the Hartree-Fock ground state~\cite{roweEquationsofMotionMethodExtended1968, khalafSoftModesMagic2020, kwanExcitonBandTopology2021, kwanMoireFractionalChern2023}. The creation operators for such excitations are defined as
\begin{equation}
    Q^\dagger_{\nu\boldsymbol{q}} = \sum_{\varphi, \boldsymbol{k}} \left(X^{\nu\boldsymbol{q}}_{\varphi}(\boldsymbol{k}) b^\dagger_{\varphi,\boldsymbol{q}}(\boldsymbol{k}) - Y^{\nu\boldsymbol{q}}_{\varphi}(\boldsymbol{k}) b^{\phantom{\dagger}}_{\varphi,-\boldsymbol{q}}(\boldsymbol{k})\right),
\end{equation}
where $\varphi$ labels pairs of particle and hole bands $(\phi_p, \phi_h)$, $b^\dagger_{\varphi,\boldsymbol{q}}(\boldsymbol{k})=\eta^\dagger_{\phi_p}(\boldsymbol{k}+\boldsymbol{q})\eta_{\phi_h}(\boldsymbol{k})$ is a creation operator for an elementary particle-hole pair with momentum $\boldsymbol{q}$ and $\eta^\dagger_{\phi}(\boldsymbol{k})$ is the creation operator for a HF eigenstate. Here, our notation allows multiple occupied (hole) bands, but in practice, we consider only a single occupied band. Below, we describe how to construct the tdHF equations from which the energies of these modes can be obtained, and show how to compute the correlation energy contributed by these modes.

\begin{figure}
    \includegraphics[width=0.95\linewidth]{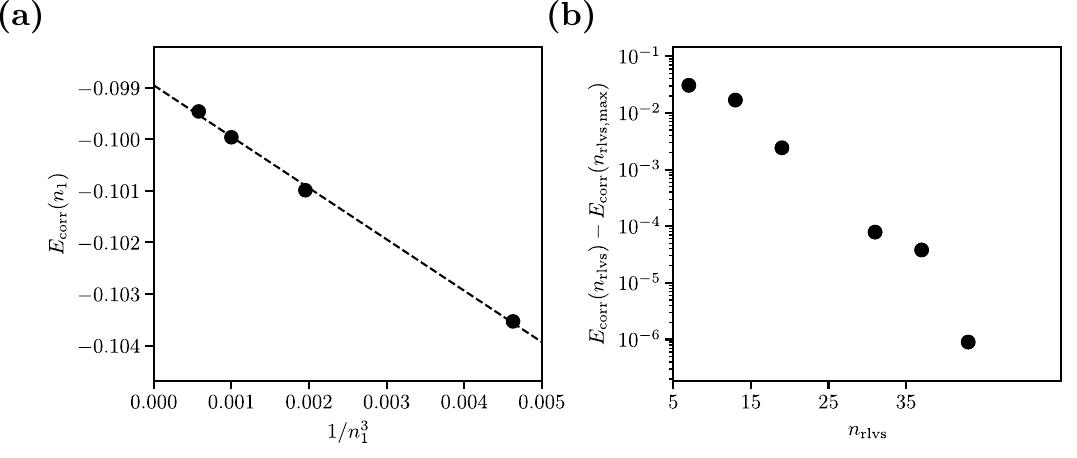}
    \caption{(a) The scaling of the correlation energy with respect to the system size $n_1$ with $n_{\Gamma}=55$. (b) The scaling of the correlation energy with respect to $n_{\Gamma}$ with $n_1=12$. Both results are computed with $\mathcal{B}A_{\text{1BZ}}=2\pi$ and $V_c/A_{\text{u.c.}}=2$. The dashed line is a linear fit in $1/n_1^3$, serving only as a guide to the eye.
    \label{si_fig:e_corr_scaling}}
\end{figure}

The correlated ground state is defined as that which is annihilated by all $Q_{\nu\boldsymbol{q}}$,
\begin{equation}
    Q_{\nu\boldsymbol{q}}\ket{0} = 0.
\end{equation}
As eigenstates of the Hamiltonian, these operators obey the Schr\"{o}dinger equation,
\begin{equation}
    \left[H, Q^\dagger_{\nu\boldsymbol{q}}\right] = \hbar\omega_{\nu\boldsymbol{q}}Q^\dagger_{\nu\boldsymbol{q}},
\end{equation}
and it can be shown that, as a result, the following equation holds for any operator $R$,
\begin{equation}
    \mel{0}{[R,[H,Q^\dagger_{\nu\boldsymbol{q}}]]}{0} = \omega_{\nu\boldsymbol{q}}\mel{0}{[R,Q^\dagger_{\nu\boldsymbol{q}}]}{0}.
\end{equation}
Taking $R=Q_{\nu\boldsymbol{q}}$ and solving the equations of motion, one obtains the tdHF equations
\begin{equation}
    \sum_{\boldsymbol{k}}
    \begin{pmatrix}
        A^{\boldsymbol{q}}(\boldsymbol{k}',\boldsymbol{k}) & B^{\boldsymbol{q}}(\boldsymbol{k}',\boldsymbol{k}) \\
        B^{-\boldsymbol{q}*}(\boldsymbol{k}',\boldsymbol{k}) & A^{-\boldsymbol{q}*}(\boldsymbol{k}',\boldsymbol{k})
    \end{pmatrix}
    \begin{pmatrix}
        X^{\nu\boldsymbol{q}}(\boldsymbol{k}) \\
        Y^{\nu\boldsymbol{q}}(\boldsymbol{k})
    \end{pmatrix}
    =
    \omega_{\nu\boldsymbol{q}}
    \begin{pmatrix}
        1 & 0 \\
        0 & -1
    \end{pmatrix}
    \begin{pmatrix}
        X^{\nu\boldsymbol{q}}(\boldsymbol{k}') \\
        Y^{\nu\boldsymbol{q}}(\boldsymbol{k}')
    \end{pmatrix},
    \label{eq:tdHF_eqs}
\end{equation}
where the particle-hole indices have been dropped for conciseness. Here $\omega_{\nu\boldsymbol{q}}$ are the energies of the excitations and the $A$ and $B$ matrices are defined as
\begin{equation}
    A_{(\varphi',\varphi)}^{\vb{q}}(\vb{k}', \vb{k}) = \mel{HF}{[b^{\phantom{\dagger}}_{\varphi',\vb{q}}(\vb{k}'), [H, b^\dagger_{\varphi,\vb{q}}(\vb{k})]]}{HF}
\end{equation}
and
\begin{equation}
    B_{(\varphi',\varphi)}^{\vb{q}}(\vb{k}',\vb{k}) = -\mel{HF}{[b^{\phantom{\dagger}}_{\varphi',\vb{q}}(\vb{k}'), [H, b^{\phantom{\dagger}}_{\varphi,-\vb{q}}(\vb{k})]]}{HF}.
\end{equation}
Obtaining these equations involves taking the quasi-boson approximation, in which commutators of the particle-hole annihilation and creation operators are replaced by their expectation values taken in the HF ground state,
\begin{equation}
    \left[b^{\phantom{\dagger}}_{\varphi,\boldsymbol{q}}(\boldsymbol{k}), b^\dagger_{\varphi',\boldsymbol{q}}(\boldsymbol{k}')\right] \approx \mel{\text{HF}}{\left[b^{\phantom{\dagger}}_{\varphi,\boldsymbol{q}}(\boldsymbol{k}), b^\dagger_{\varphi',\boldsymbol{q}}(\boldsymbol{k}')\right]}{\text{HF}} = \delta_{\phi_h\phi_h'}\delta_{\phi_p\phi_p'}\delta(\boldsymbol{k}-\boldsymbol{k}')\delta(\boldsymbol{q}-\boldsymbol{q}').
\end{equation}
It was recently shown that this approximation yields accurate correlation energies for the two-dimensional electron gas, so we expect it to perform well in this setting as well~\cite{wolfQuasibosonApproximationYields2024,jain2025elementary}.

Using the energies $\omega_{\nu\boldsymbol{q}}$ obtained by solving Eq.~\eqref{eq:tdHF_eqs}, one can write down an effective bosonic Hamiltonian for the particle-hole excitations given by
\begin{equation}
    H_{\text{B}} = \sum_{\nu\vb{q}} \omega_{\nu\vb{q}} Q_{\nu\vb{q}}^\dagger Q_{\nu\vb{q}} + E_{\text{corr}}.
\end{equation}
The correlation energy contributed by the zero-point motion of these collective modes is given by
\begin{equation}
    E_{\text{corr}} = \frac{1}{2}\sum_{\omega_{\nu\vb{q}}>0}\omega_{\nu\vb{q}} - \frac{1}{2}\sum_{\vb{q}}\text{Tr}A^{\vb{q}},
\end{equation}
where the first term is the sum of the energies of all the collective modes, and the second is the energies that the bare particle-hole excitations would have were the residual interactions between them not considered. This is the correlation energy we utilize in the main text.

\begin{figure}
    \includegraphics[width=0.55\linewidth]{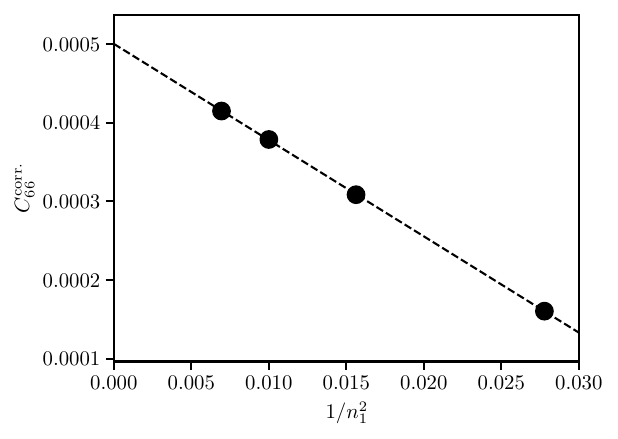}
    \caption{The scaling of the contribution of the correlation energy to the stiffness with respect to the system size $n_1$ with $\mathcal{B}A_{\text{1BZ}}=2\pi$, $V_c/A_{\text{u.c.}}=2$, and $n_{\Gamma}=55.$ The dashed line is a linear fit in $1/n_1^2$, serving only as a guide to the eye.
    \label{si_fig:c_66_corr_scaling}}
\end{figure}

It is important to consider how the correlation energy scales with both system size and the momentum cutoff. In Fig~\ref{si_fig:e_corr_scaling}, we plot the correlation energy of the triangular lattice AHC with $\mathcal{B}A_{\text{1BZ}}=2\pi$ and $V_c/A_{\text{u.c.}}=2$ as a function of $n_1$ and $n_{\Gamma}$. We find that the correlation energy scales as $1/n_1^3$ with system size and converges exponentially with the number of reciprocal lattice vectors included. It is prohibitively expensive to converge the correlation energy well with respect to system size, so we must be careful interpreting the correlation energy stiff $C_{66}^{\text{corr.}}$. In Fig.~\ref{si_fig:c_66_corr_scaling} we plot $C_{66}^{\text{corr.}}$ as a function of $n_1$ and see that it scales like $1/n_1^2$. At $n_1=12$, the system size used to compute the stiffnesses in the main text, we see we have captured roughly 80\% of the correlation energy stiffness, so our results are qualitatively unchanged by the finite system size and our conclusions hold.

\section{Variational ansatz: perturbative energy calculation}
\label{si_sec:perturbative_ansatz_energy}
Here we employ a variational ansatz for the parent band model that takes the form of a Slater determinant of single-particle states~\cite{tan2024parent, tan2024wavefunction},
\begin{equation}
    \ket{\psi^{\mathcal{C}}_{\boldsymbol{k}}} = \mathcal{N}_{\boldsymbol{k}} \sum_{\boldsymbol{g}}  e^{ -\frac{|\boldsymbol{k}+\boldsymbol{g}|^2}{4 \xi^2} -i \pi \mathcal{C}[\frac{\boldsymbol{k} \times \boldsymbol{g}}{A_{1\text{BZ}}}+ \omega(\boldsymbol{g})]} e^{i(\boldsymbol{k}+\boldsymbol{g}) \cdot \boldsymbol{r}} \ket{s_{\boldsymbol{k}+\boldsymbol{g}}}, \label{Eq_ansatz_def}
\end{equation}
where $\boldsymbol{k}$ is the crystal momentum, $\boldsymbol{g}$ enumerates the reciprocal lattice vectors, $\mathcal{C}$ is the Chern number, and $A_{\text{1BZ}}$ is the area of the first Brillouin zone. Here $e^{i \pi \omega(\boldsymbol{g})}$ is $-1$ if $\boldsymbol{g}/2$ is a RLV and $1$ otherwise, $\mathcal{N}_{\boldsymbol{k}}$ is a normalization function, and $\xi$ is a variational parameter that controls the spread of the wavefunction in momentum space. For conciseness, we define the function
\begin{equation}
    U_{\boldsymbol{g}}(\mathbf{k})= e^{ -\frac{|\boldsymbol{k}+\boldsymbol{g}|^2}{4 \xi^2} -i \pi \mathcal{C}[\frac{\boldsymbol{k} \times \boldsymbol{g}}{A_{\text{1BZ}}}+ \omega(\boldsymbol{g})]} e^{i(\boldsymbol{k}+\boldsymbol{g}) \cdot \boldsymbol{r}},
\end{equation}
such that $\ket{\psi^{\mathcal{C}}_{\boldsymbol{k}}} = \mathcal{N}_{\boldsymbol{k}} \sum_{\boldsymbol{g}}  U_{\boldsymbol{g}}(\mathbf{k})\ket{s_{\boldsymbol{k}+\boldsymbol{g}}}$ and $|\mathcal{N}_{\mathbf{k}}|^{-2} = A\sum_{\boldsymbol{g}}|U_{\boldsymbol{g}}(\mathbf{k})|^2$, with $A$ the area of the sample.

We calculate the variational energy of the ansatz wavefunction by making a strong-interaction expansion. The authors of Refs.~\cite{tan2024wavefunction, tan2024parent} used this approach to estimate the energy, and we will follow their method closely. However, their zeroth-order result does not depend on the lattice and so carries no information about the mechanical properties of the anomalous Hall crystal. To approximate these quantities, we need the next terms in the strong-interaction expansion. Our main result for this section is an approximate expression for the total energy per particle of the anomalous Hall crystal state:
\begin{align}
    \frac{E}{N} \approx& \frac{\xi^2}{m} - \frac{V_{c}}{4 \sqrt{\alpha \pi}} +   \frac{1}{2} \frac{\xi^2}{m} \sum_{\text{shortest } \boldsymbol{R}}  \xi^2 |\boldsymbol{R}|^2 e^{- \xi^2|\boldsymbol{R}|^2} \notag \\
    &- \frac{V_{c}}{4 \sqrt{\alpha \pi}}  \sum_{\text{shortest }\boldsymbol{R}} e^{-\xi^2|\boldsymbol{R}|^2 }\left[3 +e^{-\frac{|\boldsymbol{R}|^2}{2 \alpha }} I_0\left(\frac{|\boldsymbol{R}|^2}{2 \alpha }\right) -4 e^{-\frac{|\boldsymbol{R}|^2}{8 \alpha }}I_0\left(\frac{|\boldsymbol{R}|^2}{8 \alpha }\right)\right], \label{Equation_perturbative_energy}
\end{align} 
where $I_0$ is the modified Bessel function of the first kind and $\alpha=\frac{1}{ \xi^2}+ \frac{4\mathcal{C} \pi}{ A_{1\text{BZ}}}$ with positive $\mathcal{C}$. This expression holds when the Berry curvature of the parent band precisely matches the Chern number of the descendant band defined by the ansatz in Eq.~\eqref{Eq_ansatz_def}: $\mathcal{B} A_{\text{1BZ}}/2\pi = \mathcal{C}$. Eq.~\eqref{Equation_perturbative_energy} gives a first-order expansion of the energy per particle, in the small parameter $e^{- \xi^2 a^2}$, where $a$ is the lattice constant of the crystal. This small parameter decreases with interaction strength for the optimized ansatz, as is known from the zeroth-order expansion \cite{tan2024parent, tan2024wavefunction}, so we expect the perturbative expression to be valid for large interaction strengths. Because the zeroth-order term does not depend on the lattice structure, any properties like stiffness will decay rapidly with interaction strength. In the first-order term, the energy depends on the lattice structure through the sum over the shortest lattice vectors $\boldsymbol{R}$. When we compute the stiffness of the crystal, we must slightly deform the lattice, which changes the lattice vectors and results in a change in energy. We note that during this deformation, there will be lattice vectors that are very close in length to the shortest vectors. In this case, these vectors are also included in the sum.

In the remainder of this section, we explain how the energy is derived using a small-parameter expansion. We first examine the form factor for the descendant band, which is defined as
\begin{equation}
	F(\boldsymbol{k}+ \boldsymbol{q}, \boldsymbol{k})= \braket{\psi_{ \lceil \boldsymbol{k}+\boldsymbol{q} \rceil}^{\mathcal{C}}| e^{ i \boldsymbol{q} \cdot \boldsymbol{r}}| \psi_{\boldsymbol{k}}^{\mathcal{C}} },
\end{equation} 
where $\lceil \boldsymbol{v} \rceil$ folds a vector $\boldsymbol{v}$ into the first Brillouin zone, and the states are in the descendant band defined by the ansatz (Eq.~\eqref{Eq_ansatz_def}). Here $\boldsymbol{k}$ is in the first Brillouin zone, while $\boldsymbol{q}$ is a general momentum transfer. Defining $\boldsymbol{g}_0=  \boldsymbol{k}+\boldsymbol{q}-\lceil \boldsymbol{k}+\boldsymbol{q} \rceil$, which is a reciprocal lattice vector (RLV), and $\eta(\boldsymbol{g}_0)= e^{i \pi (\omega(\boldsymbol{g}_0)-1)}$, this form factor is given by \cite{tan2024wavefunction}
\begin{align}
	F(\boldsymbol{k}+ \boldsymbol{q}, \boldsymbol{k})=& A \mathcal{N}_{\boldsymbol{k}+\boldsymbol{q}} \mathcal{N}_{\boldsymbol{k}} f(\boldsymbol{k})e^{-(\frac{\mathcal{B}}{4} + \frac{1}{4 \xi^2}) |\boldsymbol{q}|^2} (\eta(\boldsymbol{g}_0))^\mathcal{C} e^{i \mathcal{C} \pi \frac{(\boldsymbol{k}+\boldsymbol{q}) \times \boldsymbol{g}_0 +\boldsymbol{k} \times \boldsymbol{q}}{A_{1\text{BZ}}}},
\end{align}
for
\begin{align}
	f(\boldsymbol{k}) = \sum_{\boldsymbol{g}} e^{-i \frac{\delta}{2} \boldsymbol{q} \times (\boldsymbol{k}+\boldsymbol{g}) } e^{- \frac{|\boldsymbol{k}+\boldsymbol{g}|^2}{2 \xi^2}} e^{- \frac{\boldsymbol{q} \cdot(\boldsymbol{k}+\boldsymbol{g})}{2 \xi^2}},
\end{align}
where $\delta=\mathcal{B}- \frac{2 \mathcal{C} \pi}{A_{1\text{BZ}}}$. This can be cast into a form that converges more rapidly at large $\xi$ by taking a Fourier series expansion for $f(\boldsymbol{k})$, resulting in an expression in terms of a sum over lattice vectors $\boldsymbol{R}$ \cite{tan2024wavefunction}:

\begin{equation}
	f(\boldsymbol{k})= \frac{2 \pi \xi^2}{A_{1\text{BZ}}} e^{\frac{|\boldsymbol{q}|^2 (1- \delta^2 \xi^4)}{8 \xi^2}} \sum_{\boldsymbol{R}}  e^{ - \frac{|\boldsymbol{R}|^2 \xi^2}{2}} e^{i (\frac{\boldsymbol{q}}{2}+\boldsymbol{k}) \cdot \boldsymbol{R}} e^{\frac{\xi^2 \delta \boldsymbol{R} \times \boldsymbol{q}}{2}}. 
\end{equation}
The form factor is then given by
\begin{align}
	F(\boldsymbol{k}+ \boldsymbol{q}, \boldsymbol{k})=& A  \frac{2 \pi \xi^2}{A_{1\text{BZ}}} \mathcal{N}_{\boldsymbol{k}+\boldsymbol{q}} \mathcal{N}_{\boldsymbol{k}} (\eta(\boldsymbol{g}_0))^\mathcal{C} e^{-|\boldsymbol{q}|^2 (\frac{(1+  \delta \xi^2)^2}{8 \xi^2}+ \frac{\mathcal{C} \pi}{2 A_{1\text{BZ}}})} e^{i \mathcal{C} \pi \frac{(\boldsymbol{k}+\boldsymbol{q}) \times \boldsymbol{g}_0 +\boldsymbol{k} \times \boldsymbol{q}}{A_{1\text{BZ}}}}  \sum_{\boldsymbol{R}}  e^{ - \frac{|\boldsymbol{R}|^2 \xi^2}{2}} e^{i (\frac{\boldsymbol{q}}{2}+\boldsymbol{k}) \cdot \boldsymbol{R}} e^{\frac{\xi^2 \delta \boldsymbol{R} \times \boldsymbol{q}}{2}}. \label{Equation_form_factor_ansatz}
\end{align}

From now on, we will examine the case where $\delta=0$, meaning that the parent Berry flux through the Brillouin zone is equal to $2\pi \mathcal{C}$. In this case
\begin{align}
	F(\boldsymbol{k}+ \boldsymbol{q}, \boldsymbol{k})=&   \frac{2 \pi \xi^2 A}{A_{1\text{BZ}}} \mathcal{N}_{\boldsymbol{k}+\boldsymbol{q}} \mathcal{N}_{\boldsymbol{k}} (\eta(\boldsymbol{g}_0))^\mathcal{C} e^{-\frac{ \alpha|\boldsymbol{q}|^2}{8}}  e^{i \mathcal{C} \pi \frac{(\boldsymbol{k}+\boldsymbol{q}) \times \boldsymbol{g}_0 +\boldsymbol{k} \times \boldsymbol{q}}{A_{1\text{BZ}}}} \sum_{\boldsymbol{R}}  e^{ - \frac{|\boldsymbol{R}|^2 \xi^2}{2}} e^{i (\frac{\boldsymbol{q}}{2}+\boldsymbol{k}) \cdot \boldsymbol{R}} . \label{Equation_form_factor_ansatz_no_delta}
\end{align}

The interaction energy for the many-body ansatz wavefunction is given in terms of the form factor as \cite{tan2024wavefunction}
\begin{align}
\braket{H_{\text{int}}}&=\frac{1}{2A} \sum_{\boldsymbol{k}_1, \boldsymbol{k}_2 \in \text{BZ}}  \sum_{\boldsymbol{q}} V(- \boldsymbol{q}) F(\boldsymbol{k}_1 - \boldsymbol{q}, \boldsymbol{k}_1)  F(\boldsymbol{k}_2 + \boldsymbol{q}, \boldsymbol{k}_2) \braket{c_{\lceil \boldsymbol{k}_1- \boldsymbol{q}\rceil} ^{\dagger}c_{ \lceil\boldsymbol{k}_2 + \boldsymbol{q}\rceil}^{\dagger}c_{\boldsymbol{k}_2 } c_{\boldsymbol{k}_1} }_{D}, \label{Eq_interaction_descendant_band}
\end{align}
where the subscript $D$ on the expectation value indicates that the creation and annihilation operators belong to the descendant band. This energy can then be split into Hartree and Fock terms by noting that the expectation value is non-zero only when $\boldsymbol{k}_2= \lceil \boldsymbol{k}_2+ \boldsymbol{q} \rceil $ (Hartree) or when $\boldsymbol{k_2} = \lceil \boldsymbol{k}_1- \boldsymbol{q} \rceil$ (Fock), because the state is a Slater determinant. In the first case, $\boldsymbol{q}$ must be a RLV, which we denote by $\boldsymbol{g}$. As a result, the interaction energy can be written as
\begin{align}
	\braket{H_{\text{int}}}&=\frac{1}{2A} \sum_{\boldsymbol{k}_1, \boldsymbol{k}_2 \in \text{BZ}}  \sum_{\boldsymbol{g}} V(- \boldsymbol{g}) F(\boldsymbol{k}_1 - \boldsymbol{g}, \boldsymbol{k}_1)  F(\boldsymbol{k}_2 + \boldsymbol{g}, \boldsymbol{k}_2)  - \frac{1}{2A} \sum_{\boldsymbol{k}_1 \in \text{BZ}}  \sum_{\boldsymbol{q}} V(- \boldsymbol{q}) |F(\boldsymbol{k}_1 - \boldsymbol{q}, \boldsymbol{k}_1)|^2  \\
	&= \braket{H_{\text{Hartree}}} - \braket{H_{\text{Fock}}}.
\end{align}
Following the approach used in Refs. \cite{tan2024parent, tan2024wavefunction}, we remove the long-ranged part of the Coulomb interaction by excluding $\boldsymbol{q} = 0$ from the sum (and do the same for the sum over $\boldsymbol{g}$ in the Hartree term).

\subsection{Correction to the Fock term}

We now compute approximations to the various terms in the variational energy, starting with the Fock term:
\begin{align*}
	E_{\text{Fock}}&= - \frac{1}{2A} \sum_{\boldsymbol{k} \in \text{BZ}}  \sum_{\boldsymbol{q}} V(- \boldsymbol{q}) |F(\boldsymbol{k} - \boldsymbol{q}, \boldsymbol{k})|^2 = - \frac{1}{2A} \sum_{\boldsymbol{k} \in \text{BZ}}  \sum_{\boldsymbol{q}} V( \boldsymbol{q}) |F(\boldsymbol{k} + \boldsymbol{q}, \boldsymbol{k})|^2.
\end{align*}
We can substitute our expression for the form factors from Eq.~\eqref{Equation_form_factor_ansatz_no_delta}, to obtain
\begin{align*}
	E_{\text{Fock}}	&= - \frac{1}{2A} \sum_{\boldsymbol{k} \in \text{BZ}}  \sum_{\boldsymbol{q}} V( \boldsymbol{q})  \left(\frac{2 \pi \xi^2 A}{A_{1\text{BZ}}}\right)^2 \mathcal{N}_{\boldsymbol{k}+\boldsymbol{q}}^2 \mathcal{N}_{\boldsymbol{k}}^2  e^{-\frac{\alpha |\boldsymbol{q}|^2}{4}}   \bigg|\sum_{\boldsymbol{R}}  e^{ - \frac{|\boldsymbol{R}|^2 \xi^2}{2}}e^{i (\frac{\boldsymbol{q}}{2}+\boldsymbol{k}) \cdot \boldsymbol{R}} \bigg|^2. 
\end{align*}

The normalization factors can be written as a series that converges rapidly at large $\xi$, using a Fourier expansion. We have
\begin{align}
\mathcal{N}_{\boldsymbol{k}}^{-2}&= A \sum_{\boldsymbol{g}} e^{-\frac{|\boldsymbol{k}+\boldsymbol{g}|^2}{2 \xi^2}} = A \frac{2 \pi \xi^2}{A_{1\text{BZ}} }  \sum_{\boldsymbol{R}}  e^{i \boldsymbol{k} \cdot \boldsymbol{R}} e^{- \frac{\xi^2|\boldsymbol{R}|^2}{2}}. \label{Equation_normalization_lattice_sum}
\end{align}
Using this expression, the Fock energy is given by
\begin{align}
	E_{\text{Fock}}	&= - \frac{1}{2A}  \sum_{\boldsymbol{q}} V( \boldsymbol{q}) e^{-\frac{\alpha |\boldsymbol{q}|^2}{4}} \sum_{\boldsymbol{k} \in \text{BZ}}  \frac{\bigg|\sum_{\boldsymbol{R}}  e^{ - \frac{|\boldsymbol{R}|^2 \xi^2}{2}}e^{i (\frac{\boldsymbol{q}}{2}+\boldsymbol{k}) \cdot \boldsymbol{R}} \bigg|^2}{ \sum_{\boldsymbol{R}'}  e^{i \boldsymbol{k} \cdot \boldsymbol{R}'} e^{- \frac{\xi^2|\boldsymbol{R}'|^2}{2}}  \sum_{\boldsymbol{R}''}  e^{i(\boldsymbol{k}+ \boldsymbol{q}) \cdot \boldsymbol{R}''} e^{- \frac{\xi^2|\boldsymbol{R}''|^2}{2}}}. \label{Equation_perturb_Fock_1}
\end{align}

We are interested in the case where $\exp(-a^2 \xi^2/2) \ll1$, where $a$ is the lattice constant ($a$ is the smaller lattice constant if the primitive lattice vectors have different lengths), which is realized in the strong interaction limit. $\exp(-a^2 \xi^2/2)$ then serves as a small parameter, allowing for a perturbative expansion. The zeroth-order term, which is computed in Ref. \cite{tan2024wavefunction}, can be found by taking $\boldsymbol{R}=\boldsymbol{R}'=\boldsymbol{R}''=0$. To compute the first correction to the Fock energy, we first write

\begin{align}
& \sum_{\boldsymbol{k} \in \text{BZ}}  \frac{\bigg|\sum_{\boldsymbol{R}}  e^{ - \frac{|\boldsymbol{R}|^2 \xi^2}{2}}e^{i (\frac{\boldsymbol{q}}{2}+\boldsymbol{k}) \cdot \boldsymbol{R}} \bigg|^2}{ \sum_{\boldsymbol{R}'}  e^{i \boldsymbol{k} \cdot \boldsymbol{R}'} e^{- \frac{\xi^2|\boldsymbol{R}'|^2}{2}}  \sum_{\boldsymbol{R}''}  e^{i(\boldsymbol{k}+ \boldsymbol{q}) \cdot \boldsymbol{R}''} e^{- \frac{\xi^2|\boldsymbol{R}''|^2}{2}}} \notag \\
& \hspace{4cm}=  \sum_{\boldsymbol{k} \in \text{BZ}}  \frac{\bigg|1+\sum_{\boldsymbol{R}\neq 0}  e^{ - \frac{|\boldsymbol{R}|^2 \xi^2}{2}} e^{i (\frac{\boldsymbol{q}}{2}+\boldsymbol{k}) \cdot \boldsymbol{R}} \bigg|^2}{ \left(1+\sum_{\boldsymbol{R}' \neq 0}  e^{i \boldsymbol{k} \cdot \boldsymbol{R}'} e^{- \frac{\xi^2|\boldsymbol{R}'|^2}{2}}\right) \left(1+  \sum_{\boldsymbol{R}'' \neq 0}  e^{i(\boldsymbol{k}+ \boldsymbol{q}) \cdot \boldsymbol{R}''} e^{- \frac{\xi^2|\boldsymbol{R}''|^2}{2}}\right)} \notag \\
&\hspace{4cm}\approx \sum_{\boldsymbol{k}} \left[ 1+ 2\sum_{\boldsymbol{R}\neq 0}  e^{ - \frac{|\boldsymbol{R}|^2 \xi^2}{2}} e^{i (\frac{\boldsymbol{q}}{2}+\boldsymbol{k}) \cdot \boldsymbol{R}}  + \sum_{\boldsymbol{R}_1, \boldsymbol{R}_2 \neq 0} e^{ - \frac{(|\boldsymbol{R}_1|^2 +|\boldsymbol{R}_2|^2) \xi^2}{2}}e^{i (\frac{\boldsymbol{q}}{2}+\boldsymbol{k}) \cdot (\boldsymbol{R}_1 - \boldsymbol{R}_2)} \right] \notag\\
& \hspace{5cm}\times \left[1- \sum_{\boldsymbol{R}' \neq 0}  e^{i \boldsymbol{k} \cdot \boldsymbol{R}'} e^{- \frac{\xi^2|\boldsymbol{R}'|^2}{2}} + \left(\sum_{\boldsymbol{R}' \neq 0}  e^{i \boldsymbol{k} \cdot \boldsymbol{R}'} e^{- \frac{\xi^2|\boldsymbol{R}'|^2}{2}}\right)^2 +...\right] \notag \\ & \hspace{5.5cm} \times  \left[1- \sum_{\boldsymbol{R}'' \neq 0}  e^{i(\boldsymbol{k}+ \boldsymbol{q}) \cdot \boldsymbol{R}''} e^{- \frac{\xi^2|\boldsymbol{R}''|^2}{2}} + \left(\sum_{\boldsymbol{R}'' \neq 0}  e^{i(\boldsymbol{k}+ \boldsymbol{q}) \cdot \boldsymbol{R}''} e^{- \frac{\xi^2|\boldsymbol{R}''|^2}{2}}\right)^2+...\right] \label{Equation_Fock_expansion}
\end{align}

At first, it may seem that the first correction will be a first-order term, which comes from taking the zeroth-order contribution from two of the terms in squared brackets and a first-order contribution from the remaining term. However, this is not the case. To see this, note that expanding the product gives an expression of the form
\begin{align*}
\sum_{\boldsymbol{k} \in \text{BZ}}&  \frac{\bigg|\sum_{\boldsymbol{R}}  e^{ - \frac{|\boldsymbol{R}|^2 \xi^2}{2}}e^{i (\frac{\boldsymbol{q}}{2}+\boldsymbol{k}) \cdot \boldsymbol{R}} \bigg|^2}{ \sum_{\boldsymbol{R}'}  e^{i \boldsymbol{k} \cdot \boldsymbol{R}'} e^{- \frac{\xi^2|\boldsymbol{R}'|^2}{2}}  \sum_{\boldsymbol{R}''}  e^{i(\boldsymbol{k}+ \boldsymbol{q}) \cdot \boldsymbol{R}''} e^{- \frac{\xi^2|\boldsymbol{R}''|^2}{2}}}\\
&= \sum_{\boldsymbol{k}}\left(1+ \sum_{\boldsymbol{R}_1 \neq 0} A_1^{\boldsymbol{q}}(\boldsymbol{R}_1) e^{ - \frac{|\boldsymbol{R}_1|^2 \xi^2}{2}} e^{i \boldsymbol{k} \cdot \boldsymbol{R}_1} + \sum_{\boldsymbol{R}_1, \boldsymbol{R}_2 \neq 0}A_2^{\boldsymbol{q}}(\boldsymbol{R}_1, \boldsymbol{R}_2)e^{ - \frac{ \xi^2}{2}(|\boldsymbol{R}_1|^2+|\boldsymbol{R}_1|^2)} e^{i \boldsymbol{k} \cdot (\boldsymbol{R}_1+\boldsymbol{R}_2)} +...\right) \\
&=  \sum_{\boldsymbol{k}} \left(1+ \sum_{n=1}^{\infty} \sum_{\set{\boldsymbol{R}_1,...,\boldsymbol{R}_n \neq 0}} A_n^{\boldsymbol{q}}(\set{\boldsymbol{R}_1,...,\boldsymbol{R}_n}) e^{ - \frac{  \xi^2}{2}\sum_{i=1}^n|\boldsymbol{R}_i|^2  } e^{i \boldsymbol{k} \cdot \sum_{i=1}^n \boldsymbol{R}_i}\right ),
\end{align*}
where the $A_n^{\boldsymbol{q}}$ are some functions that we have not yet computed, but which do not include the small parameter $\exp(-a^2\xi^2)$. The first-order terms are proportional to $\sum_{\boldsymbol{k}}e^{i \boldsymbol{k} \cdot \boldsymbol{R}}$ for some non-zero lattice vector $\boldsymbol{R}$. In the thermodynamic limit, this oscillatory term vanishes when summed over $\boldsymbol{k}$ if $\boldsymbol{R}$ is non-zero. Instead, the simplest contributing term is at second-order and involves two lattice vectors $\boldsymbol{R}_1$ and $\boldsymbol{R}_2$ such that $\boldsymbol{R}_1+\boldsymbol{R}_2=0$. More generally, we obtain contributions from the $n$th order terms for which the sum of the $n$ lattice vectors is zero.

To determine the relative sizes of the different contributions, we examine the Gaussian factors. These decay exponentially with the sum of squared lengths of the lattice vectors. This means that the largest such term is a second-order term involving $\boldsymbol{R}_1=- \boldsymbol{R}_2$, such that $\boldsymbol{R}_1$ is one of the shortest lattice vectors (excluding the zero length one, which gives the zeroth-order contribution). In this case, the Gaussian factor is $\exp( - a^2 \xi^2)$, where $a$ is the lattice constant. The next largest terms, which we will not include, are either second-order contributions involving the second shortest lattice vectors or higher-order terms involving the shortest lattice vectors. Taking the square lattice as an example, both of these contributions are suppressed by the factor $\exp( - 2a^2 \xi^2)$, compared to the $\exp( - a^2 \xi^2)$ factor on the first contribution. This allows us to estimate when our leading term in the correction is sufficient to estimate the lattice-dependent component of the energy. We are using units where the length of the primitive RLV for the triangular lattice is unity. This means that the lattice constant for the square lattice of the same density is $a= 2 \pi \sqrt{\frac{2}{\sqrt{3}}}$, meaning we need $\xi > 0.22$ or so for a correction of order 0.1, or $\xi>0.32$ to get a correction of order $0.01$. In the case of the triangular lattice, the next largest terms instead result from a third-order process involving three of the shortest lattice vectors, meaning that the term is suppressed by $\exp( - \frac{3}{2} a^2 \xi^2)$. With $a = \frac{4 \pi}{\sqrt{3}}$ for the triangular lattice, this term can be ignored compared to the first contribution when $ \xi>0.30$ (for a relative contribution of order $0.1$) or $\xi >0.42$ (for a relative contribution of order $0.01$). Given that there will be more terms contributing at higher order, a larger $\xi$ may be required for good convergence. Note that this estimate for the convergence is in terms of the variational parameter $\xi$ rather than a physical parameter such as $V_c$. We discuss how the regime of applicability depends on the interaction strength, as well as the Chern number, in Sec. \ref{Section_perturb_convergence}.

Having determined which terms to consider, we now compute the leading-order correction for the expression in Eq.~\eqref{Equation_Fock_expansion}. We obtain one term by taking the second-order contribution from the first squared bracket and the zeroth-order contribution from the other brackets:
\begin{align*}
T_1&=\sum_{\boldsymbol{k}} \sum_{\text{shortest }\boldsymbol{R}} e^{-\frac{|\boldsymbol{R}|^2}{2} \xi^2} e^{i (\frac{\boldsymbol{q}}{2}+\boldsymbol{k}) \cdot \boldsymbol{R}} \sum_{\text{shortest }\boldsymbol{R}'} e^{-\frac{|\boldsymbol{R}'|^2}{2} \xi^2} e^{-i (\frac{\boldsymbol{q}}{2}+\boldsymbol{k}) \cdot \boldsymbol{R}'} \delta_{\boldsymbol{R}, \boldsymbol{R}'} =N \sum_{\text{shortest }\boldsymbol{R}} e^{-|\boldsymbol{R}|^2 \xi^2}. 
\end{align*}
We obtain a similar term by using the second-order contributions from the other two squared brackets instead:
\begin{align*}
	T_2&= 2N \sum_{\text{shortest }\boldsymbol{R}} e^{-|\boldsymbol{R}|^2 \xi^2}.
\end{align*}
Then, we have a cross term between the linear parts of the last two squared brackets

\begin{align*}
	T_3&= N \sum_{\text{shortest }\boldsymbol{R}} e^{-|\boldsymbol{R}|^2 \xi^2} e^{-i \boldsymbol{q} \cdot \boldsymbol{R}} =N \sum_{\text{shortest }\boldsymbol{R}} e^{-|\boldsymbol{R}|^2 \xi^2} \cos(\boldsymbol{q} \cdot \boldsymbol{R}),
\end{align*}
where we used inversion symmetry to take the real part of the exponential. Finally, we have a cross term between the linear part of the first squared bracket and the linear parts of the other two:

\begin{align*}
	T_4&= -2N \sum_{\text{shortest }\boldsymbol{R}} e^{-|\boldsymbol{R}|^2 \xi^2} e^{i \frac{\boldsymbol{q} \cdot \boldsymbol{R}}{2}} (1+e^{-i \boldsymbol{q} \cdot \boldsymbol{R}}) = -4N \sum_{\text{shortest }\boldsymbol{R}} e^{-|\boldsymbol{R}|^2 \xi^2} \cos(\frac{\boldsymbol{q} \cdot \boldsymbol{R}}{2}).
\end{align*}

Substituting these contributions, which approximate Eq.~\eqref{Equation_Fock_expansion}, into the expression for the Fock energy given in Eq.~\eqref{Equation_perturb_Fock_1}, we find
\begin{align}
	E_{\text{Fock}}	
	& \approx - \frac{N}{2A}  \sum_{\boldsymbol{q}} V( \boldsymbol{q}) e^{-\frac{\alpha |\boldsymbol{q}|^2}{4}} \bigg(1 +  \sum_{\text{shortest }\boldsymbol{R}} e^{-|\boldsymbol{R}|^2 \xi^2} \left[3 + \cos(\boldsymbol{q} \cdot \boldsymbol{R}) -4\cos(\frac{\boldsymbol{q} \cdot \boldsymbol{R}}{2})\right] \bigg). 
\end{align}
Using the double-angle formula for cosine, we have 
$$3 + \cos(\boldsymbol{q} \cdot \boldsymbol{R}) -4\cos(\frac{\boldsymbol{q} \cdot \boldsymbol{R}}{2}) = 2\left(1- \cos(\frac{\boldsymbol{q} \cdot \boldsymbol{R}}{2}) \right)^2,$$
which is always non-negative, indicating that the correction enhances the magnitude of the Fock term.

In the infinite system size limit, we can convert the sum over $\boldsymbol{q}$ to an integral. We must be careful about $\boldsymbol{q}=0$, which is excluded from the sum. This happens naturally in the integral, with the contribution from $\boldsymbol{q}=0$ vanishing. This is because the $1/|\boldsymbol{q}|$ factor from the Coulomb interaction cancels with a $|\boldsymbol{q}|$ factor in the integral measure in polar coordinates. The resulting integrand is non-divergent at the origin, and so the contribution from the single point at the origin is zero. We can therefore replace the sum over $\boldsymbol{q} \neq 0$ with an integral
$$\sum_{\boldsymbol{q} \neq 0} \rightarrow \frac{A}{4 \pi^2} \int d^2q = \frac{A}{4 \pi^2} \int dq d \theta q,$$
such that 
\begin{align*}
	E_{\text{Fock}}	& \approx - \frac{N}{8 \pi^2}  \int_0^{\infty} dq \int_0^{2 \pi} d\theta qV( q) e^{-\frac{\alpha q^2}{4 } } \left(1 +  \sum_{\text{shortest }\boldsymbol{R}} e^{-|\boldsymbol{R}|^2 \xi^2} \left[3 + \cos(q|\boldsymbol{R}| \cos \theta) -4 \cos(\frac{q|\boldsymbol{R}| \cos\theta}{2})\right] \right),
\end{align*}
where we assume a rotationally symmetric interaction, and we align our axis with $\boldsymbol{R}$ for every term in the sum. Using the Coulomb interaction, $V(q)= V_{c}/q$, we get

\begin{align*}
	E_{\text{Fock}}	& \approx - \frac{N}{8 \pi^2}  \int_0^{\infty} dq \int_0^{2 \pi} d\theta V_{c} e^{-\frac{\alpha q^2}{4}} \bigg(1 +  \sum_{\text{shortest }\boldsymbol{R}} e^{-|\boldsymbol{R}|^2 \xi^2} \left[3 + \cos(q|\boldsymbol{R}| \cos \theta) -4\cos(\frac{q|\boldsymbol{R}| \cos\theta}{2})\right] \bigg)\\
    &= E^{(0)}_{\text{Fock}} + E^{(1)}_{\text{Fock}}.
\end{align*}

We first consider the zeroth-order term, which gives us
\begin{align*}
	E_{\text{Fock}}^{(0)}&=  - \frac{N}{8 \pi^2}  \int_0^{\infty} dq \int_0^{2 \pi} d \theta V_{c} e^{-\frac{\alpha q^2}{4}} = - \frac{N V_{c}}{4 \sqrt{\alpha \pi}} . 
\end{align*}
This agrees with the interaction term computed in Refs.~\cite{tan2024wavefunction, tan2024parent} for $\mathcal{C}=1$ and $\delta=0$.

Next, we look at the first-order correction to the Fock energy. For the $3e^{-|\boldsymbol{R}|^2 \xi^2}$ term, there is no additional $\boldsymbol{q}$ dependence, so the integral immediately follows from the result above, giving the contribution 
$$- \frac{3N V_{c}}{4 \sqrt{ \alpha \pi}} \sum_{\text{shortest }\boldsymbol{R}} e^{-|\boldsymbol{R}|^2 \xi^2}. $$
For the cosine terms in the correction, we must examine the integral
$$\int_0^{\infty} dq \ e^{- \frac{1}{4}\alpha q^2} \cos(q\lambda)=\sqrt{\frac{\pi}{\alpha}} e^{-\frac{\lambda^2}{\alpha}},$$
with $\lambda = |\boldsymbol{R}| \cos \theta$ or $\lambda=(|\boldsymbol{R}|/2)  \cos \theta$. Our first correction to the Fock term is
\begin{align*}
E_{\text{Fock}}^{(1)}&=- \frac{3N V_{c}}{4 \sqrt{ \alpha \pi}}  \sum_{\text{shortest }\boldsymbol{R}} e^{-|\boldsymbol{R}|^2 \xi^2} - \frac{N V_{c}}{8 \pi^2}   \sqrt{\frac{\pi}{\alpha}} \sum_{\text{shortest }\boldsymbol{R}} e^{-|\boldsymbol{R}|^2 \xi^2}  \int_0^{2 \pi} d \theta \  \left(e^{-\frac{|\boldsymbol{R}|^2 \cos^2 \theta}{\alpha}} -4 e^{-\frac{|\boldsymbol{R}|^2 \cos^2 \theta}{4\alpha}}\right).
\end{align*}

Next, we use the double-angle formula to write $\cos^2 \theta = \frac{1}{2}(\cos (2\theta) +1)$. Then we swap the integration variable from $\theta$ to $\phi = 2 \theta$, obtaining
\begin{align*}
\int_0^{2 \pi} d \theta \ \left(e^{-\frac{|\boldsymbol{R}|^2 \cos^2 \theta}{\alpha}} -4 e^{-\frac{|\boldsymbol{R}|^2 \cos^2 \theta}{4\alpha}}\right) &= \frac{1}{2} \int_0^{4\pi} d \phi \  \left(e^{-\frac{|\boldsymbol{R}|^2}{2 \alpha }} e^{-\frac{|\boldsymbol{R}|^2 \cos \phi}{2 \alpha }}  -4 e^{-\frac{|\boldsymbol{R}|^2}{8 \alpha }}e^{-\frac{|\boldsymbol{R}|^2 \cos \phi}{8 \alpha }}\right)\\
&= \int_0^{2\pi} d \phi \ \left(e^{-\frac{|\boldsymbol{R}|^2}{2 \alpha }} e^{-\frac{|\boldsymbol{R}|^2 \cos \phi}{2 \alpha }}  -4 e^{-\frac{|\boldsymbol{R}|^2}{8 \alpha }}e^{-\frac{|\boldsymbol{R}|^2 \cos \phi}{8 \alpha }} \right)\\
&= 2 \int_{\pi}^{2\pi} d \phi \ \left(e^{-\frac{|\boldsymbol{R}|^2}{2 \alpha }} e^{-\frac{|\boldsymbol{R}|^2 \cos \phi}{2 \alpha }}  -4 e^{-\frac{|\boldsymbol{R}|^2}{8 \alpha }}e^{-\frac{|\boldsymbol{R}|^2 \cos \phi}{8 \alpha }}\right),
\end{align*}
where we used the fact that $\cos \phi$ is mirrored about $\phi = \pi$. Then we shift the integration variable by $\pi$, resulting in a minus sign on the cosines, to get

\begin{align*}
	\int_0^{2 \pi}  d \theta \ \left(e^{-\frac{|\boldsymbol{R}|^2 \cos^2 \theta}{\alpha}} -4 e^{-\frac{|\boldsymbol{R}|^2 \cos^2 \theta}{4\alpha}}\right) &=  2 \int_{0}^{\pi} d \phi \ \left(e^{-\frac{|\boldsymbol{R}|^2}{2 \alpha }} e^{+\frac{|\boldsymbol{R}|^2 \cos \phi}{2 \alpha }}  -4 e^{-\frac{|\boldsymbol{R}|^2}{8 \alpha }}e^{+\frac{|\boldsymbol{R}|^2 \cos \phi}{8 \alpha }}\right).
\end{align*}
This can be related to an integral representation of a Bessel function \cite{Bateman1953}
\begin{equation}
	I_0(z) = \frac{1}{\pi} \int_0^{\pi} d \theta e^{z \cos \theta},
\end{equation}
where $I_0$ is a modified Bessel function of the first kind. Therefore, 
\begin{align*}
	\int_0^{2 \pi} d \theta \ \left(e^{-\frac{|\boldsymbol{R}|^2 \cos^2 \theta}{\alpha}} -4 e^{-\frac{|\boldsymbol{R}|^2 \cos^2 \theta}{4\alpha}}\right) &= 2 \pi \left[e^{-\frac{|\boldsymbol{R}|^2}{2 \alpha }} I_0\left(\frac{|\boldsymbol{R}|^2}{2 \alpha }\right) -4 e^{-\frac{|\boldsymbol{R}|^2}{8 \alpha }}I_0\left(\frac{|\boldsymbol{R}|^2}{8 \alpha }\right)\right].
\end{align*}

As a result, the first correction to the Fock term is
\begin{align}
	E_{\text{Fock}}^{(1)}
	&= - \frac{N V_{c}}{4 \sqrt{ \alpha \pi}} \sum_{\text{shortest }\boldsymbol{R}} e^{-|\boldsymbol{R}|^2 \xi^2} \left[3 +e^{-\frac{|\boldsymbol{R}|^2}{2 \alpha }} I_0\left(\frac{|\boldsymbol{R}|^2}{2 \alpha }\right) -4 e^{-\frac{|\boldsymbol{R}|^2}{8 \alpha }}I_0\left(\frac{|\boldsymbol{R}|^2}{8 \alpha }\right)\right] \label{Equation_Fock_correction}.
\end{align}

We note that the contribution from each lattice vector is heavily suppressed according to its length. Because of this, the Fock term will favor lattices with smaller lattice constants. As an example, the Fock term would favor the square lattice over the triangular lattice and would also promote lattices where the primitive lattice vectors have different lengths. However, as we shall see in the next section, this effect is countered by the kinetic term.

\subsection{Correction to kinetic term}

The other significant contribution to the energy comes from the kinetic term. Refs.~\cite{tan2024parent, tan2024wavefunction} also give an expression for this contribution in the strong-interaction limit. Once again, they restrict to the leading-order term, which is independent of the AHC lattice. In this section, we calculate the next contribution to determine the effect of the lattice shape. Using the quadratic dispersion of the parent band, the expectation value of the kinetic energy for the $\boldsymbol{k}$ ansatz state is
\begin{align}
	E(\boldsymbol{k})&= \frac{\sum_{\boldsymbol{g}} \frac{|\boldsymbol{k}+\boldsymbol{g}|^2}{2m} e^{ -\frac{|\boldsymbol{k}+\boldsymbol{g}|^2}{2 \xi^2}}}{\sum_{\boldsymbol{g}}  e^{ -\frac{|\boldsymbol{k}+\boldsymbol{g}|^2}{2 \xi^2}}}.
\end{align}

We can gain some intuition about how the kinetic energy depends on the lattice by considering the weak-interaction limit, where $\xi$ is very small. In this case, the Gaussian factor ensures that the ansatz state at crystal momentum $\boldsymbol{k}$ is comprised almost entirely of the parent band state at $\boldsymbol{k}+\boldsymbol{g}$ such that $|\boldsymbol{k}+\boldsymbol{g}|$ is minimized. This results in significant occupation of states only in the Wigner-Seitz cell version of the Brillouin zone, with the total kinetic energy being the sum of the dispersion over the Wigner-Seitz cell (because the parent band states with a given crystal momentum after band-folding must have a total occupation of one). A Wigner-Seitz cell with a smaller average $|\boldsymbol{k}+\boldsymbol{g}|^2$ will have lower kinetic energy, meaning that a lattice with a nearly circular Wigner-Seitz cell would be preferred. This favors the triangular lattice, which has a hexagonal cell, over the square lattice, which has a square cell. It also disfavors dilation, increasing the energy of the rectangular lattice over the square lattice. This rule is a general one since it does not depend on the Berry curvature or band geometry of the parent band, although it could be affected by the dispersion.

Although this effect is most pronounced for low $\xi$, we are more interested in the high $\xi$ limit. To access this, we perform the same Poisson summation that we employed for the Fock term. Firstly, from Eq.~\eqref{Equation_normalization_lattice_sum} we know that 
$$\sum_{\boldsymbol{g}}  e^{ -\frac{|\boldsymbol{k}+\boldsymbol{g}|^2}{2 \xi^2}} =\frac{2 \pi \xi^2}{A_{1\text{BZ}} }  \sum_{\boldsymbol{R}}  e^{i \boldsymbol{k} \cdot \boldsymbol{R}} e^{- \frac{\xi^2|\boldsymbol{R}|^2}{2}}.$$
We can then obtain $\sum_{\boldsymbol{g}} \frac{|\boldsymbol{k}+\boldsymbol{g}|^2}{2m} e^{ -\frac{|\boldsymbol{k}+\boldsymbol{g}|^2}{2 \xi^2}}$ from this expression by taking a derivative with respect to $\frac{1}{\xi^2}$ on both sides:
\begin{align*}
	&\frac{d}{d (1/\xi^2)}\sum_{\boldsymbol{g}}  e^{ -\frac{|\boldsymbol{k}+\boldsymbol{g}|^2}{2 \xi^2}}= \frac{d}{d (1/\xi^2)} \frac{2 \pi \xi^2}{A_{1\text{BZ}} }  \sum_{\boldsymbol{R}}  e^{i \boldsymbol{k} \cdot \boldsymbol{R}} e^{- \frac{\xi^2|\boldsymbol{R}|^2}{2}}\\
	& \implies \frac{|\boldsymbol{k}+\boldsymbol{g}|^2}{2m} e^{ -\frac{|\boldsymbol{k}+\boldsymbol{g}|^2}{2 \xi^2}}= \frac{2 \pi \xi^4}{mA_{1\text{BZ}} } \sum_{\boldsymbol{R}}  e^{i \boldsymbol{k} \cdot \boldsymbol{R}} e^{- \frac{\xi^2|\boldsymbol{R}|^2}{2}} \left(1- \frac{\xi^2|\boldsymbol{R}|^2}{2}\right). 
\end{align*}
Therefore, the kinetic energy of a single electron is given by 
\begin{align}
	E(\boldsymbol{k})&= \frac{\xi^2}{m} \frac{ \sum_{\boldsymbol{R}}  e^{i \boldsymbol{k} \cdot \boldsymbol{R}} e^{- \frac{\xi^2|\boldsymbol{R}|^2}{2}} \left(1- \frac{\xi^2|\boldsymbol{R}|^2}{2}\right)}{\sum_{\boldsymbol{R}}  e^{i \boldsymbol{k} \cdot \boldsymbol{R}} e^{- \frac{\xi^2|\boldsymbol{R}|^2}{2}}} =\frac{\xi^2}{m} \left(1-  \frac{ \sum_{\boldsymbol{R}}  e^{i \boldsymbol{k} \cdot \boldsymbol{R}} e^{- \frac{\xi^2|\boldsymbol{R}|^2}{2}}  \frac{\xi^2|\boldsymbol{R}|^2}{2}}{\sum_{\boldsymbol{R}}  e^{i \boldsymbol{k} \cdot \boldsymbol{R}} e^{- \frac{\xi^2|\boldsymbol{R}|^2}{2}}} \right). 
\end{align}

The zeroth-order term is $\xi^2/m$, which is independent of $\boldsymbol{k}$ and agrees with the result from Ref. \cite{tan2024parent}. Now, we wish to compute the correction to the total kinetic energy
\begin{equation}
	E_{\text{kin.}} = \sum_{\boldsymbol{k}} E(\boldsymbol{k})= \sum_{\boldsymbol{k}} \frac{\xi^2}{m} \left(1-  \frac{ \sum_{\boldsymbol{R}}  e^{i \boldsymbol{k} \cdot \boldsymbol{R}} e^{- \frac{\xi^2|\boldsymbol{R}|^2}{2}}  \frac{\xi^2|\boldsymbol{R}|^2}{2}}{\sum_{\boldsymbol{R}}  e^{i \boldsymbol{k} \cdot \boldsymbol{R}} e^{- \frac{\xi^2|\boldsymbol{R}|^2}{2}}} \right).\label{Eq_kinetic_pre_expansion}
\end{equation}

It is important to include the sum over $\boldsymbol{k}$ before we do the expansion. This is because, just as we saw for the Fock term, the first-order term would look like $e^{i \boldsymbol{k} \cdot\boldsymbol{R}}$, but this is destroyed by the sum over $\boldsymbol{k}$ unless $\boldsymbol{R}=0$. This means that the leading contribution to the correction will actually come from the second-order terms. We start by writing 
\begin{align*}
	\frac{ \sum_{\boldsymbol{R}}  e^{i \boldsymbol{k} \cdot \boldsymbol{R}} e^{- \frac{\xi^2|\boldsymbol{R}|^2}{2}}  \frac{\xi^2|\boldsymbol{R}|^2}{2}}{\sum_{\boldsymbol{R}}  e^{i \boldsymbol{k} \cdot \boldsymbol{R}} e^{- \frac{\xi^2|\boldsymbol{R}|^2}{2}}} &=  	\frac{  \sum_{\boldsymbol{R} \neq 0}  e^{i \boldsymbol{k} \cdot \boldsymbol{R}} e^{- \frac{\xi^2|\boldsymbol{R}|^2}{2}}  \frac{\xi^2|\boldsymbol{R}|^2}{2}}{1+ \sum_{\boldsymbol{R} \neq 0}  e^{i \boldsymbol{k} \cdot \boldsymbol{R}} e^{- \frac{\xi^2|\boldsymbol{R}|^2}{2}}} \\
	& \approx  \sum_{\boldsymbol{R} \neq 0}  e^{i \boldsymbol{k} \cdot \boldsymbol{R}} e^{- \frac{\xi^2|\boldsymbol{R}|^2}{2}}  \frac{\xi^2|\boldsymbol{R}|^2}{2} \left[1 - \sum_{\boldsymbol{R}' \neq 0}  e^{i \boldsymbol{k} \cdot \boldsymbol{R}'} e^{- \frac{\xi^2|\boldsymbol{R}'|^2}{2}} + \left(\sum_{\boldsymbol{R}' \neq 0}  e^{i \boldsymbol{k} \cdot \boldsymbol{R}'} e^{- \frac{\xi^2|\boldsymbol{R}'|^2}{2}}\right)^2 +... \right] .
\end{align*}

Requiring the oscillatory component to vanish, the first contribution must involve $\boldsymbol{R}=-\boldsymbol{R'}$, with $\boldsymbol{R}$ among the shortest lattice vectors. This term is of order $\xi^2 a^2 e^{- \xi^2 a^2}$, where $a$ is the lattice constant. For the square lattice, the next term comes either from including the next-shortest lattice vectors, with length $\sqrt{2}a$, or by taking the fourth-order term involving only the shortest lattice vectors. The next term is of order $4 \xi^2 a^2 e^{- 2\xi^2 a^2}$, so the realm of applicability is determined by $4e^{- \xi^2 a^2}$. The exponential decay of this next term is the same as for the Fock term, although the $|\boldsymbol{R}|^2$ factor in front of the expression for the kinetic term correction slows the decay with $|\boldsymbol{R}|$. This means that the region of applicability for our expansion likely starts at slightly higher $\xi$ than for the Fock term correction.

Using the smallest term, with $\boldsymbol{R}= -\boldsymbol{R}'$ so that it contributes after the sum over $\boldsymbol{k}$, we get
\begin{equation}
	E_{\text{kin.}} \approx  N \frac{\xi^2}{m} +   N \frac{\xi^4}{2m} \sum_{\text{shortest } \boldsymbol{R}}  |\boldsymbol{R}|^2 e^{- \xi^2|\boldsymbol{R}|^2} =E^{(0)}_{\text{kin.}}+E^{(1)}_{\text{kin.}}.
\end{equation}
We see that this correction gives an energetic cost to smaller $|\boldsymbol{R}|$, which disfavors the square lattice compared to the triangular lattice. This behavior is opposite to the Fock term, so the two energetic terms compete.

\subsection{Hartree term}
Unlike for the kinetic and Fock terms, Refs.~\cite{tan2024parent, tan2024wavefunction} do not give an explicit expression for the Hartree term. This is because it is heavily suppressed compared to the other terms. However, the Hartree term may be significant compared to the corrections to the other terms that we have considered so far. In this section, we will show that the Hartree term is negligible even in this context. The Hartree term is given by

\begin{align}
	\braket{H_{\text{Hartree}}}&=\frac{1}{2A} \sum_{\boldsymbol{k}_1, \boldsymbol{k}_2 \in \text{BZ}}  \sum_{\boldsymbol{g} \neq 0} V(- \boldsymbol{g}) F(\boldsymbol{k}_1 - \boldsymbol{g}, \boldsymbol{k}_1)  F(\boldsymbol{k}_2 + \boldsymbol{g}, \boldsymbol{k}_2). 
\end{align}
Using our expression for the form factor (Eq.~\eqref{Equation_form_factor_ansatz_no_delta}), we have
\begin{align*}
	F(\boldsymbol{k}_2 + \boldsymbol{g}, \boldsymbol{k}_2) =&  (\eta(\boldsymbol{g}))^\mathcal{C} e^{-\frac{\alpha|\boldsymbol{g}|^2}{8} } e^{2i \mathcal{C} \pi \frac{\boldsymbol{k}_2 \times \boldsymbol{g} }{A_{1\text{BZ}}}}    \frac{\sum_{\boldsymbol{R}}  e^{- \frac{|\boldsymbol{R}|^2 \xi^2}{2}} e^{i (\frac{\boldsymbol{g}}{2}+\boldsymbol{k}_2) \cdot \boldsymbol{R}} }{\sum_{\boldsymbol{R}}  e^{ - \frac{|\boldsymbol{R}|^2 \xi^2}{2}} e^{i \boldsymbol{k}_2 \cdot \boldsymbol{R}}},
\end{align*}
where we used $\boldsymbol{q}=\boldsymbol{g}$ and $\boldsymbol{g}_0(\boldsymbol{k}_2+\boldsymbol{g})=\boldsymbol{g}$.

Because the only place $\boldsymbol{k}_2$ enters the Hartree term is in this form factor, we can sum over $\boldsymbol{k}_2$ in the Brillouin zone here. This looks quite similar to the expression we had for the Fock term, and we can expand it in a similar way. However, whereas for the Fock term we needed our expansion over the $\boldsymbol{R}$ to have no net oscillatory term, in this case, we have an oscillatory pre-factor $\exp(2i \mathcal{C} \pi \frac{\boldsymbol{k}_2 \times \boldsymbol{g} }{A_{1\text{BZ}}})$ which must be canceled out. This means that we get a large decay factor from the $\boldsymbol{R}$ terms when $\boldsymbol{g}$ is large, as well as the existing prefactor (for $\mathcal{C} \neq 0$). This results in the Hartree term being very heavily suppressed, even compared to the correction to the Fock term. This is even more pronounced for higher $\mathcal{C}$ because the required oscillatory component becomes a larger lattice vector.

 The largest contribution will come from the smallest reciprocal lattice vectors $\boldsymbol{g}$ (note that we exclude $\boldsymbol{g}=0$, which corresponds to the long-ranged component of the Coulomb force). We start by expanding:
\begin{align*}
\sum_{\boldsymbol{k}_2}	F(\boldsymbol{k}_2 + \boldsymbol{g}, \boldsymbol{k}_2) & =  (\eta(\boldsymbol{g}))^\mathcal{C} e^{-\frac{\alpha|\boldsymbol{g}|^2}{8} } \sum_{\boldsymbol{k}_2}  e^{2i \mathcal{C} \pi \frac{\boldsymbol{k}_2 \times \boldsymbol{g} }{A_{1\text{BZ}}}}  \left(1+\sum_{\boldsymbol{R} \neq 0}  e^{- \frac{|\boldsymbol{R}|^2 \xi^2}{2}} e^{i (\frac{\boldsymbol{g}}{2}+\boldsymbol{k}_2) \cdot \boldsymbol{R}} \right) \\
	& \hspace{1cm} \times \left[1 -\sum_{\boldsymbol{R} \neq 0}  e^{- \frac{|\boldsymbol{R}|^2 \xi^2}{2}} e^{i \boldsymbol{k}_2 \cdot \boldsymbol{R}} +\left(\sum_{\boldsymbol{R} \neq 0}  e^{- \frac{|\boldsymbol{R}|^2 \xi^2}{2}} e^{i \boldsymbol{k}_2 \cdot \boldsymbol{R}}\right)^2 +...\right].
\end{align*}
We consider the term involving the 1 in the first bracket:
\begin{align*}
    S_1:=\sum_{\boldsymbol{k}_2}  e^{2i \mathcal{C} \pi \frac{\boldsymbol{k}_2 \times \boldsymbol{g} }{A_{1\text{BZ}}}}   \left[1 -\sum_{\boldsymbol{R} \neq 0}  e^{- \frac{|\boldsymbol{R}|^2 \xi^2}{2}} e^{i \boldsymbol{k}_2 \cdot \boldsymbol{R}} +\left(\sum_{\boldsymbol{R} \neq 0}  e^{- \frac{|\boldsymbol{R}|^2 \xi^2}{2}} e^{i \boldsymbol{k}_2 \cdot \boldsymbol{R}}\right)^2 +...\right].
\end{align*}
We can write this as
\begin{align*}
	S_1&=\sum_{\boldsymbol{k}_2}  e^{2i \mathcal{C} \pi \frac{\boldsymbol{k}_2 \times \boldsymbol{g} }{A_{1\text{BZ}}}}   \sum_{n=0}^{\infty} (-1)^n \sum_{\boldsymbol{R}_1,..\boldsymbol{R}_n} e^{ - \sum_{j=1}^n\frac{|\boldsymbol{R}_j|^2 \xi^2}{2}} e^{i \boldsymbol{k}_2 \cdot \sum_{j=1}^n \boldsymbol{R}_j} \delta\left(\sum_j \boldsymbol{R}_j, -2\mathcal{C} \frac{\pi}{A_{1\text{BZ}}} \boldsymbol{\varepsilon} \boldsymbol{g}\right),
\end{align*}
where $\boldsymbol{\varepsilon}$ is the unit antisymmetric matrix and the Kronecker delta is from the requirement that the overall oscillatory component be trivial for the sum over $\boldsymbol{k}$ to be nonzero. Note that $2\mathcal{C} \frac{\pi}{A_{1\text{BZ}}} \boldsymbol{\varepsilon} \boldsymbol{g}$ is always a lattice vector. Indeed $f( \boldsymbol{g}) = \frac{2\pi}{A_{1\text{BZ}}} \boldsymbol{\varepsilon} \boldsymbol{g}$ defines an invertible map from the reciprocal lattice to the direct lattice, with $f^{-1}(\boldsymbol{a})= - \frac{A_{1\text{BZ}}}{2\pi} \boldsymbol{\varepsilon} \boldsymbol{a}$. As a result, $2\mathcal{C} \frac{\pi}{A_{1\text{BZ}}} \boldsymbol{\varepsilon} \boldsymbol{g}$ is a lattice vector with minimum length $\mathcal{C}a$, where $a$ is the lattice constant. Then
\begin{align*}
	S_1	&= \sum_{\boldsymbol{k}_2}    \sum_{n=0}^{\infty} (-1)^n \sum_{\boldsymbol{R}_1,..\boldsymbol{R}_n} e^{ - \sum_{j=1}^n\frac{|\boldsymbol{R}_j|^2 \xi^2}{2}}  \delta\left(\sum_j \boldsymbol{R}_j, -2\mathcal{C} \frac{\pi}{A_{1\text{BZ}}} \boldsymbol{\varepsilon} \boldsymbol{g}\right).
\end{align*}

We can think of this as a weighted sum over paths made from lattice vectors, with the Kronecker delta enforcing that the end-point of the path is $-2\mathcal{C} \frac{\pi}{A_{1\text{BZ}}} \boldsymbol{\varepsilon} \boldsymbol{g}.$ The weight depends on the sum of squared lengths for the segments. Because of this, the weight is higher if we take many small steps rather than a single segment that reaches the end-point. For general $\boldsymbol{g}$, we may need to consider many different paths with the same weight. However, the situation is simpler when we consider only $\boldsymbol{g}$ that are among the shortest reciprocal lattice vectors. Consider first the square lattice. Then the shortest RLV are $\frac{2 \pi}{a} (\pm 1,0)$ and $\frac{2 \pi}{a} (0,\pm 1)$, where $a$ is the lattice constant. Taking $\boldsymbol{G}_1 = \frac{2 \pi}{a} ( 1,0)$ as an example, we have

\begin{align*}
-2\mathcal{C} \frac{\pi}{A_{1\text{BZ}}} \boldsymbol{\varepsilon} \boldsymbol{G}_1 &= -\mathcal{C} \frac{4 \pi^2}{a A_{1\text{BZ}}} \begin{pmatrix} 0 &1 \\ -1 & 0 \end{pmatrix} \begin{pmatrix} 1 \\0 \end{pmatrix} = \begin{pmatrix} 0 \\\mathcal{C}a \end{pmatrix},
\end{align*}
where we used the fact that the Brillouin zone area is just $\frac{4 \pi^2}{a^2}$ for a square lattice. We see that this is along one of the primitive lattice directions, with a length of $\mathcal{C}$. 

For a more general lattice, it is still true that if $\boldsymbol{g}$ is one of the shortest reciprocal lattice vectors, then $f(\boldsymbol{g})=\frac{2\pi}{A_{1\text{BZ}}} \boldsymbol{\varepsilon} \boldsymbol{g}$ is one of the shortest lattice vectors. This is because the invertible map $f(\boldsymbol{g})$ gives a lattice vector with length proportional to $|\boldsymbol{g}|$, so the shortest reciprocal lattice vectors give the shortest lattice vectors (and the map is invertible, so all lattice vectors are reached by the map). As a result, $-\frac{2\mathcal{C}\pi}{A_{1\text{BZ}}} \boldsymbol{\varepsilon} \boldsymbol{g}$ is parallel to one of the shortest lattice vectors, $\boldsymbol{A}$, but with length equal to $\mathcal{C}$ times the length $a$ of that vector. The highest weighted path that reaches this vector is then made from $\mathcal{C}$ copies of $\boldsymbol{A}$, with the sum of squared lengths equal to $\mathcal{C} a^2$. The Gaussian factor attached to this path is then $\exp(-\frac{\mathcal{C}a^2 \xi^2}{2})$. If $\mathcal{C} >1$, this is significantly better than the term involving only one vector, which would have a sum of squared lengths $\mathcal{C}^2 a^2$. Only including the largest term for $\boldsymbol{g}$, we get the approximation for $S_1$ as
\begin{equation}
S_1 \approx N(-1)^\mathcal{C} e^{-\frac{\mathcal{C}a^2 \xi^2}{2}}
\end{equation}

Next, we consider the term involving the other part of the numerator:
\begin{align*}
	S_2&=\sum_{\boldsymbol{k}_2} \sum_{\boldsymbol{R} \neq 0}  e^{ - \frac{|\boldsymbol{R}|^2 \xi^2}{2}} e^{i (\frac{\boldsymbol{g}}{2}+\boldsymbol{k}_2) \cdot \boldsymbol{R}}
\left[1 -\sum_{\boldsymbol{R} \neq 0}  e^{ - \frac{|\boldsymbol{R}|^2 \xi^2}{2}} e^{i \boldsymbol{k}_2 \cdot \boldsymbol{R}} +\left(\sum_{\boldsymbol{R} \neq 0}  e^{ - \frac{|\boldsymbol{R}|^2 \xi^2}{2}} e^{i \boldsymbol{k}_2 \cdot \boldsymbol{R}}\right)^2 +...\right]\\
&= \sum_{n=1}^{\infty} (-1)^{n-1} \sum_{\boldsymbol{R}_1,..\boldsymbol{R}_n} e^{ - \sum_{j=1}^n\frac{|\boldsymbol{R}_j|^2 \xi^2}{2} } e^{i \frac{\boldsymbol{g}}{2} \cdot \boldsymbol{R}_1} \delta\left(\sum_j \boldsymbol{R}_j, -2\mathcal{C} \frac{\pi}{A_{1\text{BZ}}} \epsilon \boldsymbol{g}\right).
\end{align*}

The same logic as before applies. This time, we have the phase factor $\exp(i \frac{\boldsymbol{g}}{2} \cdot \boldsymbol{R}_1)$. In our leading term, $\boldsymbol{R}_1$ is orthogonal to $\boldsymbol{g}$, so the phase factor is 1. Then, because we have $(-1)^{n-1}$ rather than $(-1)^n$, this cancels with the leading term from $S_1$ giving us zero. As a result, $S_1+S_2$ decays faster than $\exp(-\frac{\mathcal{C}a^2 \xi^2}{2})$. The Hartree term, which includes two copies of the form factor (one for $\boldsymbol{k}_1$ and one for $\boldsymbol{k}_2$), therefore decays faster than $\exp(-\mathcal{C}a^2 \xi^2)$ even before we consider the other factors. Therefore, it decays faster than the first corrections to the Fock and kinetic terms and should only be included if further corrections to those terms are also used.

\subsection{Convergence of the perturbative expansion}
\label{Section_perturb_convergence}

\begin{figure}
    \centering
    \includegraphics[width=0.8\linewidth]{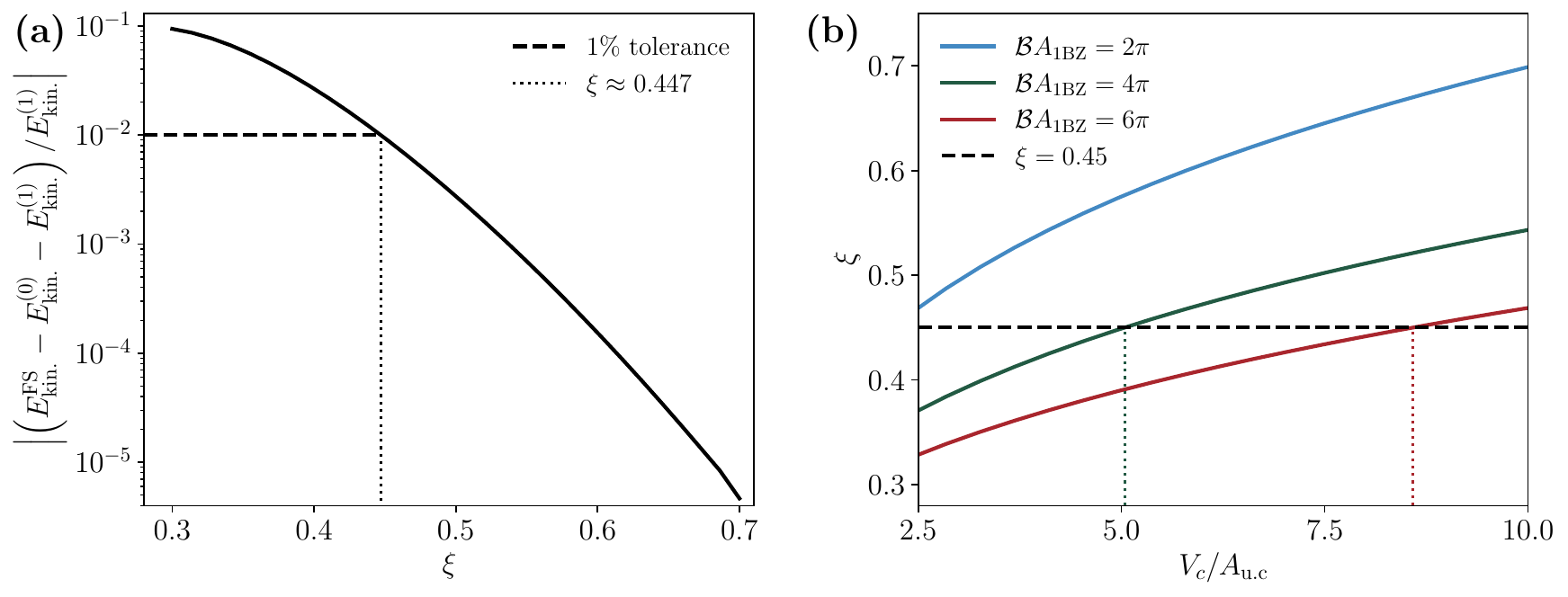}
    \caption{(a) The relative size of the correction to the kinetic energy not included in the perturbative expression, compared to the correction that is included. The ``true'' kinetic energy is calculated using Eq.~\eqref{Eq_kinetic_pre_expansion} with the Brillouin zone represented by a 500 by 500 grid and summing over lattice vectors to a radius of 10 times the lattice constant. (b) The dependence of the variational parameter $\xi$ on interaction strength from the perturbative approach. We expect the expansion to work well above $\xi \approx 0.45$.}
    \label{fig:XiInteraction}
\end{figure}

In this section, we discuss the convergence of the perturbative expansion in slightly more detail. We estimated that the general region of convergence for the triangular lattice should be $\xi >0.42$, which we round up to $\xi >0.45$. We can check this rough estimate by using the kinetic energy for the triangular lattice, which we expect to have the slowest convergence of the terms that we have considered so far. The kinetic energy can be calculated numerically for large system sizes and large cutoffs. By comparing the difference between this value, $E^{\text{FS}}_{\text{kin.}}$ and the perturbative calculation $E^{(0)}_{\text{kin.}} + E^{(1)}_{\text{kin.}}$, then dividing it by the first-order correction $E^{(1)}_{\text{kin.}}$ in the perturbative calculation, we can estimate the relative strength of the uncalculated higher order terms in the perturbative expansion. We compare the strength of these terms to the first correction rather than the entire kinetic energy because the zeroth-order contribution does not depend on the lattice. Accordingly, it does not affect quantities like the stiffness. As shown in Fig. \ref{fig:XiInteraction}a, the untreated terms are of the order 0.01 compared to the first correction above $\xi \approx 0.45$, roughly agreeing with our previous estimate.

So far, we have estimated when the perturbative expansion should work well in terms of the variational parameter. However, we should also know what values of interaction strength this corresponds to. In Fig. \ref{fig:XiInteraction}b, we plot the optimized variational parameter as a function of $V_c$ for different values of the Berry curvature and compare this to $\xi_c=0.45$, above which we expect the leading-order expansion to be accurate. As we see from the plot, the expansion should work well for $\mathcal{B}A_{1\text{BZ}}=2\pi$ for all interaction strengths that we consider. On the other hand, the expansion is only likely to give highly accurate answers for $\mathcal{B}A_{1\text{BZ}}=4\pi$ above $V_c/A_{\text{u.c}} \approx 5$ and for $\mathcal{B}A_{1\text{BZ}}=6\pi$ above $V_c/A_{\text{u.c}} \approx 8.5$.

\section{Variation of the kinetic energy with a Mexican hat dispersion}

In the main text, we have mentioned that having a dispersion with a minimum at a finite momentum $|\boldsymbol{k}|=k_0$ as
\begin{align}
    \mathcal{E}(\boldsymbol{k})=D\left(\left(\frac{|\boldsymbol{k}|}{k_0}\right)^4 - \left(\frac{|\boldsymbol{k}|}{k_0}\right)^2\right)
\end{align}
may lead to a mechanical instability of the triangular lattice electronic crystal. Indeed, if the minimum $k_0$ sits just outside of the first Brillouin zone, the electronic crystal can reduce its kinetic energy by distorting the triangular lattice to more significantly occupy the states with low kinetic energy.

\begin{figure}
    \centering
    \includegraphics[width=0.65\linewidth]{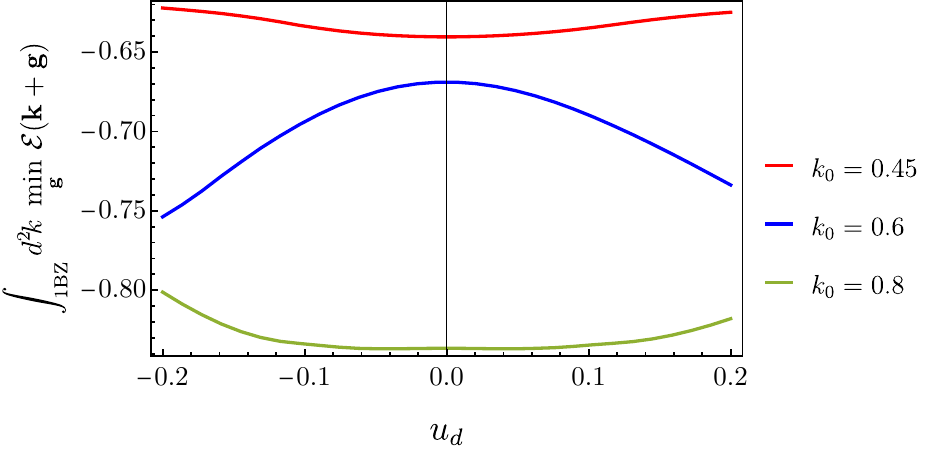}
    \vspace{-2mm}
    \caption{Variation of the kinetic energy assuming we have a single electron per crystal momentum $\boldsymbol{k}$ in the 1BZ $\int_{\text{1BZ}}d^2k\min_{\boldsymbol{g}}\mathcal{E}(\boldsymbol{k}+\boldsymbol{g})$ as we vary the shape of the underlying lattice with $u_d$. The variation of the kinetic energy is shown for different values of $k_0$ where the dispersion is $\mathcal{E}(\boldsymbol{k})=D\left(\left(|\boldsymbol{k}|/k_0\right)^4 - \left(|\boldsymbol{k}|/k_0\right)^2\right)$ with $D=1$. We note that, as in the main text, the shortest reciprocal lattice vectors have unit length $|\boldsymbol{G}_{1,2}|=1$ for $u_d=0$.
    \label{si_fig:min_kinetic_energy_mexican_hat}}
\end{figure}

To substantiate this idea, we show in Fig.~\ref{si_fig:min_kinetic_energy_mexican_hat} the variation of the kinetic energy for distorted crystals as a function of $u_d$. We assume that we have a single electron per crystal momentum $\boldsymbol{k}$ in the first BZ that occupies the lowest kinetic energy state, such that
\begin{align}
    E_{\text{kin}} = \int_{\text{1BZ}}d^2k\min_{\boldsymbol{g}}\left[\mathcal{E}(\boldsymbol{k}+\boldsymbol{g})\right].
\end{align}
Taking the the shortest reciprocal lattice vectors to have unit length $|\boldsymbol{G}_{1,2}|=1$ for $u_d=0$, we see in Fig.~\ref{si_fig:min_kinetic_energy_mexican_hat} that if the minimum of the dispersion sits inside the first Brillouin zone (e.g., $k_0=0.45$), the kinetic energy of the crystal is minimized for a triangular lattice ($u_d=0$). In contrast, if we have a dispersion minimum that sits just outside the first Brillouin zone (e.g., $k_0=0.6$ as in the main text), the crystal can reduce its kinetic energy by distorting the triangular lattice to more significantly occupy states with small dispersion. As argued in the main text, it is exactly this reduction of the kinetic energy that is driving the mechanical instability of the AHC with the Mexican hat dispersion and in R5G for a strong displacement field.

\section{Rhombohedral pentalayer graphene} \label{si_sec:Multilayer_Rhombohedral_Graphene}

For pentalayer rhombohedral graphene, we follow the modeling used in Ref.~\cite{dong2024theory}. For this work to be self-contained, we briefly review and summarize the construction below. 

\subsection{Microscopic model} \label{si_subsec:Multilayer_Rhombohedral_Graphene_microscopic_model}

\subsubsection{Moiré lattice}\label{si_subsubsec:Multilayer_Rhombohedral_Graphene_microscopic_model_moire_lattice}

The initial graphene reciprocal lattice basis vectors are 
\begin{subequations}
    \begin{align}
        \boldsymbol{G}_1 &= \frac{4\pi}{\sqrt{3} a_{\text{G}}}\left( 0,  1\right) \\
        \boldsymbol{G}_2 &= \frac{4\pi}{\sqrt{3} a_{\text{G}}}\left( -\sqrt{3}, 1 \right).
    \end{align}
\end{subequations}
The associated real space basis vectors respect $\boldsymbol{A}_i \cdot \boldsymbol{G}_j = 2 \pi \delta_{ij}$. The real space basis vectors of the hBN substrate are obtained as
\begin{align}
    \boldsymbol{A}_j^{\prime}=M R[\theta] \boldsymbol{A}_j ; \quad M=\frac{1}{1+\varepsilon} I,
\end{align}
where $R[\theta]$ is a counter-clockwise rotation matrix and $\varepsilon=(a_{\text{G}}/a_{\text{hBN}}-1)\approx -0.01698$ is the lattice mismatch. We note that the lattice constants of monolayer graphene and hBN are $a_{\text{G}}=0.246 \mathrm{~nm}$ and $a_{\text{hBN}}=0.25025 \mathrm{~nm}$, respectively.

The moiré reciprocal lattice vectors obtained by stacking graphene on top of hBN with a twist angle $\theta$ are
\begin{align}
    \boldsymbol{G}_i^{\text{mBZ}} = \boldsymbol{G}_i - \boldsymbol{G}_i' = (1 - M^{-1}R[\theta])\boldsymbol{G}_i \approx \varepsilon \boldsymbol{G}_i - \theta \hat{z} \cross \boldsymbol{G}_i,
\end{align}
where the approximation holds for small twist angle and lattice mismatch. 
It is also convenient to define the reciprocal lattice vector $\boldsymbol{G}_3^{\text{mBZ}} =\boldsymbol{G}_2^{\text{mBZ}}  - \boldsymbol{G}_1^{\text{mBZ}} $,  $\boldsymbol{G}_4^{\text{mBZ}} =-\boldsymbol{G}_1^{\text{mBZ}}$,  $\boldsymbol{G}_5^{\text{mBZ}} =-\boldsymbol{G}_2^{\text{mBZ}}$, and $\boldsymbol{G}_6^{\text{mBZ}} =-\boldsymbol{G}_3^{\text{mBZ}} $. The twist angle of $0.77^{\circ}$ reported in experiments~\cite{lu2024fractional}, leads to a moiré lattice constant of $\lambda_M = |\boldsymbol{A}_i^{\text{mBZ}}| \approx a_{\text{G}} / \sqrt{\varepsilon^2+\theta^2} \approx 11.4 \mathrm{~nm}$.

\subsubsection{Moiré rhombohedral graphene Hamiltonian}

The moiré rhombohedral graphene Hamiltonian is
\begin{align}
    H = H_{\text{R5G}} + H_{M} + H_{C},
\end{align}
where $H_{\text{R5G}}$ is the continuum kinetic term, $H_{M}$ the moiré potential from the hBN substrate that is acting on the bottom graphene layer, and $H_{C}$ the Coulomb potential. The continuum kinetic term is obtained by expanding the rhombohedral pentalayer graphene Hamiltonian about the $\boldsymbol{K}$ and $\boldsymbol{K}'$ valleys
\begin{align}
    H_{\text{R5G}} &= \sum_{\boldsymbol{k}} \sum_{\alpha, \beta, \ell, \ell'} \sum_{\eta, \sigma} c^\dagger_{\boldsymbol{k},\alpha \ell \eta \sigma} \left[ h^{\eta}_{\text{R5G}}(\boldsymbol{k}) \right]_{(\alpha \ell),(\beta \ell')} c_{\boldsymbol{k},\beta \ell' \eta \sigma}, 
\end{align}
where the different indices label the $\alpha,\beta\in\{A,B\}$ sublattices, $\ell,\ell'\in\{1,2,3,4,5\}$ layers, $\sigma\in\{\uparrow, \downarrow\}$ spin, and $\eta\in\{\boldsymbol{K},\boldsymbol{K}'\}$ valley degrees of freedom. In the $(\alpha\ell)\in\{ (A,1), (B,1), (A,2), (B,2), \ldots, (A,5), (B,5) \}$ basis, the $h^{\eta}_{\text{R5G}}$ matrix takes the form
\begin{align}
    \left[ h^{\eta}_{\text{R5G}}(\boldsymbol{k}) \right] &= \mqty(
    h_{1}^{(0)} & h^{(1)} & h^{(2)} & 0_{2\times 2} & 0_{2\times 2} \\
    h^{(1)\dagger} & h_{2}^{(0)} & h^{(1)} & h^{(2)} & 0_{2\times 2} \\
    h^{(2)\dagger} & h^{(1)\dagger} & h_{3}^{(0)} & h^{(1)} & h^{(2)} \\
    0_{2\times 2} & h^{(2)\dagger} & h^{(1)\dagger} & h_{4}^{(0)} & h^{(1)} \\
    0_{2\times 2} & 0_{2\times 2} & h^{(2)\dagger} & h^{(1)\dagger} & h_{5}^{(0)} \\
    ),
\end{align}
where the intralayer term splits into a kinetic $h^{(0)}(\boldsymbol{k})$, inversion symmetric potential $h^{\text{ISP}}_{\ell}$ and displacement field $h^{\text{D}}_{\ell}$ parts
\begin{align}
    h_{\ell}^{(0)}(\boldsymbol{k}) =  h^{(0)}(\boldsymbol{k}) + h^{\text{ISP}}_{\ell} + h^{\text{D}}_{\ell}. 
\end{align}
The layer-dependent inversion symmetric potential is 
\begin{align}
    h_{1}^{\text{ISP}} = \mqty(0 & 0 \\ 0 & \delta), \quad
    h_{2}^{\text{ISP}} = h_{3}^{\text{ISP}} = h_{4}^{\text{ISP}} = \mqty(u_a & 0 \\ 0 & u_a), \quad
    h_{5}^{\text{ISP}} = \mqty(\delta & 0 \\ 0 & 0).
\end{align}
The effect of the displacement field is modeled as a constant potential difference between different layers 
\begin{align}
    h_{\ell}^{D} &= U_d \left(3 - \ell\right) \mathds{1}_{2 \cross 2}.
\end{align}
The intralayer kinetic and inter-layer coupling terms are
\begin{subequations}    
\begin{align}
    h^{(0)}(\boldsymbol{k}) &= \mqty(0 & v_0^* \\ v_0 & 0), \\
    h^{(1)}(\boldsymbol{k}) &= \mqty(v_4^* & v_3 \\ \gamma_1 & v_4^*), \\
    h^{(2)}(\boldsymbol{k}) &= \mqty(0 & \gamma_2/2 \\ 0 & 0),
\end{align}
\end{subequations}
with the shorthand notation $v_i \equiv \sqrt{3} \gamma_i/2 \left( \pm k_x + i k_y \right)$, where $k_{x,y}$ are small momentum components expanded around $\boldsymbol{K}$ or $\boldsymbol{K}'$ and the sign $\pm$ depends on the valley of interest. The hopping parameters are taken from DFT on  rhombohedral-stacked trilayer graphene~\cite{zhang2010band}, and the on-site potentials are in agreement with those of rhombohedral-stacked tetralayer graphene~\cite{Park2023topological}
\begin{subequations}
    \begin{align}
        \gamma_0 &= 2600~\text{meV} \\
        \gamma_1 &= 356.1~\text{meV} \\
        \gamma_2 &= -15~\text{meV} \\
        \gamma_3 &= -293~\text{meV} \\
        \gamma_4 &= -144~\text{meV} \\
        \delta &= 12.2~\text{meV} \\
        u_a &= -16.4~\text{meV}.
    \end{align}
\end{subequations}

The moiré potential term is
\begin{align}
    H_M = \sum_{\boldsymbol{k}} \sum_{i=1}^{6} \sum_{\alpha, \beta} \sum_{\eta, \sigma}  c^\dagger_{\boldsymbol{k} + \boldsymbol{G}_i^{\text{mBZ}} , \alpha 1 \eta \sigma} [V_{M}^{\eta}(\boldsymbol{G}_i^{\text{mBZ}})]_{\alpha,\beta}  c^\dagger_{\boldsymbol{k}, \beta 1 \eta \sigma}.
\end{align}
It only acts on the bottom graphene layer (i.e., $\ell=1$), and only the first harmonics are kept~\cite{dong2024theory}. In the $\boldsymbol{K}$-valley, the $V_{M}^{\boldsymbol{K}}$ matrix in the $\{(A,1),(B,1)\}$ subspace takes the form~\cite{jung2014abinitio, jung2015origin, dong2024theory}
\begin{align}
[V_M^{\boldsymbol{K}}\left(\boldsymbol{G}_i^{\text{mBZ}} \right)] & = \mqty( 
    V_{A A}\left(\boldsymbol{G}_i^{\text{mBZ}} \right) & V_{A B}\left(\boldsymbol{G}_i^{\text{mBZ}} \right) \\
    V_{A B}\left(-\boldsymbol{G}_i^{\text{mBZ}} \right)^* & V_{B B}\left(\boldsymbol{G}_i^{\text{mBZ}} \right)
)
\end{align}
with 
\begin{subequations}
    \begin{align}
        & V_{A A / B B}\left(\boldsymbol{G}_{1,3,5}^{\text{mBZ}} \right)=\left[V_{A A / B B}\left(\boldsymbol{G}_{2,4,6}^{\text{mBZ}} \right)\right]^*=C_{A A / B B} e^{-i \phi_{A A / B B}} \\
        & V_{A B}\left(\boldsymbol{G}_1^{\text{mBZ}} \right)=\left[V_{A B}\left(\boldsymbol{G}_4^{\text{mBZ}} \right)\right]^*=C_{A B} e^{2 \pi i / 3} e^{-i \phi_{A B}} \\
        & V_{A B}\left(\boldsymbol{G}_5^{\text{mBZ}} \right)=\left[V_{A B}\left(\boldsymbol{G}_6^{\text{mBZ}} \right)\right]^*=C_{A B} e^{-2 \pi i / 3} e^{-i \phi_{A B}} \\
        & V_{A B}\left(\boldsymbol{G}_3^{\text{mBZ}} \right)=\left[V_{A B}\left(\boldsymbol{G}_2^{\text{mBZ}} \right)\right]^*=C_{A B} e^{-i \phi_{A B}},
    \end{align} 
\end{subequations}
and
\begin{subequations}
    \begin{align}
        C_{AA} &= -14.88~\text{meV} \\
        C_{BB} &= 12.09~\text{meV} \\
        C_{AB} &= 11.34~\text{meV} \\
        \phi_{AA} &= 50.19^\circ \\
        \phi_{BB} &= -46.64^\circ \\
        \phi_{AB} &= 19.60^\circ.
    \end{align}
\end{subequations}

Finally, the Coulomb interaction is 
\begin{align}
    H_C  = \frac{1}{2 A} \sum_{\boldsymbol{q}} V_c^{\mathrm{sc}}(\boldsymbol{q}) :\rho_{\boldsymbol{q}} \rho_{-\boldsymbol{q}}: = \frac{1}{2 A} \sum_{\boldsymbol{k}, \boldsymbol{k}^{\prime}, \boldsymbol{q}} \sum_{\substack{\alpha,\ell,\eta,\sigma\\ \beta,\ell',\eta',\sigma'}} V_c^{\mathrm{sc}}(\boldsymbol{q}) c_{\boldsymbol{k}+\boldsymbol{q}, \alpha \ell \eta \sigma}^{\dagger} c_{\boldsymbol{k}^{\prime}-\boldsymbol{q}, \beta \ell' \eta' \sigma'}^{\dagger} c_{\boldsymbol{k}^{\prime}, \beta \ell' \eta' \sigma'} c_{\boldsymbol{k}, \alpha \ell \eta \sigma},
\end{align}
where we have a dual-gated screened interaction
\begin{align}
    V_c^{\mathrm{sc}}(\boldsymbol{q})= \frac{e^2 \tanh \left(|\boldsymbol{q}| d_s\right)}{2 \epsilon_0 \epsilon|\boldsymbol{q}|},
\end{align}
with a gate distance of $d_s=30~\text{nm}$.

\subsubsection{Band basis}

Due to the large number of bands (coming from the sublattices, layers, and reciprocal lattice vectors within the momentum cutoff considered), the Hartree-Fock calculation will be done by only considering the $n_{\text{bands}}$ lowest conduction bands~\cite{dong2024theory, kwan2023moire, dong2024anomalous, zhou2024fractional, guo2024theory}. To do so, it is first convenient to work in a band basis obtained by diagonalizing the quadratic part of the Hamiltonian  
\begin{align}
    H_{\text{Kin}} &= H_{\text{R5G}} + H_{M} = \sum_{\boldsymbol{k}}  \sum_{\eta,\sigma} c_{\boldsymbol{k}, \boldsymbol{g} \alpha \ell \eta \sigma}^\dagger [h_{\text{Kin}}^{\eta}(\boldsymbol{k})]_{(\boldsymbol{g}\alpha\ell),(\boldsymbol{g}'\beta\ell')} c_{\boldsymbol{k}, \boldsymbol{g}' \beta \ell' \eta \sigma},
\end{align}
where $c_{\boldsymbol{k}+\boldsymbol{g}, \alpha \ell \eta \sigma}\equiv c_{\boldsymbol{k}, \boldsymbol{g} \alpha \ell \eta \sigma}$ with $\boldsymbol{g}=m\boldsymbol{G}_1^{\text{mBZ}}  + n\boldsymbol{G}_2^{\text{mBZ}} $ ($m,n\in\mathbb{Z}$). 
The eigenstates of $H_{\text{Kin}}$ in valley $\eta$ with momentum $\boldsymbol{k}$ and energy $\xi^{\eta}_{\boldsymbol{k},m}$ are denoted by 
\begin{align}
    \ket{\psi_{\boldsymbol{k},m\eta\sigma}} = \psi^\dagger_{\boldsymbol{k},m\eta\sigma} \ket{0} = \sum_{\boldsymbol{g},\alpha,\ell} \mu^{\eta}_{\boldsymbol{g}\alpha\ell,m}(\boldsymbol{k}) c^\dagger_{\boldsymbol{k},\boldsymbol{g}\alpha\ell\eta\sigma} \ket{0},
\end{align}
where $m$ is a band index. We work in periodic gauge  $\mu^{\eta}_{\boldsymbol{g}-\boldsymbol{g}' \alpha\ell, n}\left(\boldsymbol{k}+\boldsymbol{g}'\right)= \mu^\eta_{\boldsymbol{g}, \alpha, n}(\boldsymbol{k})$ such that $\psi_{\boldsymbol{k}+\boldsymbol{g}, n \eta \sigma}^{\dagger} = \psi_{\boldsymbol{k}, n \eta \sigma}^{\dagger}$. Explicitly, the $\mu^{\eta}_{\boldsymbol{g}\alpha\ell,m}(\boldsymbol{k})$ matrix is defined by
\begin{align}
    \sum_{\boldsymbol{g},\alpha,\ell} \sum_{\boldsymbol{g}',\beta,\ell'} \mu^{\eta *}_{\boldsymbol{g}\alpha\ell,m}(\boldsymbol{k}) \left[ h_{\text{Kin}}^{\eta}(\boldsymbol{k})\right]_{(\boldsymbol{g}\alpha\ell),(\boldsymbol{g}'\beta\ell')} \mu^\eta_{\boldsymbol{g}'\beta\ell',n}(\boldsymbol{k}) = \delta_{mn}\xi^{\eta}_{\boldsymbol{k},m}.
\end{align}

In this band basis, the density operator is
\begin{align}
    \rho_{\boldsymbol{q}} = \sum_{\boldsymbol{k}} \sum_{\boldsymbol{g},\alpha,\ell} \sum_{\eta,\sigma} c_{\boldsymbol{k}+\boldsymbol{q},\boldsymbol{g}\alpha\ell\eta\sigma}^\dagger c_{\boldsymbol{k},\boldsymbol{g}\alpha\ell\eta\sigma} = \sum_{\boldsymbol{k}}\sum_{m,n}\sum_{\eta,\sigma} \psi_{\boldsymbol{k} + \boldsymbol{q},m\eta\sigma}^{\dagger}  \Lambda_{mn}^{\eta}(\boldsymbol{k} + \boldsymbol{q}, \boldsymbol{k}) \psi_{\boldsymbol{k},n\eta\sigma},
\end{align}
where we have introduced the form factors
\begin{align}
    \Lambda_{mn}^{\eta}(\boldsymbol{k},\boldsymbol{q}) &= \sum_{\boldsymbol{g},\ell,\alpha} \mu^{\eta *}_{\boldsymbol{g}\alpha\ell,m}(\boldsymbol{k}) \mu^{\eta}_{\boldsymbol{g}\alpha\ell,n}(\boldsymbol{q}).
\end{align}
As such, the Coulomb interaction can be written as
\begin{align}
    H_{C} &= \frac{1}{2 A} \sum_{\boldsymbol{k}, \boldsymbol{k}^{\prime}, \boldsymbol{q}} \sum_{m,n,o,p} \sum_{\eta,\sigma,\eta',\sigma'} V_c^{\mathrm{sc}}(\boldsymbol{q}) \Lambda^{\eta}_{mn}(\boldsymbol{k} + \boldsymbol{q}, \boldsymbol{k}) \Lambda^{\eta'}_{op}(\boldsymbol{k}' - \boldsymbol{q}, \boldsymbol{k}') 
    \psi_{\boldsymbol{k}+\boldsymbol{q}, m \eta \sigma}^{\dagger} \psi_{\boldsymbol{k}^{\prime}-\boldsymbol{q}, o \eta' \sigma'}^{\dagger} \psi_{\boldsymbol{k}^{\prime}, p \eta' \sigma'} \psi_{\boldsymbol{k}, n \eta \sigma}.
\end{align}

\subsection{Hartree-Fock calculations} \label{si_subsec:Multilayer_Rhombohedral_Graphene_HF}

For the Hartree-Fock calculations, we assume a spin- and valley-polarized state. To simplify the notation, we will suppress spin and valley indices in the following. Performing a mean-field decoupling (similarly to the parent band model) leads to the Hartree and Fock terms
\begin{subequations} \label{eq:hartree_fock_terms_r5g}
    \begin{align}
        H_H &= \frac{1}{A} \sum_{\boldsymbol{k}, \boldsymbol{k}', \boldsymbol{g}} \sum_{m,n,o,p} V_{c}^{\text{sc}}(\boldsymbol{g}) \Lambda_{mn}(\boldsymbol{k}+\boldsymbol{g}, \boldsymbol{k}) \Lambda_{op}\left(\boldsymbol{k}'-\boldsymbol{g}, \boldsymbol{k}'\right) \mathcal{P}_{op}\left(\boldsymbol{k}'\right) \psi_{\boldsymbol{k},m}^{\dagger} \psi_{\boldsymbol{k},n} \label{eq:hartree_term_r5g} \\
        H_F &= -\frac{1}{A} \sum_{\boldsymbol{k}, \boldsymbol{k}', \boldsymbol{g}} \sum_{m,n,o,p} V_{c}^{\text{sc}}\left(\boldsymbol{k}-\boldsymbol{k}'+\boldsymbol{g}\right) \Lambda_{mn}\left(\boldsymbol{k}+\boldsymbol{g}, \boldsymbol{k}'\right) \Lambda_{op}\left(\boldsymbol{k}'-\boldsymbol{g}, \boldsymbol{k}\right) \mathcal{P}_{o n}\left(\boldsymbol{k}'\right) \psi_{\boldsymbol{k},m}^{\dagger} \psi_{\boldsymbol{k},p}, \label{eq:fock_term_r5g}
    \end{align}
\end{subequations}
where the density matrix is $\mathcal{P}_{mn}(\boldsymbol{k})=\langle\psi_{\boldsymbol{k},m}^{\dagger} \psi_{\boldsymbol{k},n}\rangle$. We keep the $n_{\text{bands}}$ lowest conduction bands. Some ambiguity exists in restricting HF calculations to low-energy bands~\cite{kwan2023moire, dong2024anomalous}. In our case, we implement the projection to the lowest conduction bands by restricting the band summation in Eq.~\eqref{eq:hartree_fock_terms_r5g} to the corresponding indices. This would correspond to the ``charge neutrality scheme'' in Refs.~\cite{kwan2023moire, huang2024self}. To solve self-consistently for $\mathcal{P}_{mn}(\boldsymbol{k})=\expval{\psi_{\boldsymbol{k},m}^{\dagger} \psi_{\boldsymbol{k},n}}$, we discretize the first Brillouin zone, introduce a momentum cutoff, and use periodic Pulay mixing in the same way as the parent band model (see Sec.~\ref{si_subsec:Hartree_fock_numerics}).

\subsection{Existence of the AHC} \label{si_subsec:Multilayer_Rhombohedral_Graphene_existence_ahc}

\begin{figure}
\includegraphics[width=0.99\linewidth]{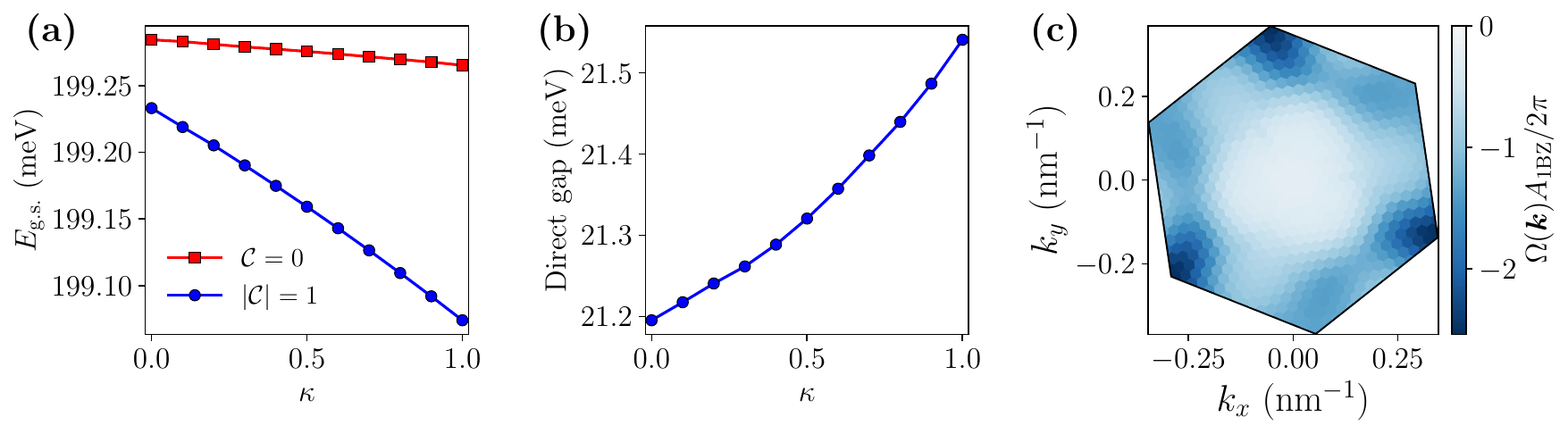}
\caption{Evolution of the (a) ground state energy per conduction electron of the $|\mathcal{C}|=1$ Chern and trivial ($\mathcal{C}=0$) insulators and (b) of the $|\mathcal{C}|=1$ Chern insulator direct band gap between the physical ($\kappa=1$) and moiréless ($\kappa=0$) regimes. (c) Berry curvature distribution of the $|\mathcal{C}|=1$ AHC's first conduction band when $\kappa=0$. Simulations are for $\epsilon=8.07$, $U_d=-36$~meV and $\theta=0.77^\circ$ with $n_1=25$ and $n_{\text{bands}}=4$. \label{si_fig:existence_of_ahc}}
\end{figure}

We briefly summarize the argument for the existence of an AHC in HF calculation of R5G that was highlighted in previous work~\cite{dong2024anomalous, zhou2024fractional, dong2024theory}. HF calculations of the system at unity filling with respect to the moiré unit cell and in a strong displacement field that polarizes the conduction electrons away from the moiré potential show that interaction leads to spin and valley polarization and an isolated fully-filled Chern $|\mathcal{C}|=1$ band~\cite{dong2024anomalous, dong2024theory, zhou2024fractional, guo2024theory, kwan2023moire, Huang2025, huang2024self}. Since the conduction electrons are polarized away from the hBN, one may naturally wonder if the underlying moiré potential is required to stabilize the Chern insulator. To investigate this point, the ground state of the model $H = H_{\text{R5G}} + \kappa H_{M} + H_{C}$ can be tracked as the moiré potential is completely removed ($\kappa=0$) to see if the Chern insulator remains stable. Fig.~\ref{si_fig:existence_of_ahc}(a) shows that the $|\mathcal{C}|=1$ Chern insulator remains lower in energy than the trivial $\mathcal{C}=0$ insulator as one interpolates between the physical ($\kappa=1$) and moiréless ($\kappa=0$) limits. The Chern insulator also remains gapped in the continuum limit when the moiré potential is removed, as illustrated in Fig.~\ref{si_fig:existence_of_ahc}(b). This indicates that HF predicts an AHC ground state that spontaneously breaks translation symmetry when $\kappa=0$. We note here that the HF calculations presented in Fig.~\ref{si_fig:existence_of_ahc} assume spin and valley polarization and only allow the system to spontaneously break translation symmetry with the same direction and periodicity as the moiré lattice even when it is completely removed.

\subsection{Lattice deformations} \label{si_subsec:Multilayer_Rhombohedral_Graphene_lattice_deformations}

In this subsection, we describe how to parameterize distortions for R5G/hBN. One subtlety is that the lattice vectors now depend on the specific twist angle considered (see~\ref{si_subsubsec:Multilayer_Rhombohedral_Graphene_microscopic_model_moire_lattice}). The real space basis vectors (i.e., satisfying $\boldsymbol{A}^{\text{mBZ}}_i\cdot\boldsymbol{G}^{\text{mBZ}}_j=2\pi\delta_{ij}$) are
\begin{subequations}
    \begin{align}
        \boldsymbol{A}_{1}^{\text{mBZ}} &= \frac{a}{2\sqrt{3}(\varepsilon^2+\theta^2)} \left(\sqrt{3}\varepsilon + 3\theta, 3\varepsilon-\sqrt{3}\theta \right) = \frac{\lambda_M}{2\sqrt{3}\sqrt{\varepsilon^2+\theta^2}} \left(\sqrt{3}\varepsilon + 3\theta, 3\varepsilon-\sqrt{3}\theta \right) \\
        \boldsymbol{A}_{2}^{\text{mBZ}} &= \frac{a}{\varepsilon^2 + \theta^2} \left( -\varepsilon, \theta \right) = \frac{\lambda_M}{\sqrt{\varepsilon^2 + \theta^2}} \left( -\varepsilon, \theta \right),
    \end{align}
\end{subequations}
where the moiré length is $\lambda_M = a/\sqrt{\theta^2 + \varepsilon^2}$. The initial (undistorted) moiré lattice sites are then
\begin{align}
    \boldsymbol{R} &= m \boldsymbol{A}_{1}^{\text{mBZ}} + n \boldsymbol{A}_{2}^{\text{mBZ}},
\end{align}
where $m,n\in\mathbb{Z}$. We distort this lattice by applying a displacement $\boldsymbol{u}(\boldsymbol{r})$, such that it can be expressed using new basis vectors $\tilde{\boldsymbol{A}}_{1}^{\text{mBZ}}$ and $\tilde{\boldsymbol{A}}_{2}^{\text{mBZ}}$ as 
\begin{align}
    \boldsymbol{R} &= m \boldsymbol{A}_1^{\text{mBZ}} + n \boldsymbol{A}_2^{\text{mBZ}} + \boldsymbol{u}(\boldsymbol{r}) = m \tilde{\boldsymbol{A}}_1^{\text{mBZ}} + n \tilde{\boldsymbol{A}}_2^{\text{mBZ}}.
\end{align}
We consider deformations for which the new basis vectors can be written as
\begin{subequations}
\begin{align}
    \tilde{\boldsymbol{A}}_1^{\text{mBZ}} &= a' \eta' R[\phi] \boldsymbol{A}_{1}^{\text{mBZ}}\\
    \tilde{\boldsymbol{A}}_2^{\text{mBZ}} &= a' \boldsymbol{A}_{2}^{\text{mBZ}},
\end{align}
\end{subequations}
where 
\begin{align}
    R[\phi] &= \mqty(\cos\phi & -\sin\phi \\ \sin\phi & \cos\phi).
\end{align}

\begin{figure}
\includegraphics[width=1.00\linewidth]{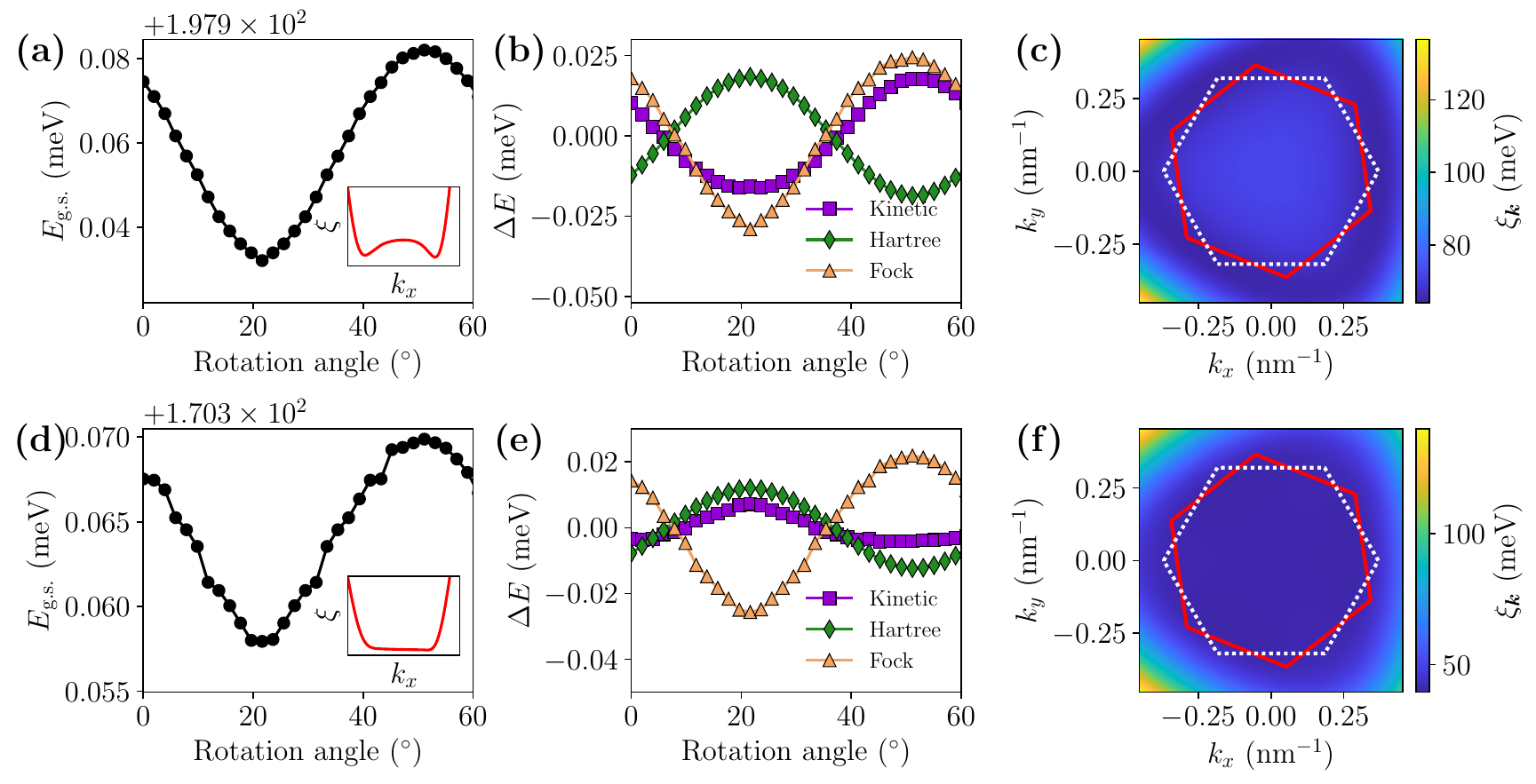}
\caption{(a) Evolution of triangular lattice AHC ground state energy per conduction electron and (b) corresponding kinetic, Hartree and Fock terms as a function of rotation angle starting from the moiré potential orientation in a strong displacement field $U_d=-36$~meV. A similar evolution of the (a) total, (b) kinetic, Hartree, and Fock energies in a weak displacement field $U_d=-20$~meV. Free dispersion of the first conduction band in (c) strong ($U_d=-36$~meV) and (f) weak ($U_d=-20$~meV) displacement fields. The initial mBZ (full red line) and mBZ for the AHC in the optimal orientation (dotted white line) are illustrated in both cases. A line cut of the free conduction band dispersion is also shown as an inset in panels (a) and (d)  for the strong and weak displacement field cases, respectively. Results are for $\epsilon=8.07$ with $n_1=23$ and $n_{\text{bands}}=7$. \label{si_fig:rotation_angle_sweep_data_ahc}}
\end{figure}

Let us first consider a shear deformation along the $\boldsymbol{A}_2^{\text{mBZ}}$ direction. In this case,
\begin{align}
    \boldsymbol{u}(\boldsymbol{r}) &= u_s \left(\boldsymbol{r}\cdot\boldsymbol{A}_2^{\text{mBZ},\perp}\right) \frac{\boldsymbol{A}_{2}^{\text{mBZ}}}{|\boldsymbol{A}_{2}^{\text{mBZ}}|},
\end{align}
where 
\begin{align}
    \frac{\boldsymbol{A}_{2}^{\text{mBZ}}}{|\boldsymbol{A}_{2}^{\text{mBZ}}|} =  \left( \frac{-\varepsilon}{\sqrt{\varepsilon^2 + \theta^2}} , \frac{\theta}{\sqrt{\varepsilon^2 + \theta^2}} \right)
\end{align}
and 
\begin{align}
    \boldsymbol{A}_2^{\text{mBZ},\perp} =  \left( \frac{\theta}{\sqrt{\varepsilon^2 + \theta^2}} , \frac{\varepsilon}{\sqrt{\varepsilon^2 + \theta^2}} \right)
\end{align}
is a normalized vector perpendicular to $\boldsymbol{A}_2^{\text{mBZ}}$ (i.e., $\boldsymbol{A}_2^{\text{mBZ}} \cdot \boldsymbol{A}_2^{\text{mBZ},\perp} = 0$). Such a deformation can be parametrized by
\begin{subequations}
    \begin{align}
        a' &= 1 \\
        \eta' &= \sqrt{1 - \frac{\sqrt{3}}{2} u_s + \frac{3}{4} u_s^2} \\
        \phi &= \text{atan2} \left( \frac{4}{\sqrt{3}} - u_s, \sqrt{3} u_s\right).
    \end{align}
\end{subequations}
Similarly, for an area-preserving dilation
\begin{subequations}
    \begin{align}
        \tilde{\boldsymbol{A}}_1^{\text{mBZ}} &= (1+u_d) \boldsymbol{A}_1^{\text{MBZ}} \\
        \tilde{\boldsymbol{A}}_2^{\text{mBZ}} &= (1+u_d)^{-1} \boldsymbol{A}_2^{\text{MBZ}},
    \end{align}
\end{subequations}
we have 
\begin{subequations}
    \begin{align}
        a' &= (1+u_d)^{-1} \\
        \eta' &= (1+u_d)^{2} \\
        \phi &= 0.
    \end{align}
\end{subequations}

\begin{figure}
\includegraphics[width=1.00\linewidth]{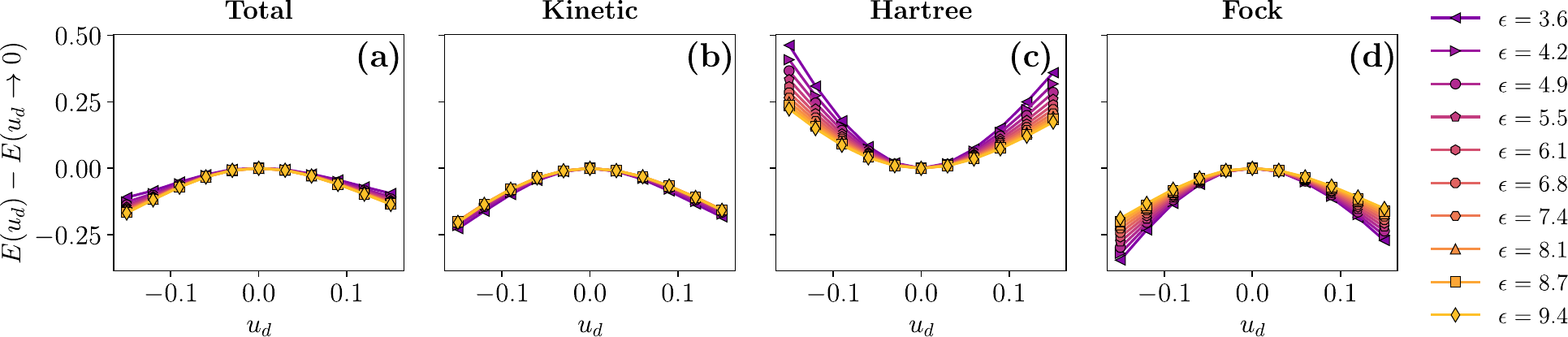}
\caption{Variation of the (a) total, (b) kinetic, (c) Hartree, and (d) Fock energy per conduction electrons in R5G as a function of the distortion strength for area-preserving dilations. Results are obtained for $n_{\text{bands}}=7$ and $n_{1}=23$ with $U_d=-36$~meV. \label{si_fig:energy_R5G_distortion}}
\end{figure}

\subsection{Optimal orientation of the AHC} \label{si_subsec:Multilayer_Rhombohedral_Graphene_optimal_orientation}

The above HF calculations (in Sec.~\ref{si_subsec:Multilayer_Rhombohedral_Graphene_existence_ahc}) assumed that the AHC crystallizes in the same direction as the original moiré lattice (i.e., the AHC lattice is described by the basis vectors $\boldsymbol{A}^{\text{mBZ}}_1$ and $\boldsymbol{A}^{\text{mBZ}}_2$). 
However, when the moiré potential is turned off (or when it is sufficiently weak), there are no \emph{a priori} reasons for this to be true. For instance, if one still assumes a triangular lattice with the same periodicity as the moiré potential, the new AHC lattice could be spanned by the basis vectors $R[\varphi]\boldsymbol{A}^{\text{mBZ}}_1$ and $R[\varphi]\boldsymbol{A}^{\text{mBZ}}_2$, where $\varphi$ is a rotation angle. It is important to note that these different crystallization orientations (i.e., different $\varphi$) are not equivalent considering the $C_3$ symmetry of the dispersion that is induced by the trigonal warping terms in the kinetic Hamiltonian (Fig.~\ref{si_fig:rotation_angle_sweep_data_ahc}(c) and (f)). For instance, Fig.~\ref{si_fig:rotation_angle_sweep_data_ahc}(a) and (d) show the evolution of the triangular lattice AHC energy per conduction electron as a function of the rotation angle starting from the moiré potential aligned configuration in a strong ($U_{d}=-36$~meV) and weak ($U_{d}=-20$~meV) displacement fields, respectively. The $|\mathcal{C}|=1$ AHC crystallizes in the same direction in both cases as determined by the Fock term and its dominant variation (see Fig.~\ref{si_fig:rotation_angle_sweep_data_ahc}(b) and (e)). It should be noted that although the rotation angle minimizing the total energy also minimizes the kinetic energy in the strong displacement field case (Fig.~\ref{si_fig:rotation_angle_sweep_data_ahc}(b)), it does not in the weak field case (Fig.~\ref{si_fig:rotation_angle_sweep_data_ahc}(e)). This stems from the presence of local minima in the free dispersion for $U_{d}=-36$~meV, compared to a flat free dispersion at $U_{d}=-20$~meV (see Fig.~\ref{si_fig:rotation_angle_sweep_data_ahc}(c) and (d) and insets of panels (a) and (d)).

\subsection{Elastic properties of the AHC} \label{si_subsec:Multilayer_Rhombohedral_Graphene_elastic_properties}

\begin{figure}
\includegraphics[width=1.00\linewidth]{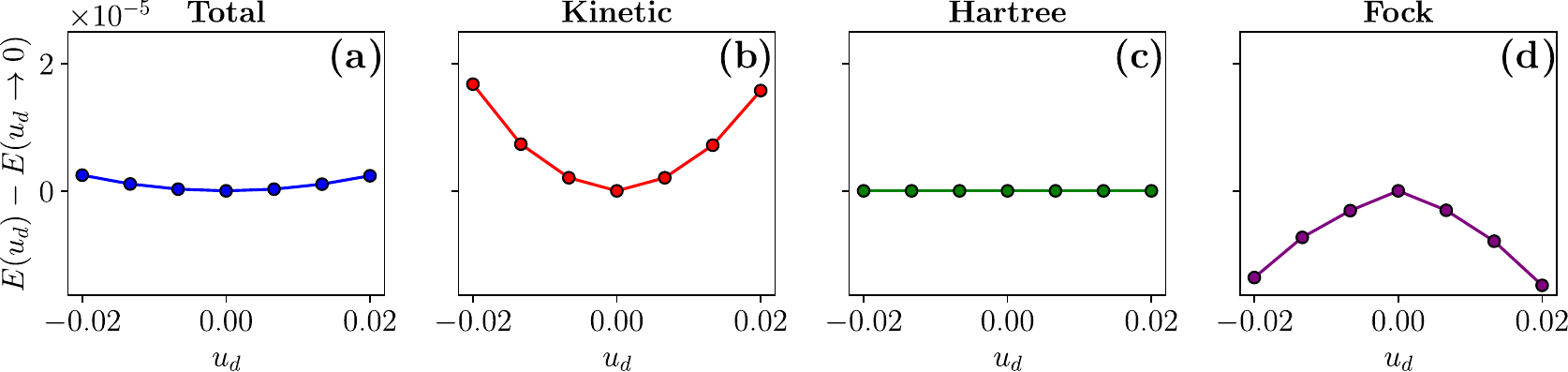}
\caption{Variation of the (a) total, (b) kinetic, (c) Hartree, and (d) Fock energy per electron in the parent band model as a function of the distortion strength for area-preserving dilations. Results are obtained for $V_{c}/A_{\text{u.c.}}=2.42$ by extrapolating the ansatz energy to the infinite system size limit. \label{si_fig:energy_parent_band_distortion}}
\end{figure}

To evaluate the AHC stiffness in R5G, we start from the triangular lattice orientation with minimal energy (Sec.~\ref{si_subsec:Multilayer_Rhombohedral_Graphene_optimal_orientation}). From this configuration, we apply shear and dilation deformations {with strengths in the range $u_s\in[-0.25,0.25]$ and $u_d\in[-0.15,0.15]$, respectively. We then fit the ground state energy variation to a second-order polynomial to extract the second-order derivative. The energy variation from which the stiffness reported in the main text has been deduced is shown in Fig.~\ref{si_fig:energy_R5G_distortion}. One can observe the negative shear and dilation stiffnesses (Fig.~\ref{si_fig:energy_R5G_distortion}(a)), indicating the presence of a mechanical instability. The variation of the different energy contributions should also be noted: the Fock and kinetic energies have a negative concavity, whereas the Hartree term has a positive concavity. This should be contrasted with the stable triangular lattice AHC found in the parent band model shown in Fig.~\ref{si_fig:energy_parent_band_distortion}. There, the kinetic and Fock terms are concave up and down, respectively (the Hartree term variation is negligible). By comparison, the triangular lattice instability of the AHC in R5G then appears to be driven by the kinetic energy. Indeed, despite a much larger Hartree energy variation that favors the stability of the triangular lattice, the triangular AHC is unstable in R5G and not in the parent band because of this opposite concavity in the kinetic energy variation.

Let us try to develop a simple conceptual understanding of these different behaviors. The upward concavity of the Hartree term in R5G is relatively simple to explain. It is well-established that a triangular network of charges minimizes the electrostatic energy in two dimensions. The Hartree energy will then increase as we distort the lattice, starting from the most stable triangular configuration. 

The upward variation of the kinetic energy in the parent band as a function of distortion can also be relatively easily understood. The kinetic energy of a fermionic system with a quadratic dispersion $|\boldsymbol{k}|^2/(2m)$ is minimized by filling the lowest kinetic energy states to obtain a rotationally invariant Fermi surface in momentum space. If, instead, one fills the first Brillouin zone of a two-dimensional lattice, the lattice with the minimal kinetic energy is the triangular lattice since its $D_{6}$ symmetric first Brillouin zone is the one that most closely approaches a circularly symmetric Fermi surface with the same density. The kinetic energy will then increase as the triangular lattice is deformed. Of course, this is a much-simplified argument since the diagonal part of the density matrix ultimately enters the calculation of the kinetic energy in HF. The momentum space occupation is not just a simple filling of the triangular lattice first Brillouin zone but extends much beyond that. However, this simple intuition should still apply since the momentum space occupation will remain invariant under the $C_6$ point group operations and should thus (assuming similar spreads of the momentum space occupation at a given interaction strength for different lattices) approximate the most closely a rotationally invariant disk that minimizes the kinetic energy. For R5G in a strong displacement field, this intuition does not hold anymore since the dispersion has local minima (see Fig.~\ref{si_fig:rotation_angle_sweep_data_ahc}(c)). A distorted triangular lattice that more heavily populates these local minima may then be more energetically favorable from a kinetic standpoint. 

The above intuition then suggests that the triangular lattice AHC in R5G may be made stable by reducing the displacement field to remove the local minima in the free dispersion (Fig.~\ref{si_fig:rotation_angle_sweep_data_ahc}(f)). We substantiate this intuition by performing a similar analysis of the triangular AHC in R5G with $U_d=-20$~meV. Fig.~\ref{si_fig:energy_R5G_distortion_weak_field} shows that, indeed, the triangular lattice is now mechanically stable and that the kinetic energy has an upward concavity.

\begin{figure}
\includegraphics[width=1.00\linewidth]{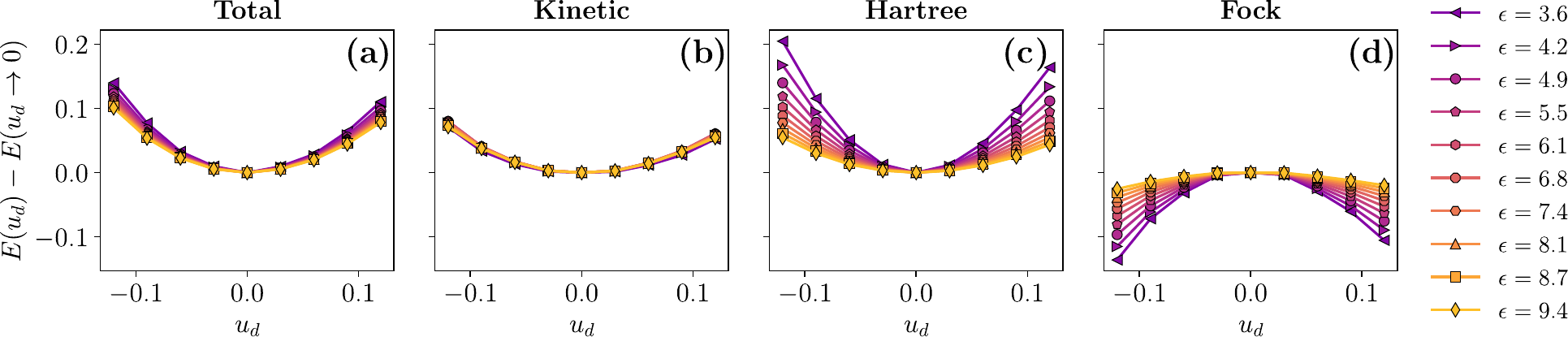}
\caption{Variation of the (a) total, (b) kinetic, (c) Hartree, and (d) Fock energy per particle in R5G as a function of the distortion strength for area-preserving dilations. Results are obtained for $n_{\text{bands}}=7$ and $n_{1}=23$ with $U_d=-20$~meV. \label{si_fig:energy_R5G_distortion_weak_field}}
\end{figure}
\bibliography{apssamp}